# Security Thresholds of Multicarrier Continuous-Variable Quantum Key Distribution

Laszlo Gyongyosi

[1] Quantum Technologies Laboratory, Department of Telecommunications
*Budapest University of Technology and Economics*
2 Magyar tudosok krt, Budapest, *H*-1117, Hungary
[2] MTA-BME Information Systems Research Group
*Hungarian Academy of Sciences*
7 Nador st., Budapest, *H*-1051, Hungary

gyongyosi@hit.bme.hu

**Abstract**

We prove the secret key rate formulas and derive security threshold parameters of multicarrier continuous-variable quantum key distribution (CVQKD). In a multicarrier CVQKD scenario, the Gaussian input quantum states of the legal parties are granulated into Gaussian subcarrier CVs (continuous-variables). The multicarrier communication formulates Gaussian sub-channels from the physical quantum channel, each dedicated to the transmission of a subcarrier CV. The Gaussian subcarriers are decoded by a unitary CV operation, which results in the recovered single-carrier Gaussian CVs. We derive the formulas through the AMQD (adaptive multicarrier quadrature division) scheme, the SVD-assisted (singular value decomposition) AMQD, and the multiuser AMQD-MQA (multiuser quadrature allocation). We prove that the multicarrier CVQKD leads to improved secret key rates and higher tolerable excess noise in comparison to single-carrier CVQKD. We derive the private classical capacity of a Gaussian sub-channel and the security parameters of an optimal Gaussian collective attack in the multicarrier setting. We reveal the secret key rate formulas for one-way and two-way multicarrier CVQKD protocols, assuming homodyne and heterodyne measurements and direct and reverse reconciliation. The results reveal the physical boundaries of physically allowed Gaussian attacks in a multicarrier CVQKD scenario and confirm that the improved transmission rates lead to enhanced secret key rates and security thresholds.

# 1 Introduction

The continuous-variable quantum key distribution (CVQKD) allows for legal parties to transmit information with unconditional security over the currently established telecommunication networks [1–20]. The CVQKD systems, in contrast to discrete variable (DV) QKD protocols, do not require single-photon sources and can be implemented by standard devices and modulation techniques, allowing an efficient signal processing in practical scenarios. The CVQKD schemes in general are based on Gaussian modulation, which is a well-applicable practical finding in the experiment. In CVQKD, the information is carried via a Gaussian-modulated position and momentum quadratures in the phase space. The Gaussian quantum states (referred to as *single-carriers* throughout) are sent through a noisy quantum channel by the sender, Alice. The quantum channel is attacked by an eavesdropper (Eve), and the receiver (Bob) gets a noisy system. Since the optimal attack against CVQKD is a Gaussian attack [2–3, 17–19], the noise of the quantum channel can be provably modeled as an additive white Gaussian noise. The security of CVQKD has been already proven against several of the most powerful optimal Gaussian collective attacks, where the eavesdropper is allowed to use quantum memory and to perform a collective (joint) measurement on her quantum register at the end of the protocol run.

Besides the attractive properties of CVQKD, in comparison with traditional telecommunication techniques [21–25], the efficiency of the protocol still requires several improvements, both in secret key rates and the maximum tolerable excess noise. In particular, with this in mind the *multicarrier* CVQKD has been recently introduced, and the AMQD (adaptive multicarrier quadrature division) scheme has been defined [4]. Specifically, while in the standard CVQKD the information is transmitted by single-carriers, in an AMQD modulation the information is granulated into Gaussian subcarrier CVs. These subcarriers divide the physical quantum link into several Gaussian sub-channels, leading to an improved transmission rate and higher tolerable excess noise overall. An important property of AMQD is that it significantly improves the transmission rates in the low-SNR regimes, which is specifically crucial in experimental CVQKD scenarios. The AMQD modulation has been extended to SVD-assisted (singular value decomposition) AMQD [6], where the SVD of the Gaussian quantum channel derives the eigenchannels of the link. Precisely, in the SVD-AMQD, the transmit information of the users are pre-coded by a corresponding unitary operation in the encoding phase, which are then sent through the eigenchannels of the Gaussian channel. The received quantum states are decoded by the inverse post-unitary operation. The SVD-assistance leads to further improved transmission rates and the maximum tolerable excess noise, which can also be exploited in multiuser CVQKD scenarios. The benefits of AMQD can also be extended to a multiuser environment. The AMQD-MQA (multiuser quadrature allocation) mechanism [5] is based on the AMQD modulation, more precisely on the sophisticated allocation mechanism of the Gaussian subcarrier CVs. In an AMQD-MQA setting, the legal parties use a common, shared physical Gaussian link in parallel. In particular, the multicarrier transmission improves the multiuser transmission rates, leading to an enhanced simultaneous communication through the Gaussian sub-channels. Here we show that the improved classical information trans-mission rates allow for the parties to establish higher secret key rates, and to extend the boundaries on the maximal amount of tolerable excess noise of single-carrier CVQKD.

We derive the *security thresholds* and prove the *secret key rate* formulas of multicarrier CVQKD. We demonstrate the results for one-way and two-way CVQKD, homodyne and heterodyne measurements, and direct and reverse reconciliation. The multicarrier CVQKD transmission is analyzed through the AMQD, the SVD-assisted AMQD, and the multiuser AMQD-MQA scheme. We study the impacts of optimal Gaussian attacks on multicarrier CVQKD and determine the *private classical capacity* [22, 26–28] of a Gaussian sub-channel. The results reveal that the improved transmission rates of multicarrier modulation lead to improved secret key rates, higher tolerable excess noise, and better security thresholds. We show that the enhanced secret key rates can be extended to a multiuser setting, allowing the legal parties a reliable simultaneous private communication over a noisy Gaussian link. For the multiuser scenario of multicarrier CVQKD, we define the private classical capacity *regions* of the users and derive the *sum* and *symmetric* private classical capacities.

This paper is organized as follows. In Section 2, the preliminary findings are summarized. Section 3 proposes the proofs of the secret key rate formulas and the security thresholds of multicarrier CVQKD. Section 4 discusses the physical boundaries of optimal Gaussian attacks. Finally, Section 5 concludes the results. Supplementary information is included in the Appendix.

## 2 Preliminaries

First we summarize briefly the notations and basic terms. For further information, see the detailed descriptions of [4–6].

### 2.1 Basic Terms and Definitions

#### 2.1.1 Multicarrier CVQKD

First we summarize the basic notations of AMQD from [4]. The following description assumes a single user, and the use of $n$ Gaussian sub-channels $\mathcal{N}_i$ for the transmission of the subcarriers, from which only $l$ sub-channels will carry valuable information.

In the single-carrier modulation scheme, the $j$-th input single-carrier state $\left|\varphi_j\right\rangle = \left|x_j + \mathrm{i} p_j\right\rangle$ is a Gaussian state in the phase space $\mathcal{S}$, with i.i.d. Gaussian random position and momentum quadratures $x_j \in \mathbb{N}\left(0, \sigma_{\omega_0}^2\right)$, $p_j \in \mathbb{N}\left(0, \sigma_{\omega_0}^2\right)$, where $\sigma_{\omega_0}^2$ is the modulation variance of the quadratures. (For simplicity, $\sigma_{\omega_0}^2$ is referred to as the single-carrier modulation variance, throughout.) Particularly, this Gaussian single-carrier is transmitted through a Gaussian quantum channel $\mathcal{N}$. In the multicarrier scenario, the information is carried by Gaussian subcarrier CVs, $\left|\phi_i\right\rangle = \left|x_i + \mathrm{i} p_i\right\rangle$, $x_i \in \mathbb{N}\left(0, \sigma_\omega^2\right)$, $p_i \in \mathbb{N}\left(0, \sigma_\omega^2\right)$, where $\sigma_\omega^2$ is the modulation variance of the subcarrier quadratures, which are transmitted through a noisy Gaussian sub-channel $\mathcal{N}_i$. Pre-

cisely, each $\mathcal{N}_i$ Gaussian sub-channel is dedicated for the transmission of one Gaussian subcarrier CV from the $n$ subcarrier CVs. (*Note*: index $l$ refers to the subcarriers, while index $j$, to the single-carriers, throughout the manuscript.) The single-carrier state $|\varphi_j\rangle$ in the phase space $\mathcal{S}$ can be modeled as a zero-mean, circular symmetric complex Gaussian random variable $z_j \in \mathcal{CN}\left(0, \sigma_{\omega_{z_j}}^2\right)$, with variance $\sigma_{\omega_{z_j}}^2 = \mathbb{E}\left[|z_j|^2\right]$, and with i.i.d. real and imaginary zero-mean Gaussian random components $\mathrm{Re}(z_j) \in \mathbb{N}\left(0, \sigma_{\omega_0}^2\right)$, $\mathrm{Im}(z_j) \in \mathbb{N}\left(0, \sigma_{\omega_0}^2\right)$.

In the multicarrier CVQKD scenario, let $n$ be the number of Alice's input single-carrier Gaussian states. Precisely, the $n$ input coherent states are modeled by an $n$-dimensional, zero-mean, circular symmetric complex random Gaussian vector

$$\mathbf{z} = \mathbf{x} + \mathrm{i}\mathbf{p} = (z_1, \ldots, z_n)^T \in \mathcal{CN}(0, \mathbf{K_z}), \tag{1}$$

where each $z_j$ can be modeled as a zero-mean, circular symmetric complex Gaussian random variable

$$z_j \in \mathcal{CN}\left(0, \sigma_{\omega_{z_j}}^2\right), \; z_j = x_j + \mathrm{i}p_j. \tag{2}$$

Specifically, the real and imaginary variables (i.e., the position and momentum quadratures) formulate $n$-dimensional real Gaussian random vectors, $\mathbf{x} = (x_1, \ldots, x_n)^T$ and $\mathbf{p} = (p_1, \ldots, p_n)^T$, with zero-mean Gaussian random variables

$$f(x_j) = \frac{1}{\sigma_{\omega_0}\sqrt{2\pi}} e^{\frac{-x_j^2}{2\sigma_{\omega_0}^2}}, \; f(p_j) = \frac{1}{\sigma_{\omega_0}\sqrt{2\pi}} e^{\frac{-p_j^2}{2\sigma_{\omega_0}^2}}, \tag{3}$$

where $\mathbf{K_z}$ is the $n \times n$ Hermitian covariance matrix of $\mathbf{z}$:

$$\mathbf{K_z} = \mathbb{E}\left[\mathbf{z}\mathbf{z}^\dagger\right], \tag{4}$$

while $\mathbf{z}^\dagger$ is the adjoint of $\mathbf{z}$.
For vector $\mathbf{z}$,

$$\mathbb{E}[\mathbf{z}] = \mathbb{E}\left[e^{\mathrm{i}\gamma}\mathbf{z}\right] = \mathbb{E}e^{\mathrm{i}\gamma}[\mathbf{z}] \tag{5}$$

holds, and

$$\mathbb{E}\left[\mathbf{z}\mathbf{z}^T\right] = \mathbb{E}\left[e^{\mathrm{i}\gamma}\mathbf{z}\left(e^{\mathrm{i}\gamma}\mathbf{z}\right)^T\right] = \mathbb{E}e^{\mathrm{i}2\gamma}\left[\mathbf{z}\mathbf{z}^T\right], \tag{6}$$

for any $\gamma \in [0, 2\pi]$. The density of $\mathbf{z}$ is as follows (if $\mathbf{K_z}$ is invertible):

$$f(\mathbf{z}) = \frac{1}{\pi^n \det \mathbf{K_z}} e^{-\mathbf{z}^\dagger \mathbf{K_z}^{-1} \mathbf{z}}. \tag{7}$$

A $n$-dimensional Gaussian random vector is expressed as $\mathbf{x} = \mathbf{As}$, where $\mathbf{A}$ is an (invertible)

linear transform from $\mathbb{R}^n$ to $\mathbb{R}^n$, and $\mathbf{s}$ is an $n$-dimensional standard Gaussian random vector $\mathbb{N}(0,1)_n$. This vector is characterized by its covariance matrix $\mathbf{K}_\mathbf{x} = \mathbb{E}\left[\mathbf{x}\mathbf{x}^T\right] = \mathbf{A}\mathbf{A}^T$, as

$$\mathbf{x} = \frac{1}{\left(\sqrt{2\pi}\right)^n \sqrt{\det\left(\mathbf{A}\mathbf{A}^T\right)}} e^{-\frac{\mathbf{x}^T \mathbf{x}}{2\left(\mathbf{A}\mathbf{A}^T\right)}}. \tag{8}$$

The Fourier transformation $F(\cdot)$ of the $n$-dimensional Gaussian random vector $\mathbf{v} = \left(v_1, \ldots, v_n\right)^T$ results in the $n$-dimensional Gaussian random vector $\mathbf{m} = \left(m_1, \ldots, m_n\right)^T$, precisely:

$$\mathbf{m} = F(\mathbf{v}) = e^{\frac{-\mathbf{m}^T \mathbf{A}\mathbf{A}^T \mathbf{m}}{2}} = e^{\frac{-\sigma_{\omega_0}^2 \left(m_1^2 + \ldots + m_n^2\right)}{2}}. \tag{9}$$

In the first step of AMQD, Alice applies the inverse FFT (fast Fourier transform) operation to vector $\mathbf{z}$ (see (1)), which results in an $n$-dimensional zero-mean, circular symmetric complex Gaussian random vector $\mathbf{d}$, $\mathbf{d} \in \mathcal{CN}(0, \mathbf{K}_\mathbf{d})$, $\mathbf{d} = \left(d_1, \ldots, d_n\right)^T$, precisely as

$$\mathbf{d} = F^{-1}(\mathbf{z}) = e^{\frac{\mathbf{d}^T \mathbf{A}\mathbf{A}^T \mathbf{d}}{2}} = e^{\frac{\sigma_{\omega_0}^2 \left(d_1^2 + \ldots + d_n^2\right)}{2}}, \tag{10}$$

where

$$d_i = x_{d_i} + \mathrm{i} p_{d_i}, \ d_i \in \mathcal{CN}\left(0, \sigma_{\omega_{d_i}}^2\right), \tag{11}$$

where $\sigma_{\omega_{d_i}}^2 = \mathbb{E}\left[\left|d_i\right|^2\right]$ and the position and momentum quadratures of $\left|\phi_i\right\rangle$ are i.i.d. Gaussian random variables

$$\mathrm{Re}(d_i) = x_{d_i} \in \mathbb{N}\left(0, \sigma_{\omega_i}^2\right), \ \mathrm{Im}(d_i) = p_{d_i} \in \mathbb{N}\left(0, \sigma_{\omega_i}^2\right), \tag{12}$$

where $\mathbf{K}_\mathbf{d} = \mathbb{E}\left[\mathbf{d}\mathbf{d}^\dagger\right]$, $\mathbb{E}[\mathbf{d}] = \mathbb{E}\left[e^{\mathrm{i}\gamma}\mathbf{d}\right] = \mathbb{E}e^{\mathrm{i}\gamma}[\mathbf{d}]$, and $\mathbb{E}\left[\mathbf{d}\mathbf{d}^T\right] = \mathbb{E}\left[e^{\mathrm{i}\gamma}\mathbf{d}\left(e^{\mathrm{i}\gamma}\mathbf{d}\right)^T\right] = \mathbb{E}e^{\mathrm{i}2\gamma}\left[\mathbf{d}\mathbf{d}^T\right]$ for any $\gamma \in [0, 2\pi]$.

The $\mathbf{T}(\mathcal{N})$ transmittance vector of $\mathcal{N}$ in the multicarrier transmission is

$$\mathbf{T}(\mathcal{N}) = \left[T_1(\mathcal{N}_1), \ldots, T_n(\mathcal{N}_n)\right]^T \in \mathcal{C}^n, \tag{13}$$

where

$$T_i(\mathcal{N}_i) = \mathrm{Re}\left(T_i(\mathcal{N}_i)\right) + \mathrm{i}\,\mathrm{Im}\left(T_i(\mathcal{N}_i)\right) \in \mathcal{C}, \tag{14}$$

is a complex variable, which quantifies the position and momentum quadrature transmission (i.e., gain) of the $i$-th Gaussian sub-channel $\mathcal{N}_i$, in the phase space $\mathcal{S}$, with real and imaginary parts

$$0 \leq \mathrm{Re}\,T_i(\mathcal{N}_i) \leq 1/\sqrt{2}, \text{ and } 0 \leq \mathrm{Im}\,T_i(\mathcal{N}_i) \leq 1/\sqrt{2}. \tag{15}$$

Particularly, the $T_i(\mathcal{N}_i)$ variable has the squared magnitude of

$$\left|T_i\left(\mathcal{N}_i\right)\right|^2 = \operatorname{Re} T_i\left(\mathcal{N}_i\right)^2 + \operatorname{Im} T_i\left(\mathcal{N}_i\right)^2 \in \mathbb{R}, \tag{16}$$

where

$$\operatorname{Re} T_i\left(\mathcal{N}_i\right) = \operatorname{Im} T_i\left(\mathcal{N}_i\right). \tag{17}$$

The Fourier-transformed transmittance of the $i$-th sub-channel $\mathcal{N}_i$ (resulted from CVQFT operation at Bob) is denoted by

$$\left|F\left(T_i\left(\mathcal{N}_i\right)\right)\right|^2. \tag{18}$$

The $n$-dimensional zero-mean, circular symmetric complex Gaussian noise vector $\Delta \in \mathcal{CN}\left(0, \sigma_\Delta^2\right)_n$ of the quantum channel $\mathcal{N}$, is evaluated as

$$\Delta = \left(\Delta_1, \ldots, \Delta_n\right)^T \in \mathcal{CN}\left(0, \mathbf{K}_\Delta\right), \tag{19}$$

where

$$\mathbf{K}_\Delta = \mathbb{E}\left[\Delta \Delta^\dagger\right], \tag{20}$$

with independent, zero-mean Gaussian random components

$$\Delta_{x_i} \in \mathbb{N}\left(0, \sigma_{\mathcal{N}_i}^2\right), \text{ and } \Delta_{p_i} \in \mathbb{N}\left(0, \sigma_{\mathcal{N}_i}^2\right), \tag{21}$$

with variance $\sigma_{\mathcal{N}_i}^2$, for each $\Delta_i$ of a Gaussian sub-channel $\mathcal{N}_i$, which identifies the Gaussian noise of the $i$-th sub-channel $\mathcal{N}_i$ on the quadrature components in the phase space $\mathcal{S}$.

The CVQFT-transformed noise vector can be rewritten as

$$F\left(\Delta\right) = \left(F\left(\Delta_1\right), \ldots, F\left(\Delta_n\right)\right)^T, \tag{22}$$

with independent components $F\left(\Delta_{x_i}\right) \in \mathbb{N}\left(0, \sigma_{F(\mathcal{N}_i)}^2\right)$ and $F\left(\Delta_{p_i}\right) \in \mathbb{N}\left(0, \sigma_{F(\mathcal{N}_i)}^2\right)$ on the quadratures, for each $F\left(\Delta_i\right)$. Precisely, it also defines an $n$-dimensional zero-mean, circular symmetric complex Gaussian random vector $F\left(\Delta\right) \in \mathcal{CN}\left(0, \mathbf{K}_{F(\Delta)}\right)$ with a covariance matrix

$$\mathbf{K}_{F(\Delta)} = \mathbb{E}\left[F\left(\Delta\right) F\left(\Delta\right)^\dagger\right], \tag{23}$$

where $\mathbf{K}_{F(\Delta)} = \mathbf{K}_\Delta$, by theory.

At a constant subcarrier modulation variance $\sigma_{\omega_i}^2$ for the $n$ Gaussian subcarrier CVs, the corresponding relation is

$$\tfrac{1}{n}\sum_{i=1}^{n} \sigma_{\omega_i}^2 = \sigma_\omega^2, \tag{24}$$

where $\sigma_{\omega_i}^2$ is the modulation variance of the quadratures of the subcarrier $\left|\phi_i\right\rangle$ transmitted by sub-channel $\mathcal{N}_i$. Assuming $l$ *good* Gaussian sub-channels from the $n$ with *constant* quadrature

modulation variance $\sigma_{\omega_i}^2$, where $\sigma_{\omega_i}^2 = 0$ for the $i$-th unused sub-channel,

$$\sum_{i=1}^{l} \sigma_{\omega_i}^2 = l\sigma_\omega^2 < n\sigma_{\omega_0}^2. \tag{25}$$

In particular, from the relation of (25), for the transmittance parameters the following relation follows at a given modulation variance $\sigma_{\omega_0}^2$, precisely,

$$|T_{AMQD}(\mathcal{N}_i)|^2 \sigma_{\omega_0}^2 > |T(\mathcal{N})|^2 \sigma_{\omega_0}^2, \tag{26}$$

where $|T(\mathcal{N})|^2$ is the transmittance of $\mathcal{N}$ in a single-carrier scenario, and

$$|T_{AMQD}(\mathcal{N}_i)|^2 = \tfrac{1}{l}\sum_{i=1}^{l} |F(T_i(\mathcal{N}_i))|^2. \tag{27}$$

For the method of the determination of these $l$ Gaussian sub-channels, see [4]. Alice's $i$-th Gaussian subcarrier is precisely expressed as follows:

$$|\phi_i\rangle = |d_i\rangle = |F^{-1}(z)\rangle. \tag{28}$$

Specifically, the result of (10) defines $n$, independent $\mathcal{N}_i$ Gaussian sub-channels, each with noise variance $\sigma_{\mathcal{N}_i}^2$, one for each subcarrier coherent state $|\phi_i\rangle$. After the CV subcarriers are transmitted through the noisy channel, Bob applies the CVQFT (continuous-variable quantum Fourier transform) unitary operation, which gives him the noisy version $|\varphi_j'\rangle = |z_j'\rangle$ of Alice's single-carrier input $z_j$. For further details and description of AMQD, see [4]. For the extension of the analysis to a multiuser setting, see the derivations of [5]. The description of the SVD-assisted multiuser AMQD CVQKD scheme can be found in [6].

### 2.1.2 Gaussian States

A Gaussian density matrix $\rho_{\mathcal{G}}$ can be characterized by a displacement operator and a correlation matrix $\mathbf{K}_{\rho_{\mathcal{G}}}$. The $\mathbf{K}_{\rho_{\mathcal{G}}}$ can be diagonalized by a corresponding matrix $\mathbf{M}_{\mathcal{S}}$ [2], such that

$$\mathbf{M}_{\mathcal{S}}^T \mathbf{K}_{\rho_{\mathcal{G}}} \mathbf{M}_{\mathcal{S}} = diag(s_1, s_1 ..., s_n, s_n), \tag{29}$$

where $diag(\cdot)$ refers to a diagonal matrix, and $s_i$-s are real elements which formulate the $\mathcal{S}_{\rho_{\mathcal{G}}}$ *symplectic spectra* (symplectic eigenspectrum of $\mathbf{K}_{\rho_{\mathcal{G}}}$) as

$$\mathcal{S}_{\rho_{\mathcal{G}}} = (s_1 ..., s_n). \tag{30}$$

Particularly, a two-mode entangled Gaussian density $\rho_{\mathcal{G}}$ with variance $\sigma_\omega^2$ can be characterized by the $\mathbf{K}_{\rho_{\mathcal{G}}}$ of

$$\mathbf{K}_{\rho_{\mathcal{G}}} = \begin{pmatrix} \sigma_\omega^2 I & \sqrt{\sigma_\omega^4 - 1} Z \\ \sqrt{\sigma_\omega^4 - 1} Z & \sigma_\omega^2 I \end{pmatrix}, \tag{31}$$

where $I$ is the identity matrix, $I = diag(1,1)$, while $Z$ refers to the Pauli $Z$ matrix, $Z = diag(1,-1)$. The $S(\cdot)$ entropy of $\rho_{\mathcal{G}}$ can be expressed as

$$S(\rho_{\mathcal{G}}) = \sum_{i=1}^{n} g(s_i), \tag{32}$$

where

$$g(s_i) = \frac{s_i+1}{2}\log_2 \frac{s_i+1}{2} - \frac{s_i-1}{2}\log_2 \frac{s_i-1}{2}. \tag{33}$$

### 2.1.3 Private Classical Capacity

The $P(\mathcal{N})$ private classical capacity of $\mathcal{N}$ in direct reconciliation (DR, i.e., from Alice to Bob) is defined as

$$P^{DR}(\mathcal{N}) = \max_{\forall \rho_i}(I(A:B) - I(A:E)), \tag{34}$$

where $I(A:B)$ is the mutual information between Alice and Bob, while $I(A:E)$ is the mutual information between Alice and Eve. In the reverse reconciliation (RR, i.e., from Bob to Alice), the corresponding capacity is

$$P^{RR}(\mathcal{N}) = \max_{\forall \rho_i}(I(A:B) - I(B:E)), \tag{35}$$

where $I(B:E)$ is the mutual information between Bob and Eve.

Assuming collective (joint) measurement in the protocol runs on Bob's and Eve's sides (also referred to as *collective CVQKD protocols*), the related DR and RR formulas are

$$P^{DR}(\mathcal{N}) = \max_{\forall \rho_i}(\chi(A:B) - \chi(A:E)) \tag{36}$$

and

$$P^{RR}(\mathcal{N}) = \max_{\forall \rho_i}(\chi(A:B) - I(B:E)), \tag{37}$$

where $\chi$ is the Holevo quantity. We use collective protocols in the evaluation of the secret key rates in Section 3.

Note that for collective protocols, the $I(B:E)$ function causes negative divergence in (37) (*except* in one-way collective CVQKD at heterodyne measurement) [2]; thus these RR rate formulas are derived via

$$P^{RR}(\mathcal{N}) = \max_{\forall \rho_i}(I(A:B) - \chi(B:E)). \tag{38}$$

## 2.2 Optimal Gaussian Attack on Multicarrier CVQKD

First we analyze the correlation of Alice and Bob's quadratures in an AMQD modulation, for a homodyne and heterodyne measurement, $M_{\text{hom}}$ and $M_{\text{het}}$. Then we characterize the correlation measures at an optimal Gaussian attack on the multicarrier transmission.

### 2.2.1 Private Classical Capacity of a Gaussian Sub-Channel

In this section we derive the private classical capacity of a Gaussian sub-channel $\mathcal{N}_i$ for direct and reverse reconciliation. The formal analysis of this section will be completed in Section 3 with the single-carrier formulas, since in an AMQD modulation the subcarriers are converted back to single-carriers at the decoder via the CVQFT operation.

Let $x_i, p_i$ refer to Alice's $i$-th Gaussian subcarrier CV $|\phi_i\rangle$, and $x'_i, p'_i$ to Bob's noisy subcarrier quadratures, $\langle x_i^2 \rangle = \langle p_i^2 \rangle = \sigma_{\omega_i}^2$ and $\langle x'^2_i \rangle = \langle p'^2_i \rangle = |F(T_i(\mathcal{N}_i))|^2 \sigma_{\omega_i}^2 + \sigma_{\mathcal{N}_i}^2$. Following the formalism of [1-2] throughout, Alice's estimators $e_{x'_i}$ and $e_{p'_i}$ on Bob's $x'_i$ and $p'_i$ quadratures are precisely evaluated as

$$e_{x'_i} = \varepsilon_{A,x'_i} x_i, \quad \varepsilon_{A,x'_i} = \frac{\langle x'_i x_i \rangle}{\langle x_i^2 \rangle}, \tag{39}$$

with a conditional variance

$$\hat{\sigma}^2_{x'_i | e_{x'_i}} = \left\langle \left( x'_i - e_{x'_i} \right)^2 \right\rangle = \langle x'^2_i \rangle - \frac{|\langle x_i x'_i \rangle|^2}{\langle x_i^2 \rangle}, \tag{40}$$

and

$$e_{p'_i} = \varepsilon_{A,p'_i} p_i, \quad \varepsilon_{A,p'_i} = \frac{\langle p'_i p_i \rangle}{\langle p_i^2 \rangle}, \tag{41}$$

with a conditional variance

$$\hat{\sigma}^2_{p'_i | e_{p'_i}} = \left\langle \left( p'_i - e_{p'_i} \right)^2 \right\rangle = \langle p'^2_i \rangle - \frac{|\langle p_i p'_i \rangle|^2}{\langle p_i^2 \rangle}. \tag{42}$$

Bob's estimators $e_{x_i}$ and $e_{p_i}$ on Alice's $x_i$ and $p_i$ quadratures are

$$e_{x_i} = \varepsilon_{B,x_i} x'_i, \quad \varepsilon_{B,x_i} = \frac{\langle x_i x'_i \rangle}{\langle x'^2_i \rangle}, \tag{43}$$

$$\hat{\sigma}^2_{x_i | e_{x_i}} = \left\langle \left( x_i - e_{x_i} \right)^2 \right\rangle = \langle x_i^2 \rangle - \frac{|\langle x'_i x_i \rangle|^2}{\langle x'^2_i \rangle}, \tag{44}$$

$$e_{p_i} = \varepsilon_{B,p_i} p'_i, \quad \varepsilon_{B,p_i} = \frac{\langle p_i p'_i \rangle}{\langle p'^2_i \rangle}, \tag{45}$$

$$\hat{\sigma}^2_{p_i | e_{p_i}} = \left\langle \left( p_i - e_{p_i} \right)^2 \right\rangle = \langle p_i^2 \rangle - \frac{|\langle p'_i p_i \rangle|^2}{\langle p'^2_i \rangle}. \tag{46}$$

In particular, the optimal Gaussian attack is performed by an $\mathcal{Cl}_{Eve}$ entangling cloner [1–4]. Using Eve's quadratures $x_{E,i}$ and $p_{E,i}$, Eve's estimators $e^E_{x_i}$ and $e^E_{p_i}$ on Alice's quadratures are as follows [1]:

$$e^E_{x_i} = \varepsilon_{E,x_i} x_{E,i}, \ \varepsilon_{E,x_i} = \frac{\langle x_i x_{E,i} \rangle}{\langle x^2_{E,i} \rangle}, \tag{47}$$

$$\hat{\sigma}^2_{x_i|e^E_{x_i}} = \left\langle \left( x_i - e^E_{x_i} \right)^2 \right\rangle = \langle x^2_i \rangle - \frac{|\langle x_{E,i} x_i \rangle|^2}{\langle x^2_{E,i} \rangle}, \tag{48}$$

$$e^E_{p_i} = \varepsilon_{E,p_i} p_{E,i}, \ \varepsilon_{E,p_i} = \frac{\langle p_i p_{E,i} \rangle}{\langle p^2_{E,i} \rangle}, \tag{49}$$

$$\hat{\sigma}^2_{p_i|e^E_{p_i}} = \left\langle \left( p_i - e^E_{p_i} \right)^2 \right\rangle = \langle p^2_i \rangle - \frac{|\langle p_{E,i} p_i \rangle|^2}{\langle p^2_{E,i} \rangle}. \tag{50}$$

Eve's estimators on Bob's quadratures are referred to as $e^E_{x'_i}, e^E_{p'_i}$, and expressed as:

$$e^E_{x'_i} = \varepsilon_{E,x'_i} x_{E,i}, \ \varepsilon_{E,x'_i} = \frac{\langle x'_i x_{E,i} \rangle}{\langle x^2_{E,i} \rangle}, \tag{51}$$

$$\hat{\sigma}^2_{x'_i|e^E_{x'_i}} = \left\langle \left( x'_i - e^E_{x'_i} \right)^2 \right\rangle = \langle x'^2_i \rangle - \frac{|\langle x_{E,i} x'_i \rangle|^2}{\langle x^2_{E,i} \rangle}, \tag{52}$$

$$e^E_{p'_i} = \varepsilon_{E,p'_i} p_{E,i}, \ \varepsilon_{E,p'_i} = \frac{\langle p'_i p_{E,i} \rangle}{\langle p^2_{E,i} \rangle}, \tag{53}$$

$$\hat{\sigma}^2_{p'_i|e^E_{p'_i}} = \left\langle \left( p'_i - e^E_{p'_i} \right)^2 \right\rangle = \langle p'^2_i \rangle - \frac{|\langle p_{E,i} p'_i \rangle|^2}{\langle p^2_{E,i} \rangle}. \tag{54}$$

Particularly, assuming that Eve attacks each $|\phi_i\rangle$ Gaussian subcarriers via an $\mathcal{Cl}_{Eve}$ entangling cloner with transmittance $T_{Eve,i}$ for the $i$-th subcarrier, the conditional variances on Alice and Bob's estimators are precisely as follows:

$$\sigma^2_{e_{x_i}|e_{x'_i}} = \langle e^2_{x_i} \rangle - \frac{|\langle x_i e_{x_i} \rangle|^2}{\langle x^2_i \rangle}, \tag{55}$$

$$\sigma^2_{e_{p_i}|e_{p'_i}} = \langle e^2_{p_i} \rangle - \frac{|\langle p_i e_{p_i} \rangle|^2}{\langle p^2_i \rangle}. \tag{56}$$

For Eve's quadratures in the RR direction

$$\sigma^2_{e_{x_i}|e^E_{x'_i}} = \langle e^2_{x_i} \rangle - \frac{|\langle x_{E,i} e_{x_i} \rangle|^2}{\langle x^2_{E,i} \rangle}, \tag{57}$$

$$\sigma^2_{e_{p_i}|e^E_{p'_i}} = \langle e^2_{p_i} \rangle - \frac{|\langle p_{E,i} e_{p_i} \rangle|^2}{\langle p^2_{E,i} \rangle}. \tag{58}$$

For the DR direction

$$\sigma^2_{e_{x'_i}|e^E_{x_i}} = \langle e^2_{x'_i} \rangle - \frac{|\langle x_{E,i} e_{x'_i} \rangle|^2}{\langle x^2_{E,i} \rangle}, \tag{59}$$

$$\sigma^2_{e_{p'_i}|e^E_{p_i}} = \langle e^2_{p'_i} \rangle - \frac{|\langle p_{E,i} e_{p'_i} \rangle|^2}{\langle p^2_{E,i} \rangle}. \tag{60}$$

Precisely, in a collective attack, Eve is assumed to use a quantum memory and to perform a joint measurement; thus the corresponding correlation function is the Holevo information $\chi(A:E)$ and $\chi(B:E)$, for the DR and RR cases, respectively. If Bob is also allowed to use collective measurement, then $\chi(A:B)$ quantifies the correlation between Alice and Bob. See Section 3.

The parameters of an entangling cloner attack in AMQD are summarized in Figure 1.

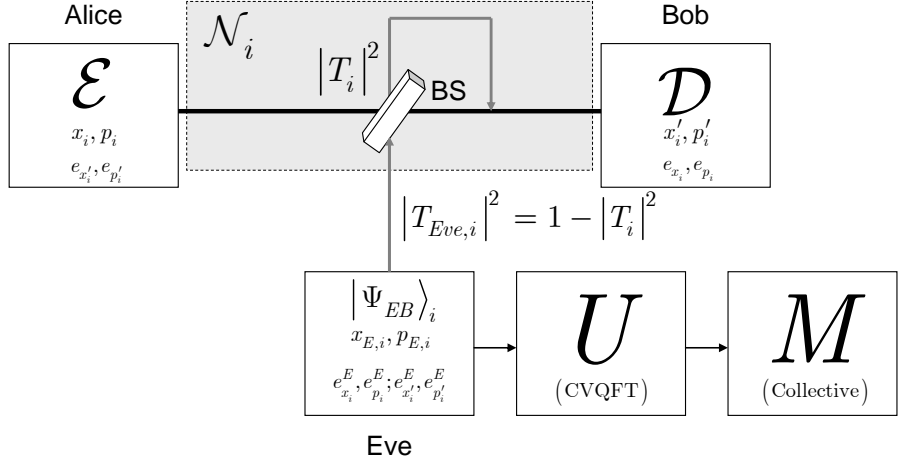

**Figure 1**. The entangling cloner attack in an AMQD modulation. Eve is equipped with a BS with transmittance $|T_i|^2 = 1 - |T_{Eve,i}|^2$ for the attacking of each Gaussian sub-channels $\mathcal{N}_i$. She puts the $l$ Gaussian subcarriers into a quantum memory, then applies the $U$ (CVQFT) unitary operation to recover the single-carrier CV. She then puts the recovered single-carrier Gaussian CV into a quantum register. At the end of the protocol run, she applies an $M$ collective measurement on her register.

For simplicity, in this section we use the functions $\chi(A:B)$, $\chi(B:E)$ and $\chi(A:E)$. From (55) and (56), the $\chi_x(A:B)$ and $\chi_p(A:B)$ between Alice and Bob are

$$\chi_x(A:B) = \tfrac{1}{2}\log_2 \frac{\langle e_{x_i}^2 \rangle}{\sigma^2_{e_{x_i}|e_{x_i'}}},$$
$$\chi_p(A:B) = \tfrac{1}{2}\log_2 \frac{\langle e_{p_i}^2 \rangle}{\sigma^2_{e_{p_i}|e_{p_i'}}}. \tag{61}$$

From (57) and (58), the Holevo information between Bob and Eve is

$$\chi_x(B:E) = \tfrac{1}{2}\log_2 \frac{\langle e_{x_i}^2 \rangle}{\sigma^2_{e_{x_i}|e_{x_i'}^E}} = \tfrac{1}{2}\log_2 \frac{\langle e_{x_i}^2 \rangle}{\langle e_{x_i}^2 \rangle - \frac{|\langle x_{E,i} e_{x_i} \rangle|^2}{\langle (x_{E,i})^2 \rangle}},$$
$$\chi_p(B:E) = \tfrac{1}{2}\log_2 \frac{\langle e_{p_i}^2 \rangle}{\sigma^2_{e_{p_i}|e_{p_i'}^E}} = \tfrac{1}{2}\log_2 \frac{\langle e_{p_i}^2 \rangle}{\langle e_{p_i}^2 \rangle - \frac{|\langle p_{E,i} e_{p_i} \rangle|^2}{\langle (p_{E,i})^2 \rangle}}. \tag{62}$$

While from (59) and (60), the Holevo quantities between Alice and Eve are

$$\chi_x(A:E) = \tfrac{1}{2}\log_2 \frac{\langle e^2_{x'_i}\rangle}{\sigma^2_{e'_{x_i}|e^E_{x_i}}} = \tfrac{1}{2}\log_2 \frac{\langle e^2_{x'_i}\rangle}{\langle e^2_{x'_i}\rangle - \frac{|\langle x_{E,i} e'_{x_i}\rangle|^2}{\langle (x_{E,i})^2\rangle}},$$

$$\chi_p(A:E) = \tfrac{1}{2}\log_2 \frac{\langle e^2_{p'_i}\rangle}{\sigma^2_{e'_{p_i}|e^E_{p_i}}} = \tfrac{1}{2}\log_2 \frac{\langle e^2_{p'_i}\rangle}{\langle e^2_{p'_i}\rangle - \frac{|\langle p_{E,i} e'_{p_i}\rangle|^2}{\langle (p_{E,i})^2\rangle}}. \qquad (63)$$

Specifically, the reverse reconciliation $S^{RR}_x$, $S^{RR}_p$ secret key rates from (61) and (62) are as follows:

$$S^{RR}_x = \chi_x(A:B) - \chi_x(B:E), \qquad (64)$$

$$S^{RR}_p = \chi_p(A:B) - \chi_p(B:E). \qquad (65)$$

The $S^{DR}_x$, $S^{DR}_p$ secret key rates in direct reconciliation are expressed from (61) and (63) as

$$S^{DR}_x = \chi_x(A:B) - \chi_x(A:E), \qquad (66)$$

$$S^{DR}_p = \chi_p(A:B) - \chi_p(A:E). \qquad (67)$$

To express these rates, we first define the $P(\mathcal{N}_i)$ private classical capacity of an $\mathcal{N}_i$ Gaussian quantum channel in the AMQD modulation.

In particular, the $\text{SNR}^*_i$ (signal to noise ratio) of the $i$-th Gaussian sub-channel $\mathcal{N}_i$ for the transmission of private classical information (i.e., for the derivation of the secret key rate) under an optimal Gaussian attack is expressed as

$$\text{SNR}^*_i = \frac{\sigma^2_{\omega_i}}{\sigma^2_{\mathcal{N}^*_i}}, \qquad (68)$$

where $\sigma^2_{\mathcal{N}^*_i}$ is precisely evaluated as

$$\sigma^2_{\mathcal{N}^*_i} = \sigma^2_{\omega_i} \left( \frac{\sigma^2_{\omega_i}|F(T_i(\mathcal{N}_i))|^2 + \sigma^2_{X_i}}{1 + \sigma^2_{X_i}\sigma^2_{\omega_i}|F(T_i(\mathcal{N}_i))|^2} - 1 \right)^{-1}, \qquad (69)$$

where

$$\sigma^2_{X_i} = \sigma^2_0 + N_i, \qquad (70)$$

and where $\sigma^2_0$ is the vacuum noise and $N_i$ is the excess noise of the Gaussian sub-channel $\mathcal{N}_i$ defined as

$$N_i = \frac{(W_i - 1)\left(|F(T_{Eve,i})|^2\right)}{1 - |F(T_{Eve,i})|^2}, \qquad (71)$$

where $W_i$ is the variance of Eve's EPR state used for the attacking of $\mathcal{N}_i$, while

$$|T_{Eve,i}|^2 = 1 - |T_i|^2 \qquad (72)$$

is the transmittance of Eve's beam splitter (BS), while $|T_i|^2$ is the transmittance of $\mathcal{N}_i$.

Particularly, from (69) the $P(\mathcal{N}_i)$ (real domain) is expressed as

$$P(\mathcal{N}_i) = \tfrac{1}{2}\log_2\left(1 + \frac{\sigma_{\omega_i}^2 |F(T_i(\mathcal{N}_i))|^2}{\sigma_{\mathcal{N}_i^*}^2}\right) = \tfrac{1}{2}\log_2\left(\frac{\sigma_{\omega_i}^2 |F(T_i(\mathcal{N}_i))|^2 + \sigma_{X_i}^2}{1+\sigma_{X_i}^2\sigma_{\omega_i}^2|F(T_i(\mathcal{N}_i))|^2}\right)$$
$$= \tfrac{1}{2}\log_2\left(\frac{\sigma_{\omega_i}^2|F(T_i(\mathcal{N}_i))|^2 + (\sigma_0^2+N_i)}{1+(\sigma_0^2+N_i)\sigma_{\omega_i}^2|F(T_i(\mathcal{N}_i))|^2}\right). \tag{73}$$

Putting the pieces together, the (real domain) private classical capacities $P^{RR}(\mathcal{N}_i)$ and $P^{DR}(\mathcal{N}_i)$ for a given Gaussian sub-channel $\mathcal{N}_i$ are precisely expressed as

$$P^{RR}(\mathcal{N}_i) = \max_{\forall i}\left\{\tfrac{1}{2}\log_2\frac{\sigma_{\omega_i}^2+\sigma_{X_i}^2}{s+\sigma_{X_i}^2} - \tfrac{1}{2}\log_2\frac{\langle e_{A_i}^2\rangle}{\langle e_{A_i}^2\rangle - \frac{|\langle E_i e_{A_i}\rangle|^2}{\langle (E_i)^2\rangle}}\right\}, \tag{74}$$

where $\langle e_{A_i}^2\rangle$ is Bob's estimator on Alice's system $A_i$, while $E_i$ is Eve's system, and

$$P^{DR}(\mathcal{N}_i) = \max_{\forall i}\left\{\tfrac{1}{2}\log_2\frac{\sigma_{\omega_i}^2+\sigma_{X_i}^2}{s+\sigma_{X_i}^2} - \tfrac{1}{2}\log_2\frac{\langle e_{B_i}^2\rangle}{\langle e_{B_i}^2\rangle - \frac{|\langle E_i e_{B_i}\rangle|^2}{\langle (E_i)^2\rangle}}\right\}, \tag{75}$$

where $\langle e_{B_i}^2\rangle$ is Alice's estimator on Bob's system $B_i$, while $E_i$ is Eve's system. The detailed secret key rate formulas for the AMQD transmission are given in an extended form in Section 3.

## 3 Secret Key Rates of AMQD and AMQD-MQA

In this section, we prove the secret key rates of AMQD [4], SVD-assisted AMQD [6], and AMQD-MQA [5] for collective CVQKD protocols, following the formalisms of [1] and [2]. The secret key rates of AMQD are evaluated for the case of $n$ Gaussian subcarriers and $l$ Gaussian sub-channels with $\nu_i < \nu_{Eve}$. The variance of Eve's EPR ancilla is $W_i$ for the attack of a sub-channel $\mathcal{N}_i$ [2–3]. For the multiuser scenario of AMQD-MQA, $K$ users are assumed for the parallel use of the Gaussian quantum channel. The subcarrier Gaussian CVs of Alice, Bob, and Eve will be referred to as $A_i$, $B_i$, and $E_i$, respectively. The single-carrier Gaussian CVs are referred to as $A_j$, $B_j$, and $E_j$.

**Theorem 1** (AMQD secret key rates). *The AMQD leads to improved secret key rates in a CVQKD scenario in comparison to a single-carrier CVQKD.*

*Proof.*
In the first part of the proof, we reveal the secret key rates of AMQD for one-way CVQKD, assuming homodyne and heterodyne measurements with RR and DR. In the second part, we reveal the secret key rates of AMQD for two-way CVQKD. Finally, we conclude the results with the analysis of the secret key rates of the SVD-assisted AMQD.

## 3.1 AMQD Secret Key Rates in One-Way CVQKD

Let $d_i = x_i + \mathrm{i} p_i$ refer to Alice's $i$-th Gaussian subcarrier CV $|\phi_i\rangle$, where the Gaussian variables $x_i$ and $p_i$ refer to the position and momentum quadratures $\tilde{x}_i$, $\tilde{p}_i$ of the phase-space Gaussian subcarrier CV $|\phi_i\rangle = \tilde{x}_i + \mathrm{i} \tilde{p}_i$. Following the formalism of [2] throughout the analysis, the quadratures particularly can be precisely rewritten as

$$\tilde{x}_i = x_i + \tilde{x}_i \big| x_i \tag{76}$$

and

$$\tilde{p}_i = p_i + \tilde{p}_i \big| p_i. \tag{77}$$

Specifically, the modulation variances of $\tilde{x}_i$ and $\tilde{p}_i$ are referred to as $\sigma^2_{\tilde{x}_i}$ and $\sigma^2_{\tilde{p}_i}$, respectively, such that

$$\begin{aligned}\sigma^2_{\tilde{x}_i} &= \sigma^2_{\tilde{p}_i} \\ &= \sigma^2_\omega - 1 + \left(\sigma^2_{\tilde{x}_i|x_i} + \sigma^2_{\tilde{p}_i|p_i}\right),\end{aligned} \tag{78}$$

where

$$\sigma^2_{\tilde{x}_i|x_i} = \sigma^2_{\tilde{p}_i|p_i} = 1, \tag{79}$$

thus (78) can be rewritten as

$$\sigma^2_{\tilde{x}_i} = \sigma^2_{\tilde{p}_i} = \sigma^2_\omega. \tag{80}$$

Let the variance of Eve's EPR ancilla used for the attack of the $i$-th Gaussian sub-channel $\mathcal{N}_i$ to be referred as $W_i$. The covariance matrix of Eve's $i$-th density matrix $\rho_{E_i} = |\Psi_{EB}\rangle\langle\Psi_{EB}|_i$ is expressed as [2]

$$\mathbf{K}_{\rho_{E_i} = |\Psi_{EB}\rangle\langle\Psi_{EB}|_i} = \begin{pmatrix} W_i I & \sqrt{W_i^2 - 1} Z \\ \sqrt{W_i^2 - 1} Z & W_i I \end{pmatrix}. \tag{81}$$

Assuming that Bob receives a noisy Gaussian subcarrier CV $|\phi'_i\rangle = \tilde{x}'_i + \mathrm{i}\tilde{p}'_i$, the Gaussian subcarrier CV can be rewritten as $d'_i = x'_i + \mathrm{i} p'_i$. The noisy subcarrier quadratures $\tilde{x}'_i$ and $\tilde{p}'_i$ of the received $|\phi'_i\rangle$ have variance $\sigma^2_{\tilde{\omega}}$ as

$$\begin{aligned}\sigma^2_{\tilde{\omega}} &= \sigma^2_{\tilde{x}'_i} = \sigma^2_{\tilde{p}'_i} \\ &= \left|F\left(T_{Eve,i}\right)\right|^2 W_i + \left|F\left(T_i\left(\mathcal{N}_i\right)\right)\right|^2 \sigma^2_\omega \\ &= \left(1 - \left|F\left(T_i\left(\mathcal{N}_i\right)\right)\right|^2\right) W_i + \left|F\left(T_i\left(\mathcal{N}_i\right)\right)\right|^2 \sigma^2_\omega.\end{aligned} \tag{82}$$

In particular, the variance $\sigma_{\tilde{\omega}_E}^2$ of Eve's subcarrier quadratures $\tilde{x}_{E,i}$ and $\tilde{p}_{E,i}$ are evaluated for the attacking of the $i$-th subcarrier Gaussian CV $\left|\phi_i'\right\rangle$ as

$$\begin{aligned}
\sigma_{\tilde{\omega}_E}^2 &= \sigma_{\tilde{x}_{E,i}}^2 = \sigma_{\tilde{p}_{E,i}}^2 \\
&= \left|F\left(T_{Eve,i}\right)\right|^2 \sigma_\omega^2 + \left|F\left(T_i\left(\mathcal{N}_i\right)\right)\right|^2 W_i \\
&= \left(1 - \left|F\left(T_i\left(\mathcal{N}_i\right)\right)\right|^2\right)\sigma_\omega^2 + \left|F\left(T_i\left(\mathcal{N}_i\right)\right)\right|^2 W_i.
\end{aligned} \quad (83)$$

Precisely, the $\sigma_{\tilde{x}_i'|x_i}^2, \sigma_{\tilde{p}_i'|p_i}^2$ conditional variances of Alice's variables and Bob's noisy subcarrier quadratures are evaluated as

$$\begin{aligned}
\sigma_{B_i|A_i}^2 &= \sigma_{\tilde{x}_i'|x_i}^2 = \sigma_{\tilde{p}_i'|p_i}^2 \\
&= \left|F\left(T_{Eve,i}\right)\right|^2 W_i + \left|F\left(T_i\left(\mathcal{N}_i\right)\right)\right|^2 \\
&= \left(1 - \left|F\left(T_i\left(\mathcal{N}_i\right)\right)\right|^2\right)W_i + \left|F\left(T_i\left(\mathcal{N}_i\right)\right)\right|^2.
\end{aligned} \quad (84)$$

Specifically, the conditional variances of Alice's variables and Eve's noisy subcarrier quadratures are

$$\begin{aligned}
\sigma_{E_i|A_i}^2 &= \sigma_{\tilde{x}_{E,i}'|x_i}^2 = \sigma_{\tilde{p}_{E,i}'|p_i}^2 \\
&= \left|F\left(T_{Eve,i}\right)\right|^2 + \left|F\left(T_i\left(\mathcal{N}_i\right)\right)\right|^2 W_i \\
&= 1 - \left|F\left(T_i\left(\mathcal{N}_i\right)\right)\right|^2 + \left|F\left(T_i\left(\mathcal{N}_i\right)\right)\right|^2 W_i.
\end{aligned} \quad (85)$$

From $\sigma_{B_i|A_i}^2$ and $\sigma_{E_i|A_i}^2$, the covariance matrices $\mathbf{K}_{\rho_{B_i}}$ and $\mathbf{K}_{\rho_{E_i}}$ of the density $\rho_{B_i} = \left|\phi_i'\right\rangle\left\langle\phi_i'\right|$ of the noisy $\left|\phi_i'\right\rangle = \tilde{x}_i' + i\tilde{p}_i'$, and Eve's density matrix $\rho_{E_i} = \left|\Psi_{EB}\right\rangle\left\langle\Psi_{EB}\right|_i$ are

$$\mathbf{K}_{\rho_{B_i}} = \sigma_{\tilde{\omega}_E}^2 I \quad (86)$$

and

$$\mathbf{K}_{\rho_{E_i}} = \begin{pmatrix} \sigma_{\tilde{\omega}_E}^2 I & \kappa_i Z \\ \kappa_i Z & W_i I \end{pmatrix}, \quad (87)$$

while $\kappa_i$ is precisely as

$$\begin{aligned}
\kappa_i &= \sqrt{\left(1 - \left|F\left(T_{Eve,i}\right)\right|^2\right)\left(W_i^2 - 1\right)} \\
&= \sqrt{\left|F\left(T_i\left(\mathcal{N}_i\right)\right)\right|^2\left(W_i^2 - 1\right)}.
\end{aligned} \quad (88)$$

From (86) and (87), the $\mathbf{K}_{\rho_{B_i E_i}}$ covariance of the $i$-th joint density matrix of Bob and Eve, $\rho_{B_i E_i}$ is precisely expressed as [2]

$$\mathbf{K}_{\rho_{B_i E_i}} = \begin{pmatrix} \mathbf{K}_{\rho_{E_i}} & \mathbf{K}_{\mu_i,\theta_i} \\ \mathbf{K}^T_{\mu_i,\theta_i} & \mathbf{K}_{\rho_{B_i}} \end{pmatrix}, \tag{89}$$

where $\mathbf{K}_{\mu_i,\theta_i}$ is

$$\mathbf{K}_{\mu_i,\theta_i} = \begin{pmatrix} \mu_i I \\ \theta_i Z \end{pmatrix} = \begin{pmatrix} \left(W_i - \sigma_\omega^2\right)\sqrt{\left(|F(T_{Eve,i})|^2 |F(T_i(\mathcal{N}_i))|^2\right)} I \\ \left[\sqrt{|F(T_{Eve,i})|^2 \left(W_i^2 - 1\right)} = \sqrt{\left(1 - |F(T_i(\mathcal{N}_i))|^2\right)\left(W_i^2 - 1\right)}\right] Z \end{pmatrix}. \tag{90}$$

Specifically, for the quadratures $x_i, p_i$ of the $i$-th Gaussian subcarrier CV, the conditional covariance matrices [2] are precisely

$$\mathbf{K}_{x'_i|x_i} = \mathbf{K}_{p'_i|p_i} = \mathbf{K}\left(1, \sigma_\omega^2\right), \tag{91}$$

$$\mathbf{K}_{E_i|x_i} = \mathbf{K}_{E_i|p_i} = \mathbf{K}\left(1, \sigma_\omega^2\right), \tag{92}$$

$$\mathbf{K}_{x'_i,p'_i|x_i,p_i} = \mathbf{K}(I), \tag{93}$$

and

$$\mathbf{K}_{E_i|x_i,p_i} = \mathbf{K}(I). \tag{94}$$

Without loss of generality, assuming $l < n$ Gaussian subcarriers for the information transmission, from the conditional variances $\sigma^2_{B_i|A_i}$ and $\sigma^2_{E_i|A_i}$ of the Gaussian subcarriers, the average conditional variances $\hat{\sigma}^2_{B_j|A_j}$ and $\hat{\sigma}^2_{E_j|A_j}$ are evaluated as

$$\hat{\sigma}^2_{B_j|A_j} = \tfrac{1}{l}\sum_{i=1}^{l} \sigma^2_{B_i|A_i}, \text{ and } \hat{\sigma}^2_{E_j|A_j} = \tfrac{1}{l}\sum_{i=1}^{l} \sigma^2_{E_i|A_i}, \tag{95}$$

where

$$\begin{aligned}
\hat{\sigma}^2_{B_j|A_j} &= \tfrac{1}{l}\sum_{i=1}^{l} \sigma^2_{\tilde{x}'_i} = \tfrac{1}{l}\sum_{i=1}^{l} \sigma^2_{\tilde{p}'_i} \\
&= \tfrac{1}{l}\sum_{i=1}^{l}\left(|F(T_{Eve,i})|^2 W_i\right) + \tfrac{1}{l}\sum_{i=1}^{l}|F(T_i(\mathcal{N}_i))|^2 \\
&= \left(1 - \tfrac{1}{l}\sum_{i=1}^{l}|F(T_i(\mathcal{N}_i))|^2\right)\tfrac{1}{l}\sum_{i=1}^{l} W_i + \tfrac{1}{l}\sum_{i=1}^{l}|F(T_i(\mathcal{N}_i))|^2 \\
&= \left(1 - \tfrac{1}{l}\sum_{i=1}^{l}|F(T_i(\mathcal{N}_i))|^2\right) W + \tfrac{1}{l}\sum_{i=1}^{l}|F(T_i(\mathcal{N}_i))|^2,
\end{aligned} \tag{96}$$

where $W = \tfrac{1}{l}\sum_l W_i$, and

$$\begin{aligned}
\hat{\sigma}^2_{E_j|A_j} &= \tfrac{1}{l}\sum_{i=1}^{l}\sigma^2_{\tilde{x}_{E,i}} = \tfrac{1}{l}\sum_{i=1}^{l}\sigma^2_{\tilde{p}_{E,i}} \\
&= \tfrac{1}{l}\sum_{i=1}^{l}\left(\left|F(T_{Eve,i})\right|^2\right) + \tfrac{1}{l}\sum_{i=1}^{l}\left(\left|F(T_i(\mathcal{N}_i))\right|^2 W_i\right) \\
&= \left(1 - \tfrac{1}{l}\sum_{i=1}^{l}\left|F(T_i(\mathcal{N}_i))\right|^2\right) + \tfrac{1}{l}\sum_{i=1}^{l}\left|F(T_i(\mathcal{N}_i))\right|^2 \tfrac{1}{l}\sum_{i=1}^{l}W_i \\
&= \left(1 - \tfrac{1}{l}\sum_{i=1}^{l}\left|F(T_i(\mathcal{N}_i))\right|^2\right) + \left(\tfrac{1}{l}\sum_{i=1}^{l}\left|F(T_i(\mathcal{N}_i))\right|^2\right)W.
\end{aligned} \qquad (97)$$

Particularly, using the covariance matrices of (91)–(94), the symplectic spectra [2] can be constructed for the Gaussian subcarriers. We omit this step and show only the symplectic spectra of the Gaussian single-carriers recovered from the Gaussian subcarriers via the CVQFT operation; because the secret key rate formulas will also use this representation.

### 3.1.1 Homodyne Measurement, Reverse Reconciliation

In the case of homodyne measurement, Bob first applies for all collected $\left|\phi_i\right\rangle$ subcarriers the CVQFT operation to recover the Gaussian single-carrier CV $\left|\varphi'_j\right\rangle = x'_j + \mathrm{i}p'_j$, which is the noisy version of Alice's Gaussian single-carrier CV $\left|\varphi_j\right\rangle = x_j + \mathrm{i}p_j$. The quadratures are referred to as $(x_j, p_j)$ and $(x'_j, p'_j)$. Bob then measures via $M_{\mathrm{hom}}$ a noisy quadrature $x'_j$ or $p'_j$ randomly from the quadrature pair $(x'_j, p'_j)$ of $\left|\varphi'_j\right\rangle$, for $j = 1,\ldots,N$.

*3.1.1.1 Private classical capacity of a Gaussian sub-channel*

First we derive the $P^{RR}(\mathcal{N}_i)$ private classical capacity of a Gaussian sub-channel $\mathcal{N}_i$. Alice's conditional variance on Bob's received subcarrier quadratures $x'_i, p'_i$ for the $i$-th Gaussian subcarrier CV are precisely as follows:

$$\begin{aligned}
\sigma^2_{e_{x_i}|e_{x'_i}} &= \left\langle e^2_{x_i}\right\rangle - \frac{\left|\left\langle x_i e_{x_i}\right\rangle\right|^2}{\left\langle x^2_i\right\rangle} \\
&= \left|F(T_i(\mathcal{N}_i))\right|^2 \sigma^2_{X_i} + \left|F(T_i(\mathcal{N}_i))\right|^2 sN_0 \\
&= \left|F(T_i(\mathcal{N}_i))\right|^2 \left(\sigma^2_{X_i} + s\right)N_0,
\end{aligned} \qquad (98)$$

where $N_0$ is the shot-noise variance and $s$ is the squeezing factor, which is $s = 1$ for coherent states and constrained into the range of $\tfrac{1}{\sigma^2_{\omega_i}} < s < \sigma^2_{\omega_i}$ [1], and

$$\begin{aligned}
\sigma^2_{e_{p_i}|e_{p'_i}} &= \left\langle e^2_{p_i}\right\rangle - \frac{\left|\left\langle p_i e_{p_i}\right\rangle\right|^2}{\left\langle p^2_i\right\rangle} \\
&= \left|F(T_i(\mathcal{N}_i))\right|^2 \left(\sigma^2_{X_i} + \tfrac{1}{s}\right)N_0,
\end{aligned} \qquad (99)$$

where $\sigma_{\omega_i}^2$ is the modulation variance of the quadratures $x_i, p_i$ of the $i$-th subcarrier $|\phi_i\rangle$. Specifically, after some calculations, the following relations follow for the conditional variances of (55) and (56):

$$\sigma_{e_{x_i}|e_{x_i'}}^2 \geq |F(T_i(\mathcal{N}_i))|^2 \left(\sigma_{X_i}^2 + \frac{1}{\sigma_{\omega_i}^2}\right) N_0, \tag{100}$$

$$\sigma_{e_{p_i}|e_{p_i'}}^2 \geq |F(T_i(\mathcal{N}_i))|^2 \left(\sigma_{X_i}^2 + \frac{1}{\sigma_{\omega_i}^2}\right) N_0, \tag{101}$$

and for Eve's estimators

$$\sigma_{e_{x_i}|e_{x_i'}^E}^2 = \frac{1}{|F(T_i(\mathcal{N}_i))|^2 \left(\sigma_{X_i}^2 + \frac{1}{\sigma_{\omega_i}^2}\right)} N_0, \tag{102}$$

$$\sigma_{e_{p_i}|e_{p_i'}^E}^2 = \frac{1}{|F(T_i(\mathcal{N}_i))|^2 \left(\sigma_{X_i}^2 + \frac{1}{\sigma_{\omega_i}^2}\right)} N_0. \tag{103}$$

Precisely, from (98) and (99), the $I_x(A:B)$ and $I_p(A:B)$ between Alice and Bob are

$$I_x(A:B) = \tfrac{1}{2}\log_2 \frac{\langle e_{x_i}^2\rangle}{|F(T_i(\mathcal{N}_i))|^2 (\sigma_{X_i}^2 + s) N_0},$$

$$I_p(A:B) = \tfrac{1}{2}\log_2 \frac{\langle e_{x_i}^2\rangle}{|F(T_i(\mathcal{N}_i))|^2 (\sigma_{X_i}^2 + \tfrac{1}{s}) N_0}, \tag{104}$$

and from (57) and (58), the Holevo quantities between Bob and Eve are

$$\chi_x(B:E) = \tfrac{1}{2}\log_2 \frac{\langle e_{x_i}^2\rangle}{\frac{1}{|F(T_i(\mathcal{N}_i))|^2\left(\sigma_{X_i}^2+\frac{1}{\sigma_{\omega_i}^2}\right)} N_0},$$

$$\chi_p(B:E) = \tfrac{1}{2}\log_2 \frac{\langle e_{p_i}^2\rangle}{\frac{1}{|F(T_i(\mathcal{N}_i))|^2\left(\sigma_{X_i}^2+\frac{1}{\sigma_{\omega_i}^2}\right)} N_0}. \tag{105}$$

The reverse reconciliation $S_x^{RR}$, $S_p^{RR}$ secret key rates from (61) and (62) are as follows:

$$S_x^{RR} = I_x(A:B) - \chi_x(B:E), \tag{106}$$

$$S_p^{RR} = I_p(A:B) - \chi_p(B:E). \tag{107}$$

Particularly, the private classical capacity $P^{RR}(\mathcal{N}_i)$ for a given $\mathcal{N}_i$ is expressed as

$$P^{RR}(\mathcal{N}_i) = \max_{\forall i}\Big(\tfrac{1}{2}\log_2 \frac{\sigma_{\omega_i}^2 + \sigma_{X_i}^2}{s+\sigma_{X_i}^2} -$$

$$\tfrac{1}{2}\log_2\Big(\big(|F(T_i(\mathcal{N}_i))|^2 \sigma_{\omega_i}^2 + |F(T_i(\mathcal{N}_i))|^2 \sigma_{X_i}^2\big)\big(|F(T_i(\mathcal{N}_i))|^2 \sigma_{X_i}^2 + |F(T_i(\mathcal{N}_i))|^2 \tfrac{1}{\sigma_{\omega_i}^2}\big)\Big)\Big)$$

$$= \max_{\forall i} \tfrac{1}{2}\log_2 \frac{1}{\Big(|F(T_i(\mathcal{N}_i))|^2\sigma_{X_i}^2 + |F(T_i(\mathcal{N}_i))|^2 \tfrac{1}{\sigma_{\omega_i}^2}\Big)\Big(|F(T_i(\mathcal{N}_i))|^2\sigma_{X_i}^2 + |F(T_i(\mathcal{N}_i))|^2 s\Big)}.$$

$$\tag{108}$$

*3.1.1.2 Secret key rate*

In the next step, the $S_{one-way}^{RR,M_{\text{hom}}}$ secret key rate is evaluated for the recovered single-carrier Gaussian CVs. Introducing the notation $\langle \cdot \rangle$ for the variance of the quadratures, for Bob's noisy quadratures (which noise arise from Eve's Gaussian attack, which will noise be corrected in the post-processing phase) $\langle x_j'^2 \rangle = \langle p_j'^2 \rangle = \frac{1}{l}\sum_l |F(T_i(\mathcal{N}_i))|^2 \sigma_{\omega_0}^2 + \sigma_{\mathcal{N}}^2$. Eve's quadratures are expressed as $\langle x_{E,j}^2 \rangle = \langle p_{E,j}^2 \rangle = \frac{1}{l}\sum_l |F(T_i(\mathcal{N}_i))|^2 \sigma_{\omega_0}^2 + \sigma_{\mathcal{N}}^2$. Bob performs $M_{\text{hom}}$, which gives him the estimator operators $e_{x_j'}$ or $e_{p_j'}$. Specifically, Bob's estimator operators $e_{x_{E,j}}^B$ and $e_{p_{E,j}}^B$ on Eve's noisy quadrature $x_{E,j}$ are precisely as follows [1]:

$$e_{x_{E,j}}^B = \varepsilon_{B,x_{E,j}} x_j', \quad \varepsilon_{B,x_{E,j}} = \frac{\langle x_{E,j} x_j' \rangle}{\langle x_j'^2 \rangle}, \tag{109}$$

$$\hat{\sigma}^2_{x_{E,j}|e_{x_{E,j}}^B} = \left\langle \left( x_{E,j} - e_{x_{E,j}}^B \right)^2 \right\rangle = \langle x_{E,j}^2 \rangle - \frac{|\langle x_j' x_{E,j} \rangle|^2}{\langle x_j'^2 \rangle}, \tag{110}$$

and since $\rho_{E_j B_j} = |\Psi_{EB}\rangle\langle\Psi_{EB}|_j$ is an entangled state,

$$e_{p_{E,j}}^B = -\varepsilon_{B,p_{E,j}} p_j', \quad \varepsilon_{B,p_{E,j}} = \frac{\langle p_{E,j} p_j' \rangle}{\langle p_j'^2 \rangle}, \tag{111}$$

$$\hat{\sigma}^2_{p_{E,j}|e_{p_{E,j}}^B} = \left\langle \left( p_{E,j} - e_{p_{E,j}}^B \right)^2 \right\rangle = \langle p_{E,j}^2 \rangle - \frac{|\langle p_j' p_{E,j} \rangle|^2}{\langle p_j'^2 \rangle}. \tag{112}$$

From the estimators $e_{x_{E,j}}^B$ and $e_{p_{E,j}}^B$, precisely the following relations bring up:

$$|\langle x_j' x_{E,j} \rangle|^2 \leq \langle x_j'^2 \rangle \langle x_{E,j}^2 \rangle - N_0^2 \frac{\langle x_j'^2 \rangle}{\langle p_{E,j}^2 \rangle}, \tag{113}$$

$$|\langle p_j' p_{E,j} \rangle|^2 \leq \langle p_j'^2 \rangle \langle p_{E,j}^2 \rangle - N_0^2 \frac{\langle p_j'^2 \rangle}{\langle x_{E,j}^2 \rangle}. \tag{114}$$

Specifically, a commutation relation argument [1] then yields

$$\left[ \left( x_{E,j} - e_{x_{E,j}}^B \right), p_{E,j} \right] = [x_{E,j}, p_{E,j}] - \varepsilon_{B,x_{E,j}}\left( x_j', p_{E,j} \right) = [x_{E,j}, p_{E,j}], \tag{115}$$

$$\left[ \left( p_{E,j} - e_{p_{E,j}}^B \right), x_{E,j} \right] = [p_{E,j}, x_{E,j}] - \varepsilon_{B,p_{E,j}}\left( p_j', x_{E,j} \right) = [p_{E,j}, x_{E,j}]. \tag{116}$$

In particular, since Eve uses an EPR state $\rho_{E_j B_j} = |\Psi_{EB}\rangle\langle\Psi_{EB}|_j$ for attacking (e.g., in an AMQD multicarrier scenario, the joint density matrix $\rho_{E_j B_j}$ models the $l$ instances of EPR state $\rho_{E_j B_j} = \rho_{E_i B_i}^{\otimes l} = |\Psi_{EB}\rangle\langle\Psi_{EB}|_i^{\otimes l}$), the following relations hold for the joint system of Bob and Eve:

$$\frac{\hat{\sigma}^2_{x_{E,j}|e^B_{x_{E,j}}}}{N_0} = \frac{1}{\frac{1}{l}\sum_l |F(T_i(\mathcal{N}_i))|^2 \sigma^2_{\omega_0} + \sigma^2_{\mathcal{N}}} \quad \text{and} \quad \frac{\hat{\sigma}^2_{p_{E,j}|e^B_{p_{E,j}}}}{N_0} = \frac{1}{\frac{1}{l}\sum_l |F(T_i(\mathcal{N}_i))|^2 \sigma^2_{\omega_0} + \sigma^2_{\mathcal{N}}}, \tag{117}$$

$$\hat{\sigma}^2_{x_{E,j}|e^B_{x_{E,j}}} = \hat{\sigma}^2_{p_{E,j}|e^B_{p_{E,j}}} = \frac{1}{\frac{1}{l}\sum_l |F(T_i(\mathcal{N}_i))|^2 \sigma^2_{\omega_0} + \sigma^2_{\mathcal{N}}} N_0, \tag{118}$$

$$\langle x'_j x_{E,j} \rangle = \sqrt{\left(\frac{1}{l}\sum_l |F(T_i(\mathcal{N}_i))|^2 \sigma^2_{\omega_0} + \sigma^2_{\mathcal{N}}\right)^2} N_0, \tag{119}$$

$$\langle p'_j p_{E,j} \rangle = -\sqrt{\left(\frac{1}{l}\sum_l |F(T_i(\mathcal{N}_i))|^2 \sigma^2_{\omega_0} + \sigma^2_{\mathcal{N}}\right)^2} N_0. \tag{120}$$

Particularly, the $M_{\text{hom}}$ leads to covariance matrices for Bob's measured quadrature $x'_j$ or $p'_j$ conditioned on Alice's (single-carrier) Gaussian variable $x_j$ or $p_j$, precisely as

$$\mathbf{K}_{x'_j|x_j} = \mathbf{K}_{p'_j|p_j} = \mathbf{K}\left(1, \sigma^2_{\omega_0}\right). \tag{121}$$

Specifically, in the case of $M_{\text{hom}}$ and reverse reconciliation, the corresponding $\mathbf{K}_{E_j|x_j}$, $\mathbf{K}_{E_j|p_j}$ conditional covariance matrices of Eve's system are precisely evaluated as

$$\mathbf{K}_{E_j|x_j} = \mathbf{K}_{E_j|p_j} = \mathbf{K}\left(1, \sigma^2_{\omega_0}\right). \tag{122}$$

Without loss of generality, assuming that $0 < \frac{1}{l}\sum_l |F(T_i(\mathcal{N}_i))|^2 < 1$ for the $l$ Gaussian sub-channels $\mathcal{N}_i$ and $\sigma^2_{\omega_0} \gg 1$ holds, the following symplectic spectra can be constructed for Bob's system, $B_j$:

$$\mathcal{S}_{x'_j} = \mathcal{S}_{p'_j} = \left(\frac{1}{l}\sum_l |F(T_i(\mathcal{N}_i))|^2 \sigma^2_{\omega_0}\right), \tag{123}$$

$$\mathcal{S}_{x'_j|x_j} = \mathcal{S}_{p'_j|p_j} = \sqrt{\hat{\sigma}^2_{B_j|A_j}\left(\frac{1}{l}\sum_l |F(T_i(\mathcal{N}_i))|^2\right) \sigma^2_{\omega_0}}, \tag{124}$$

and for Eve's system, $E_j$, precisely:

$$\mathcal{S}_{E_j} = \left(\left(1 - \frac{1}{l}\sum_l |F(T_i(\mathcal{N}_i))|^2\right)\sigma^2_{\omega_0}, \frac{1}{l}\sum_l W_i\right). \tag{125}$$

In particular, in case of a reverse reconciliation, the related symplectic spectra $\mathcal{S}_{E_j|x'_j}$ and $\mathcal{S}_{E_j|p'_j}$ are evaluated as follows [2]:

$$\mathcal{S}_{E_j|x'_j} = \mathcal{S}_{E_j|p'_j} = \left(\sqrt{\frac{1}{\frac{1}{l}\sum_l |F(T_i(\mathcal{N}_i))|^2}\left(1 - \frac{1}{l}\sum_l |F(T_i(\mathcal{N}_i))|^2\right)\sigma^2_{\omega_0} \frac{1}{l}\sum_l W_i}, 1\right). \tag{126}$$

Finally, for the joint system $\rho_{B_j E_j}$ of Bob and Eve, the $\mathcal{S}_{x'_j, E_j}, \mathcal{S}_{p'_j, E_j}$ symplectic spectra are

$$\mathcal{S}_{x'_j, E_j} = \mathcal{S}_{p'_j, E_j} = \left(\sigma^2_{\omega_0}, 1, 1\right). \tag{127}$$

Note, as it has been mentioned in Section 2.1.3, in case of RR, the secret key rate formulas $S^{RR,M_{\text{hom}}}_{one-way}$, $S^{RR,M_{\text{hom}}}_{two-way}$, $S^{RR,M_{\text{het}}}_{two-way}$ are evaluated from the difference of functions $I(A:B)$ and the $\chi(B:E)$, because $I(B:E)$ causes negative divergence in the rate formulas of (37) [2].

Particularly, using the related covariance matrices of (121)–(122), the symplectic spectra of (123)–(125) and (126)–(127), the mutual information $I(A:B)$ and Holevo quantity $\chi(B:E)$ are determined, from which the $S^{RR,M_{\text{hom}}}_{one-way}$ asymptotic secret key rate for the RR, $M_{\text{hom}}$, AMQD modulation in one-way CVQKD is as follows:

$$\begin{aligned}
S^{RR,M_{\text{hom}}}_{one-way} &= I(A:B) - \chi(B:E) \\
&= \tfrac{1}{2}\log_2 \frac{\tfrac{1}{l}\sum_l W_i}{\left(1 - \tfrac{1}{l}\sum_l |F(T_i(\mathcal{N}_i))|^2\right)\hat{\sigma}^2_{B_j|A_j}} - \left(\tfrac{\tfrac{1}{l}\sum_l W_i + 1}{2}\log_2 \tfrac{\tfrac{1}{l}\sum_l W_i + 1}{2} - \tfrac{\tfrac{1}{l}\sum_l W_i - 1}{2}\log_2 \tfrac{\tfrac{1}{l}\sum_l W_i - 1}{2}\right),
\end{aligned} \tag{128}$$

where $\hat{\sigma}^2_{B_j|A_j}$ has been derived in (96).

∎

### 3.1.2 Homodyne Measurement, Direct Reconciliation

*3.1.2.1 Private classical capacity of a Gaussian sub-channel*

Without loss of generality, using a slightly modified version of the corresponding terms of Section 2.2.1, the $P^{DR}(\mathcal{N}_i)$ private classical capacity of $\mathcal{N}_i$ is follows from (75).

*3.1.2.2 Secret key rate*

The secret key rate $S^{DR,M_{\text{hom}}}_{one-way}$ is then evaluated as follows. In case of direct reconciliation, for the variance of the quadratures, for Alice's quadratures $\langle x_j^2 \rangle = \langle p_j^2 \rangle = \sigma^2_{\omega_0}$, and Eve's quadratures are $\langle x_{E,j}^2 \rangle = \langle p_{E,j}^2 \rangle = \tfrac{1}{l}\sum_l |F(T_i(\mathcal{N}_i))|^2 \sigma^2_{\omega_0} + \sigma^2_{\mathcal{N}}$. Particularly, the systems of Alice and Eve are modeled as independent quantum systems, and using the estimator operators $e^A_{x_{E,j}}$ and $e^A_{p_{E,j}}$ on Alice's side, the following relations hold for the estimation of Eve's noisy quadrature $x_{E,j}$ at $M_{\text{hom}}$:

$$e^A_{x_{E,j}} = \varepsilon_{A, x_{E,j}} x_j, \quad \varepsilon_{A, x_{E,j}} = \frac{\langle x_{E,j} x_j \rangle}{\langle x_j^2 \rangle}, \tag{129}$$

$$e^A_{p_{E,j}} = \varepsilon_{A, p_{E,j}} p_j, \quad \varepsilon_{A, p_{E,j}} = \frac{\langle p_{E,j} p_j \rangle}{\langle p_j^2 \rangle}. \tag{130}$$

From $e^A_{x_{E,j}}$ and $e^A_{p_{E,j}}$, the following conditional variances bring up:

$$\hat{\sigma}^2_{x_{E,j}|e^A_{x_{E,j}}} = \left\langle \left(x_{E,j} - e^A_{x_{E,j}}\right)^2 \right\rangle = \left\langle x_{E,j}^2 \right\rangle - \frac{\left|\left\langle x_j x_{E,j}\right\rangle\right|^2}{\left\langle x_j^2\right\rangle}, \tag{131}$$

$$\hat{\sigma}^2_{p_{E,j}|e^A_{p_{E,j}}} = \left\langle \left(p_{E,j} - e^A_{p_{E,j}}\right)^2 \right\rangle = \left\langle p_{E,j}^2 \right\rangle - \frac{\left|\left\langle p_j p_{E,j}\right\rangle\right|^2}{\left\langle p_j^2\right\rangle}, \tag{132}$$

where

$$\begin{aligned}\left|\left\langle x_j x_{E,j}\right\rangle\right|^2 &\leq \left\langle x_j^2\right\rangle\left\langle x_{E,j}^2\right\rangle - N_0^2 \frac{\left\langle x_j^2\right\rangle}{\left\langle p_{E,j}^2\right\rangle} \\ &= \sigma^2_{\omega_0}\left(\frac{1}{l}\sum_l \left|F\left(T_i\left(\mathcal{N}_i\right)\right)\right|^2 \sigma^2_{\omega_0}\right) - \frac{N_0^2 \sigma^2_{\omega_0}}{\frac{1}{l}\sum_l \left|F\left(T_i\left(\mathcal{N}_i\right)\right)\right|^2 \sigma^2_{\omega_0} + \sigma^2_{\mathcal{N}}},\end{aligned} \tag{133}$$

$$\begin{aligned}\left|\left\langle p_j p_{E,j}\right\rangle\right|^2 &\leq \left\langle p_j^2\right\rangle\left\langle p_{E,j}^2\right\rangle - N_0^2 \frac{\left\langle p_j^2\right\rangle}{\left\langle x_{E,j}^2\right\rangle} \\ &= \sigma^2_{\omega_0}\left(\frac{1}{l}\sum_l \left|F\left(T_i\left(\mathcal{N}_i\right)\right)\right|^2 \sigma^2_{\omega_0} + \sigma^2_{\mathcal{N}}\right) - \frac{N_0^2 \sigma^2_{\omega_0}}{\frac{1}{l}\sum_l \left|F\left(T_i\left(\mathcal{N}_i\right)\right)\right|^2 \sigma^2_{\omega_0} + \sigma^2_{\mathcal{N}}}.\end{aligned} \tag{134}$$

Then the commutation relation [1] leads to

$$\left[\left(x_{E,j} - e^A_{x_{E,j}}\right), p_{E,j}\right] = \left[x_{E,j}, p_{E,j}\right] - \varepsilon_{A,x_{E,j}}\left(x_j, p_{E,j}\right) = \left[x_{E,j}, p_{E,j}\right], \tag{135}$$

$$\left[\left(p_{E,j} - e^A_{p_{E,j}}\right), x_{E,j}\right] = \left[p_{E,j}, x_{E,j}\right] - \varepsilon_{A,p_{E,j}}\left(p_j, x_{E,j}\right) = \left[p_{E,j}, x_{E,j}\right]. \tag{136}$$

Specifically, assuming the use of coherent Gaussian CVs for the transmission, a corresponding relation yields as

$$\frac{\hat{\sigma}^2_{x_{E,j}|e^A_{x_{E,j}}}}{N_0} = 1 \text{ and } \frac{\hat{\sigma}^2_{p_{E,j}|e^A_{p_{E,j}}}}{N_0} = 1, \tag{137}$$

thus

$$\hat{\sigma}^2_{x_{E,j}|e^A_{x_{E,j}}} = \hat{\sigma}^2_{p_{E,j}|e^A_{p_{E,j}}} = N_0. \tag{138}$$

For Eve's system $E_j$, the corresponding covariance matrices are evaluated as follows:

$$\mathbf{K}_{E_j|x_j} = \mathbf{K}_{E_j|p_j} = \mathbf{K}\left(1, \sigma^2_{\omega_0}\right). \tag{139}$$

In particular, for the direct reconciliation, the symplectic spectra $\mathcal{S}_{E_j|x_j}$ and $\mathcal{S}_{E_j|p_j}$ are precisely as follows [2]:

$$\mathcal{S}_{E_j|x_j} = \mathcal{S}_{E_j|p_j} = \left(\sqrt{\hat{\sigma}^2_{E_j|A_j}\left(1 - \frac{1}{l}\sum_l\left|F\left(T_i\left(\mathcal{N}_i\right)\right)\right|^2\right)\sigma^2_{\omega_0}}, \sqrt{\hat{\sigma}^2_{B_j|A_j}\frac{1}{l}\sum_l W_i\Big/\hat{\sigma}^2_{E_j|A_j}}\right). \tag{140}$$

Without loss of generality, by utilizing the covariance matrices of (121) and (139), the symplectic spectra of (123)–(125), and (139)–(140), the Holevo quantities of $\chi\left(A:B\right)$ and $\chi\left(A:E\right)$ the $S_{one-way}^{DR,M_{\text{hom}}}$ asymptotic secret key rate for the DR, $M_{\text{hom}}$, AMQD modulation in one-way CVQKD is precisely as follows:

$$\begin{aligned}S_{one-way}^{DR,M_{\text{hom}}} &= \chi(A:B) - \chi(A:E) \\ &\quad \tfrac{1}{2}\log_2 \frac{\left(\tfrac{1}{l}\sum_l |F(T_i(\mathcal{N}_i))|^2\right)\hat{\sigma}_{E_j|A_j}^2}{\left(1-\left(\tfrac{1}{l}\sum_l |F(T_i(\mathcal{N}_i))|^2\right)\right)\hat{\sigma}_{B_j|A_j}^2} \\ &\quad + \left(\frac{\sqrt{\tfrac{1}{l}\sum_l W_i \hat{\sigma}_{B_j|A_j}^2 / \hat{\sigma}_{E_j|A_j}^2}+1}{2}\log_2 \frac{\sqrt{\tfrac{1}{l}\sum_l W_i \hat{\sigma}_{B_j|A_j}^2 / \hat{\sigma}_{E_j|A_j}^2}+1}{2} - \frac{\sqrt{\tfrac{1}{l}\sum_l W_i \hat{\sigma}_{B_j|A_j}^2 / \hat{\sigma}_{E_j|A_j}^2}-1}{2}\log_2 \frac{\sqrt{\tfrac{1}{l}\sum_l W_i \hat{\sigma}_{B_j|A_j}^2 / \hat{\sigma}_{E_j|A_j}^2}-1}{2}\right) \\ &\quad - \left(\frac{\tfrac{1}{l}\sum_l W_i+1}{2}\log_2 \frac{\tfrac{1}{l}\sum_l W_i+1}{2} - \frac{\tfrac{1}{l}\sum_l W_i-1}{2}\log_2 \frac{\tfrac{1}{l}\sum_l W_i-1}{2}\right).\end{aligned}$$
(141)

∎

### 3.1.3 Heterodyne Measurement, Reverse Reconciliation

*3.1.3.1 Private classical capacity of a Gaussian sub-channel*

Particularly, similar to Section 3.1.2.1, we can use a slightly modified version of (98)–(103); thus the $P^{RR}(\mathcal{N}_i)$ private classical capacity of $\mathcal{N}_i$ is coincidences to (74).

*3.1.3.2 Secret key rate*

In case of $M_{\text{het}}$ and RR, Bob obtains both of the estimators $e_{x_{E,j}}^B, e_{p_{E,j}}^B$ via $M_{\text{het}}$ to estimate Eve's quadratures $x_{E,j}, p_{E,j}$ from his noisy $x'_j$ and $p'_j$. Eve's quadratures can be expressed as follows [1]:

$$\begin{aligned}x_{E,j} &= \left(x_{E,j} - e_{x_{E,j}}^B\right) + e_{x_{E,j}}^B, \\ p_{E,j} &= \left(p_{E,j} - e_{p_{E,j}}^B\right) + e_{p_{E,j}}^B.\end{aligned}$$
(142)

Without loss of generality, assuming that the estimators $e_{x_{E,j}}^B, e_{p_{E,j}}^B$ have variances $\tfrac{1}{l}\sum_l |F(T_i(\mathcal{N}_i))|^2 \sigma_{\omega_0}^2 + \sigma_{\mathcal{N}}^2$, the conditional variances $\hat{\sigma}_{x_{E,j}|e_{x_{E,j}}^B}^2$ and $\hat{\sigma}_{p_{E,j}|e_{p_{E,j}}^B}^2$ are precisely expressed as

$$\begin{aligned}\hat{\sigma}_{x_{E,j}|e_{x_{E,j}}^B}^2 &= \left\langle \left(x_{E,j} - e_{x_{E,j}}^B\right)^2 \right\rangle = \left\langle x_{E,j}^2 \right\rangle - \frac{|\langle x'_j x_{E,j}\rangle|^2}{\langle x'^2_j \rangle} \\ &= \left[\left(\tfrac{1}{l}\sum_l |F(T_i(\mathcal{N}_i))|^2 \sigma_{\omega_0}^2 + \sigma_{\mathcal{N}}^2\right) - \frac{\left(\tfrac{1}{l}\sum_l |F(T_i(\mathcal{N}_i))|^2 \sigma_{\omega_0}^2 + \sigma_{\mathcal{N}}^2\right)^2 - 1}{\left(\tfrac{1}{l}\sum_l |F(T_i(\mathcal{N}_i))|^2 \sigma_{\omega_0}^2 + \sigma_{\mathcal{N}}^2\right) + \mathrm{H}}\right] N_0 \\ &= \frac{\mathrm{H}\left(\tfrac{1}{l}\sum_l |F(T_i(\mathcal{N}_i))|^2 \sigma_{\omega_0}^2 + \sigma_{\mathcal{N}}^2\right) + 1}{\left(\tfrac{1}{l}\sum_l |F(T_i(\mathcal{N}_i))|^2 \sigma_{\omega_0}^2 + \sigma_{\mathcal{N}}^2\right) + \mathrm{H}} N_0,\end{aligned}$$
(143)

and

$$\hat{\sigma}^2_{p_{E,j}|e^B_{p_{E,j}}} = \left\langle \left(p_{E,j} - e^B_{p_{E,j}}\right)^2 \right\rangle = \left\langle p^2_{E,j} \right\rangle - \frac{|\langle p'_j p_{E,j}\rangle|^2}{\langle p'^2_j \rangle}$$

$$= \left[\left(\frac{1}{l}\sum_l |F(T_i(\mathcal{N}_i))|^2 \sigma^2_{\omega_0} + \sigma^2_{\mathcal{N}}\right) - \frac{\left(\frac{1}{l}\sum_l |F(T_i(\mathcal{N}_i))|^2 \sigma^2_{\omega_0} + \sigma^2_{\mathcal{N}}\right)^2 - 1}{\left(\frac{1}{l}\sum_l |F(T_i(\mathcal{N}_i))|^2 \sigma^2_{\omega_0} + \sigma^2_{\mathcal{N}}\right) + \frac{1}{\mathrm{H}}}\right] N_0 \quad (144)$$

$$= \frac{\left(\frac{1}{l}\sum_l |F(T_i(\mathcal{N}_i))|^2 \sigma^2_{\omega_0} + \sigma^2_{\mathcal{N}}\right) + \mathrm{H}}{\mathrm{H}\left(\frac{1}{l}\sum_l |F(T_i(\mathcal{N}_i))|^2 \sigma^2_{\omega_0} + \sigma^2_{\mathcal{N}}\right) + 1} N_0 = \frac{N_0^2}{\hat{\sigma}^2_{x_{E,j}|e_{x_{E,j}}}},$$

where $\mathrm{H} = \frac{1-|T_B|^2}{|T_B|^2}$, and $|T_B|^2$ is the transmittance of Bob's internal beam splitter, used for the separation of the quadratures. The related commutation relations at $M_{\mathrm{het}}$ are

$$[x'_j, p'_j] = [x_{E,j}, p_{E,j}] = 0, \quad (145)$$

$$[x'_j - e^B_{x_{E,j}}, p'_j - e^B_{p_{E,j}}] = -[x'_j, p'_j]. \quad (146)$$

Specifically, the $M_{\mathrm{het}}$ leads to a covariance matrix $\mathbf{K}_{x'_j, p'_j | x_j, p_j}$ as [2]

$$\mathbf{K}_{x'_j, p'_j | x_j, p_j} = \mathbf{K}(I). \quad (147)$$

Precisely, for Eve's system $E_j$, the corresponding covariance matrix is

$$\mathbf{K}_{E_j | x_j, p_j} = \mathbf{K}(I). \quad (148)$$

In particular, assuming that $0 < \frac{1}{l}\sum_l |F(T_i(\mathcal{N}_i))|^2 < 1$ for the $l$ Gaussian sub-channels $\mathcal{N}_i$, and $\sigma^2_{\omega_0} \gg 1$ holds, the following symplectic spectra [2] can be constructed for Bob's system:

$$\mathcal{S}_{x'_j p'_j} = \left(\frac{1}{l}\sum_l |F(T_i(\mathcal{N}_i))|^2 \sigma^2_{\omega_0}\right), \quad (149)$$

$$\mathcal{S}_{x'_j, p'_j | x_j, p_j} = \left(\hat{\sigma}^2_{B_j | A_j}\right), \quad (150)$$

and for Eve's system the symplectic spectra is

$$\mathcal{S}_{E_j} = \left(\left(1 - \frac{1}{l}\sum_l |F(T_i(\mathcal{N}_i))|^2\right)\sigma^2_{\omega_0}, \frac{1}{l}\sum_l W_i\right). \quad (151)$$

Precisely, in case of reverse reconciliation and $M_{\mathrm{het}}$, $\mathcal{S}_{E_j|x'_i,p'_j}$ is as follows:

$$\mathcal{S}_{E_j|x'_j,p'_j} = \left(\frac{1}{\frac{1}{l}\sum_l |F(T_i(\mathcal{N}_i))|^2}\left(1 - \frac{1}{l}\sum_l |F(T_i(\mathcal{N}_i))|^2 + \hat{\sigma}^2_{B_j|A_j}\right), 1\right). \quad (152)$$

Particularly, for the joint system $\rho_{B_j E_j}$ of Bob and Eve, the $\mathcal{S}_{x'_j, E_j}, \mathcal{S}_{p'_j, E_j}$ symplectic spectra are

$$\mathcal{S}_{x'_j, p'_j, E_j} = \left(\sigma^2_{\omega_0}, 1, 1\right). \quad (153)$$

Without loss of generality, using the related covariance matrices of (147)–(148), symplectic spectra of (125) and (149)–(153), the $\chi(A:B)$ and $I(B:E)$ functions, the $S_{one-way}^{RR,M_{\text{het}}}$ asymptotic secret key rate for the RR, $M_{\text{het}}$, AMQD modulation in one-way CVQKD, from (37) is precisely as follows:

$$\begin{aligned}
S_{one-way}^{RR,M_{\text{het}}} &= \chi(A:B) - I(B:E) \\
&= \log_2 \frac{1}{1-\left(\frac{1}{l}\sum_l |F(T_i(\mathcal{N}_i))|^2\right)} \\
&\quad - \left(\frac{\hat{\sigma}_{B_j|A_j}^2 + 1}{2}\log_2 \frac{\hat{\sigma}_{B_j|A_j}^2 + 1}{2} - \frac{\hat{\sigma}_{B_j|A_j}^2 - 1}{2}\log_2 \frac{\hat{\sigma}_{B_j|A_j}^2 - 1}{2}\right) \\
&\quad - \left(\frac{\frac{1}{l}\sum_l W_i + 1}{2}\log_2 \frac{\frac{1}{l}\sum_l W_i + 1}{2} - \frac{\frac{1}{l}\sum_l W_i - 1}{2}\log_2 \frac{\frac{1}{l}\sum_l W_i - 1}{2}\right).
\end{aligned} \quad (154)$$

∎

### 3.1.4 Heterodyne Measurement, Direct Reconciliation

#### 3.1.4.1 Private classical capacity of a Gaussian sub-channel

Without loss of generality, using a slightly modified version of the related formulas of Section 2.2.1, the $P^{DR}(\mathcal{N}_i)$ private classical capacity of $\mathcal{N}_i$ can be derived from (75).

#### 3.1.4.2 Secret key rate

In case of $M_{\text{het}}$ and DR, Alice uses $e_{x_{E,j}}^A, e_{p_{E,j}}^A$ to estimate Eve's quadratures $x_{E,j}, p_{E,j}$ from her $x_j$ and $p_j$. Using the differences of $\left(x_{E,j} - e_{x_{E,j}}^A\right)$ and $\left(p_{E,j} - e_{p_{E,j}}^A\right)$, Eve's quadratures are [1]

$$\begin{aligned}
x_{E,j} &= \left(x_{E,j} - e_{x_{E,j}}^A\right) + e_{x_{E,j}}^A, \\
p_{E,j} &= \left(p_{E,j} - e_{p_{E,j}}^A\right) + e_{p_{E,j}}^A.
\end{aligned} \quad (155)$$

Specifically, the variances of $\hat{\sigma}_{x_{E,j}|e_{x_{E,j}}^A}^2$ and $\hat{\sigma}_{p_{E,j}|e_{p_{E,j}}^A}^2$ are as follows:

$$\begin{aligned}
\hat{\sigma}_{x_{E,j}|e_{x_{E,j}}^A}^2 &= \left\langle \left(x_{E,j} - e_{x_{E,j}}^A\right)^2 \right\rangle \\
&= \left\langle x_{E,j}^2 \right\rangle - \frac{|\langle x_j x_{E,j} \rangle|^2}{\langle x_j^2 \rangle},
\end{aligned} \quad (156)$$

and

$$\begin{aligned}
\hat{\sigma}_{p_{E,j}|e_{p_{E,j}}^A}^2 &= \left\langle \left(p_{E,j} - e_{p_{E,j}}^A\right)^2 \right\rangle \\
&= \left\langle p_{E,j}^2 \right\rangle - \frac{|\langle p_j p_{E,j} \rangle|^2}{\langle p_j^2 \rangle} \\
&= \frac{N_0^2}{\hat{\sigma}_{x_{E,j}|e_{x_{E,j}}^A}^2}.
\end{aligned} \quad (157)$$

Particularly, the corresponding commutation relations are

$$[x_j, p_j] = [x_{E,j}, p_{E,j}] = 0, \tag{158}$$

and

$$\left[x_j - e^A_{x_{E,j}}, p_j - e^A_{p_{E,j}}\right] = -[x_j, p_j]. \tag{159}$$

Eve's corresponding covariance matrix is

$$\mathbf{K}_{E_j | x_j, p_j} = \mathbf{K}(I). \tag{160}$$

The symplectic spectra $\mathcal{S}_{E_j | x_i, p_j}$ is evaluated as follows:

$$\mathcal{S}_{E_j | x_i, p_j} = \left(\hat{\sigma}^2_{B_j | A_j}, 1\right). \tag{161}$$

In particular, using the covariance matrices of (147)–(148), symplectic spectra of (125), (149)–(151), and (161), the Holevo quantities of $\chi(A:B)$ and $\chi(A:E)$ are determined, from which the $S^{DR, M_{\text{het}}}_{one-way}$ asymptotic secret key rate for the DR, $M_{\text{het}}$, AMQD modulation in one-way CVQKD is precisely as follows:

$$\begin{aligned} S^{DR, M_{\text{het}}}_{one-way} &= \chi(A:B) - \chi(A:E) \\ &= \log_2 \frac{\frac{1}{l}\sum_l |F(T_i(\mathcal{N}_i))|^2}{1 - \frac{1}{l}\sum_l |F(T_i(\mathcal{N}_i))|^2} \\ &\quad - \left(\frac{\frac{1}{l}\sum_l W_i + 1}{2} \log_2 \frac{\frac{1}{l}\sum_l W_i + 1}{2} - \frac{\frac{1}{l}\sum_l W_i - 1}{2} \log_2 \frac{\frac{1}{l}\sum_l W_i - 1}{2}\right). \end{aligned} \tag{162}$$

∎

The derivation of the AMQD secret key rates for two-way CVQKD is included in the Appendix.

## 3.2 SVD-Assisted AMQD

The SVD-assisted AMQD modulation [6] leads to a virtually improved modulation variance

$$\sigma^2_{\omega''} = \sigma^2_\omega (1 + c) > \sigma^2_\omega, \tag{163}$$

where $c > 0$ for the subcarrier transmission, for further details and analysis see [6]. The SVD defines a logical layer above the physical layer of the multicarrier transmission, and the virtual increment of the multicarrier modulation variance from $\sigma^2_{\omega''}$ to $\sigma^2_\omega$ in (164) is precisely analogous to a decrease in Eve's transmittance $\left|F(T'_{Eve,i})\right|^2 < \left|F(T_{Eve,i})\right|^2$ at a constant multicarrier modulation variance $\sigma^2_\omega$, which after some calculations yields the following:

$$\left|F(T'_{Eve,i})\right|^2 = 1 - v_i \left(1 - \left|F(T_{Eve,i})\right|^2\right), \tag{165}$$

where $\left(1 - \left|F(T_{Eve,i})\right|^2\right) < \frac{1}{v_i}$, and $v_i > 1$ is expressed as

$$v_i = \frac{\sigma^2_{\omega''_i}}{\sigma^2_{\omega_i}}$$
$$= \frac{\nu_{Eve} - \left(\sigma^2_{\mathcal{N}} \big/ \max_{n_{\min}} \lambda^2_i\right)}{\nu_{Eve} - \left(\sigma^2_{\mathcal{N}} \big/ \max_{\forall i} \left|F\left(T_i\left(\mathcal{N}_i\right)\right)\right|^2\right)}, \tag{166}$$

where $\max_{n_{\min}} \lambda^2_i > \max_{\forall i} \left|F\left(T_i\left(\mathcal{N}_i\right)\right)\right|^2$ is precisely the largest eigenvalue of $F(\mathbf{T})F(\mathbf{T})^\dagger$ [6],

$$F(\mathbf{T})F(\mathbf{T})^\dagger = U_2 \Gamma \Gamma^T U_2^{-1}, \tag{167}$$

and where $\Gamma^T$ is the transpose of $\Gamma$ (e.g., elements of the SVD of $F(\mathbf{T})$). For further details, see [6]. In particular, the SVD-assisted rate formulas can be computed by the previously obtained AMQD secret key rate formulas, using the following for the transmittance of the $\mathcal{N}_i$ Gaussian sub-channels, without loss of generality, as

$$\begin{aligned}
\tfrac{1}{l}\sum_l \left|F\left(T'_i\left(\mathcal{N}_i\right)\right)\right|^2 &= \tfrac{1}{l}\sum_l \left(\left|F\left(T_i\left(\mathcal{N}_i\right)\right)\right|^2 v_i\right) \\
&= \tfrac{1}{l}\sum_l \left|F\left(T_i\left(\mathcal{N}_i\right)\right)\right|^2 \cdot \tfrac{1}{l}\sum_l v_i \\
&= \tfrac{1}{l}\sum_l \left|F\left(T_i\left(\mathcal{N}_i\right)\right)\right|^2 \frac{\nu_{Eve} - \left(\sigma^2_{\mathcal{N}} \big/ \max_{n_{\min}} \lambda^2_i\right)}{\nu_{Eve} - \left(\sigma^2_{\mathcal{N}} \big/ \max_{\forall i} \left|F\left(T_i\left(\mathcal{N}_i\right)\right)\right|^2\right)}.
\end{aligned} \tag{168}$$

For simplicity, we do not repeat these formulas here.

∎

Since for all formulas above $\tfrac{1}{l}\sum_l \left|F\left(T'_i\left(\mathcal{N}_i\right)\right)\right|^2 > \tfrac{1}{l}\sum_{i=1}^{l} \left|F\left(T_i\left(\mathcal{N}_i\right)\right)\right|^2 > \left|T_i\left(\mathcal{N}_i\right)\right|^2$ at a given single-carrier modulation variance $\sigma^2_{\omega_0}$, the proof is concluded.

∎

## 3.3 AMQD-MQA Secret Key Rates

In this section, we derive the private classical capacity regions of users $U_1$ and $U_2$ and the corresponding secret key rates in an AMQD-MQA multiuser scenario. First, we have to summarize the derivation of the classical capacity region of the users, from [5].

**Theorem 2** (AMQD-MQA secret key rates). *The AMQD-MQA leads to improved secret key rates in a $K \to K$ multiuser CVQKD scenario in comparison to a single-carrier $K \to K$ multiuser CVQKD scenario.*

*Proof.*
The proof assumes two users, $U_k$, $k = 1, 2$. Let $l$ be the number of good $\mathcal{N}_i$ Gaussian sub-channels. (i.e., the noise of the sub-channels is below the critical security parameter $\nu_{\text{Eve}}$, which identifies the optimal Gaussian collective attack, see the properties of AMQD and AMQD-MQA in [4] and [5]). Let the transmittance of the $i$-th sub-channel be $T_i(\mathcal{N}_i) \in \mathcal{C}$.

The outputs of $U_1$ and $U_2$ are expressed as follows:

$$y_k = F(T(\mathcal{N}))z_k + F(\Delta), \quad k = 1,2. \tag{169}$$

Specifically, the output $d$-dimensional output of the $k$-th user is

$$\mathbf{y}_k = F(\mathbf{T}(\mathcal{N}))\mathbf{z}_k + F(\Delta), \tag{170}$$

where $\mathbf{T}(\mathcal{N}) = [T_{k,1}(\mathcal{N}),...,T_{k,d}(\mathcal{N})]^T$, $\mathbf{z}_k(\mathcal{N}) = [z_{k,1},...,z_{k,d}]^T \in \mathcal{CN}(0,\mathbf{K}_{\mathbf{z}_k})$, and $\mathbf{K}_{\mathbf{z}_k}$ is the covariance matrix of the zero-mean, circular symmetric Gaussian random vector $\mathbf{z}_k \in \mathcal{CN}(0,\mathbf{K}_{\mathbf{z}_k})$ of $U_k$.

In particular, the *sum* capacity [5] is the total throughput over the $l$ sub-channels of $\mathcal{N}$ at a constant modulation variance $\sigma_\omega^2$ is as follows:

$$\begin{aligned} C_{\text{sum}}(\mathcal{N}) &= \max_{(R_1,R_2) \in C} R_1 + R_2 \\ &= \max_{\forall i} \sum_l \log_2\left(1 + \frac{\sigma_{\omega_i}^2 |F(T_i(\mathcal{N}_i))|^2}{\sigma_\mathcal{N}^2}\right). \end{aligned} \tag{171}$$

The *symmetric* capacity [5] is the maximum common rate at which both $U_1$ and $U_2$ can reliably transmit information over the $l$ sub-channels of $\mathcal{N}$, as follows:

$$C_{\text{sym}}(\mathcal{N}) = \max_{(R_{\text{sym}},R_{\text{sym}}) \in C} R_{\text{sym}} = \tfrac{1}{2} \max_{\forall i} \sum_l \log_2\left(1 + \frac{\sigma_{\omega_i}^2 |F(T_i(\mathcal{N}_i))|^2}{\sigma_\mathcal{N}^2}\right), \tag{172}$$

where $R_{\text{sym}}$ is the rate at which both $U_1$ and $U_2$ can simultaneously communicate in a reliable form.

In particular, for $K$ users $U_1,...U_K$, the sum capacity and the symmetric capacity of $\mathcal{N}$ are expressed as

$$C_{\text{sum}}(\mathcal{N}) = \max_{(R_1,...,R_K) \in C} \sum_K R_i = \max_{\forall i} \sum_l \log_2\left(1 + \frac{\sigma_{\omega_i}^2 |F(T_i(\mathcal{N}_i))|^2}{\sigma_\mathcal{N}^2}\right), \tag{173}$$

and

$$C_{\text{sym}}(\mathcal{N}) = \max_{(R_{\text{sym}},...,R_{\text{sym}}) \in C} R_{\text{sym}} = \tfrac{1}{K} \max_{\forall i} \sum_l \log_2\left(1 + \frac{\sigma_{\omega_i}^2 |F(T_i(\mathcal{N}_i))|^2}{\sigma_\mathcal{N}^2}\right). \tag{174}$$

The C capacity region [5] is the region of the rates of $(R_1, R_2)$ of $U_1$ and $U_2$, at which both users can have a simultaneous reliable communication over the quantum channel $\mathcal{N}$. The region C upper bounds the independent single transmission rates as of $U_1$ and $U_2$, which can be maxi-mized if all the $l$ sub-channels with a total constraint $l\sigma_\omega^2$ (i.e., all degrees of freedom) are dedicated to user $k$, as follows:

$$R_k \leq \max_{\forall i} \sum_l \log_2\left(1 + \frac{\sigma_{\omega_i}^2 |F(T_i(\mathcal{N}_i))|^2}{\sigma_\mathcal{N}^2}\right), \; k = 1,\ldots,K. \tag{175}$$

Particularly, if the equality holds, then only user $U_k$ is allowed to transmit over the $l$ sub-channels. These rates define the corner points $C_1 = \max_{(R_1)\in\mathcal{C}} R_1$ and $C_2 = \max_{(R_2)\in\mathcal{C}} R_2$ of $U_1$ and $U_2$.

At the corner points, the rate of the given user is maximal whereas rate of the other user is zero. On the line between the corner points, both users are allowed to simultaneously communicate at rates $R_1$ and $R_2$.

Hence, in the corner points $C_1$, $C_2$, only $U_1$, $U_2$ is allowed to transmit over the $l$ sub-channels, whereas the rate of the other user is zero. Taking the $\mathcal{H}$ convex hull of all possible independent input distributions leads to the capacity region C as

$$\mathsf{C} = \mathcal{H}\left(\bigcup_{z_1,z_2} \mathsf{C}(z_1, z_2)\right). \tag{176}$$

The inputs $z_1$ and $z_2$ are zero-mean Gaussian random variables, for which follows that all information quantities that characterize capacity region C are simultaneously maximized because the capacity region $\mathsf{C}(\mathcal{CN},\mathcal{CN})$ with zero-mean, circular symmetric complex Gaussian random distribution variables formulates the superset S of all other capacity regions (for the proof of optimality, see [5]) with arbitrary $p_{x'}$ distributions, that is,

$$\mathsf{C}(\mathcal{CN},\mathcal{CN}) = \mathsf{S}\left(\bigcup_{\forall p_{x'}} \mathsf{C}(p_{x'}, p_{x'})\right) = \mathsf{S}\bigcup_{\forall p_{x'}} \mathcal{H}\left(\bigcup_{x'_1, x'_2} \mathsf{C}(x'_1, x'_2)\right). \tag{177}$$

As follows, for the $z_k \in \mathcal{CN}\left(0, \mathbb{E}\left[|z_k|^2\right]\right)$ inputs, one obtains the capacity region C:

$$\begin{aligned} I_{\mathrm{MQA}}(z_1 : y | z_2) &= \max_{\forall x'_1} I(x'_1 : y | x'_2), \\ I_{\mathrm{MQA}}(z_2 : y | z_1) &= \max_{\forall x'_2} I(x'_2 : y | x'_1), \end{aligned} \tag{178}$$

where $I_{\mathrm{MQA}}(\cdot)$ is the corresponding correlation measure function between Alice and Bob (e.g., the $I(A:B)$ mutual information or the Holevo information $\chi(A:B)$, depending on the attributes of the actual CVQKD protocol) and

$$I_{\mathrm{MQA}}(z_1, z_2 : y) = \max_{\forall x'_1, x'_2} I(x'_1, x'_2 : y). \tag{179}$$

Particularly, for $l$ independent $d_i \in \mathcal{CN}\left(0, \sigma_{d_i}^2\right)$ subcarriers,

$$\begin{aligned}
I(F(\mathbf{d}) : \mathbf{y}) &= H_{\text{diff}}(\mathbf{y}) - H_{\text{diff}}(\mathbf{y}|F(\mathbf{d})) \\
&= \sum_{l} H_{\text{diff}}(y_i) - H_{\text{diff}}(y_i|z_i) \\
&= \sum_{l} H_{\text{diff}}(y_i) - H_{\text{diff}}(F(\Delta_i)) \\
&= \sum_{l} \log_2\left(1 + \frac{\sigma_{\omega_i}^2 |F(T_i(\mathcal{N}_i))|^2}{\sigma_{\mathcal{N}}^2}\right),
\end{aligned} \quad (180)$$

where $H_{\text{diff}}(\cdot)$ is the differential entropy. For a CV variable $x$ with probability density function $F_x$, the differential entropy is evaluated as follows:

$$H_{\text{diff}}(x) = \int_{-\infty}^{\infty} F_x(u) \log_2\left(\frac{1}{F_x(u)}\right) du. \quad (181)$$

The conditional differential entropy, at given output $y$ is

$$H_{\text{diff}}(x|y) = \int_{-\infty}^{\infty} F_{x,y}(u,v) \log_2\left(\frac{1}{F_{x|y}(u|v)}\right) du dv. \quad (182)$$

From these quantities, the continuous mutual information function is

$$I(x:y) = H_{\text{diff}}(x) - H_{\text{diff}}(x|y), \quad (183)$$

and

$$H_{\text{diff}}(F(\mathbf{d}), \mathbf{y}) = H_{\text{diff}}(F(\mathbf{d})) + H_{\text{diff}}(\mathbf{y}|F(\mathbf{d})) \leq H_{\text{diff}}(F(\mathbf{d})) + H_{\text{diff}}(\mathbf{y}). \quad (184)$$

(*Note*: If the $l$ Gaussian subcarriers are not completely mutually independent, then $\leq$ stands in the second line of (180), whereas if the $l$ subcarriers are derived from not the optimal zero-mean circular symmetric complex Gaussian random $\mathcal{CN}$ distribution, then $\leq$ stands in the third line of (180).)

The input maximization leads to the following sum capacity in an AMQD-MQA setting:

$$C_{\text{sum}}(\mathcal{N}) = \max_{\forall i} \sum_{l} \log_2\left(1 + \frac{\sigma_{\omega_i}^2 |F(T_i(\mathcal{N}_i))|^2}{\sigma_{\mathcal{N}}^2}\right), \quad (185)$$

where $\frac{1}{l}\sum_l |F(T_i(\mathcal{N}_i))|^2 > \frac{1}{l}\sum_l |T_i(\mathcal{N}_i)|^2$, with a total transmit variance constraint

$$\frac{1}{l}\sum_l \sigma_{\omega_i}^2 = \sigma_\omega^2,\ \sigma_{\omega_i}^2 \geq 0. \quad (186)$$

Hence, for the $l$ sub-channels with a constant nonzero modulation variance $\sigma_{\omega_i}^2 > 0$,

$$\sigma_\omega^2 < \sigma_{\omega_0}^2, \quad (187)$$

where $\sigma_{\omega_0}^2$ is the modulation variance of the $x_j, p_j$ quadratures of $z_i$ (i.e., single-carrier modulation variance, see Section 2.1.1).

In particular, for two users, $U_1$ and $U_2$, the C capacity region $(R_1, R_2)$ of AMQD-MQA is as follows:

$$R_1 \leq I_{\text{MQA}}(z_1 : y, F(T(\mathcal{N}))|z_2) = I_{\text{MQA}}(z_1 : y|F(T(\mathcal{N})), z_2), \\ R_2 \leq I_{\text{MQA}}(z_2 : y, F(T(\mathcal{N}))|z_1) = I_{\text{MQA}}(z_2 : y|F(T(\mathcal{N})), z_1), \tag{188}$$

and

$$R_1 + R_2 \leq I_{\text{MQA}}(z_1, z_2 : y, F(T(\mathcal{N}))) = I_{\text{MQA}}(z_1, z_2 : y|F(T(\mathcal{N}))). \tag{189}$$

Particularly, for user $U_k$,

$$R_k^{\text{MQA}} \leq \max_{\forall i} \sum_l \log_2\left(1 + \frac{\sigma_{\omega_i}^2 |F(T_i(\mathcal{N}_i))|^2}{\sigma_\mathcal{N}^2}\right), \ k \in 1, \ldots K, \tag{190}$$

the sum rate (the overall throughput rate of the users) $R_{\text{sum}}^{\text{MQA}}$ is calculated as,

$$R_{\text{sum}}^{\text{MQA}} = \sum_K R_k^{\text{MQA}} \leq \max_{\forall i} \sum_l \log_2\left(1 + \frac{\sigma_{\omega_i}^2 |F(T_i(\mathcal{N}_i))|^2}{\sigma_\mathcal{N}^2}\right), \tag{191}$$

and the symmetric rate (the common rate at which all users can have a simultaneous reliable communication) $R_{\text{sym}}^{\text{MQA}}$ is calculated as

$$R_{\text{sym}}^{\text{MQA}} \leq \tfrac{1}{K} \max_{\forall i} \sum_l \log_2\left(1 + \frac{\sigma_{\omega_i}^2 |F(T_i(\mathcal{N}_i))|^2}{\sigma_\mathcal{N}^2}\right). \tag{192}$$

For $K$ users $U_1, \ldots U_K$, from the results of (190) and (191) trivially follows $R_k^{\text{MQA}}$ and $R_{\text{sym}}^{\text{MQA}}$. The results on the C capacity region of $(R_1, R_2)$ of $U_1$ and $U_2$ in AMQD-MQA are summarized as follows. The corner points $C_1$ and $C_2$ identify the maximal rates at with a single user can communicate. The line between the two corner points represents that trade-off between the rates of users $U_1$ and $U_2$, at which simultaneously reliable transmission is possible. The transmission is realized through $l$ subcarriers, each having a constant modulation variance $\sigma_\omega^2$ per subcarrier quadrature components. The two users communicate over the Gaussian quantum channel with rates $R_1$ and $R_2$. At the corner points $C_1$ and $C_2$, only one user is allowed to transmit and all degrees of freedom is allocated to that user.

Specifically, from the derivation of the C classical capacity region, the private classical capacity region P is derived as follows. In particular, the private classical capacity region P is precisely defined as

$$\text{P} = \mathcal{H}\left(\bigcup_{z_1, z_2} \text{P}(z_1, z_2)\right), \tag{193}$$

where

$$P(C\mathbb{N}, C\mathbb{N}) = S\left[\bigcup_{\forall p_{x'}} P(p_{x'}, p_{x'})\right] = S\bigcup_{\forall p_{x'}} \mathcal{H}\left[\bigcup_{x'_1, x'_2} P(x'_1, x'_2)\right]. \quad (194)$$

For two users, $U_1$ and $U_2$, the P region of secret key rates $(S_1, S_2)$ of AMQD-MQA is as follows:

$$\begin{aligned} S_1 &\leq I_{\text{MQA}}(z_1 : y, F(T(\mathcal{N}))|z_2) - I_{\text{MQA}}(y : E_1) \\ &= I_{\text{MQA}}(z_1 : y|F(T(\mathcal{N})), z_2) - I_{\text{MQA}}(y : E_1), \\ S_2 &\leq I_{\text{MQA}}(z_2 : y, F(T(\mathcal{N}))|z_1) - I_{\text{MQA}}(y : E_2) \\ &= I_{\text{MQA}}(z_2 : y|F(T(\mathcal{N})), z_1) - I_{\text{MQA}}(y : E_2), \end{aligned} \quad (195)$$

where $I_{\text{MQA}}(\cdot)$ is the related correlation measure function at an optimal Gaussian collective attack between Bob and Eve, or Alice and Eve (e.g., $I(B:E)$, $\chi(B:E)$ or $I(A:E)$, $\chi(A:E)$ depending on the direction of the reconciliation and the measurement type of the actual CVQKD protocol), $E_k$ is Eve's variable in the attacking of user's $U_k$ transmission, and

$$\begin{aligned} S_1 + S_2 &\leq I_{\text{MQA}}(z_1, z_2 : y, F(T(\mathcal{N}))) - (I_{\text{MQA}}(y : E_1) + I_{\text{MQA}}(y : E_2)) \\ &= I_{\text{MQA}}(z_1, z_2 : y|F(T(\mathcal{N}))) - (I_{\text{MQA}}(y : E_1) + I_{\text{MQA}}(y : E_2)). \end{aligned} \quad (196)$$

Thus, for a single user $U_k$, the secret key rate $S_k^{\text{MQA}}$ is precisely

$$S_k^{\text{MQA}} \leq \max_{\forall i}\left(\sum_l \log_2\left(1 + \frac{\sigma_{\omega_i}^2 |F(T_i(\mathcal{N}_i))|^2}{\sigma_{\mathcal{N}}^2}\right) - I_{\text{MQA}}(y : E_k)\right), \; k \in 1, \ldots K. \quad (197)$$

Specifically, for $K$ users, the *sum secret key rate* (the overall secret key rate throughput of the users) $S_{\text{sum}}^{\text{MQA}}$ is precisely calculated as

$$S_{\text{sum}}^{\text{MQA}} = \sum_K S_k^{\text{MQA}} \leq \max_{\forall i}\left(\sum_l \log_2\left(1 + \frac{\sigma_{\omega_i}^2 |F(T_i(\mathcal{N}_i))|^2}{\sigma_{\mathcal{N}}^2}\right) - \sum_K I_{\text{MQA}}(y : E_k)\right), \quad (198)$$

where $\sum_K I_{\text{MQA}}(y : E_k)$ is the amount of the eavesdropped information of the $K$ users, quantified by $I(B:E)$, $\chi(B:E)$ or $I(A:E)$, $\chi(A:E)$ functions for each $U_k$, depending on the attributes of the CVQKD protocol.

The *symmetric secret key rate* (the common rate at which all $K$ users can have a reliable simultaneous secret communication) $S_{\text{sym}}^{\text{MQA}}$ is calculated as

$$S_{\text{sym}}^{\text{MQA}} \leq \tfrac{1}{K}\left(\max_{\forall i}\left(\sum_l \log_2\left(1 + \frac{\sigma_{\omega_i}^2 |F(T_i(\mathcal{N}_i))|^2}{\sigma_{\mathcal{N}}^2}\right) - \sum_K I_{\text{MQA}}(y : E_k)\right)\right). \quad (199)$$

For $K$ users $U_1, \ldots U_K$, the $P_{\text{sum}}(\mathcal{N})$ *sum private classical capacity* and the $P_{\text{sym}}(\mathcal{N})$ *symmetric private classical capacity* of $\mathcal{N}$ are precisely expressed as

$$P_{\text{sum}}(\mathcal{N}) = \max_{(S_1,\ldots,S_K)\in \text{P}} \sum_K S_i$$
$$= \max_{\forall i}\left[\sum_l \log_2\left(1 + \frac{\sigma_{\omega_i}^2 |F(T_i(\mathcal{N}_i))|^2}{\sigma_{\mathcal{N}}^2}\right) - \sum_K I_{\text{MQA}}(y:E_k)\right] \quad (200)$$
$$= \max_{\forall i}\left[\sum_l \left(\log_2\left(1 + \frac{\sigma_{\omega_i}^2 |F(T_i(\mathcal{N}_i))|^2}{\sigma_{\mathcal{N}_i^*}^2}\right)\right)\right],$$

where $\sigma_{\mathcal{N}_i^*}^2$ was shown in (69), and for $P_{\text{sym}}(\mathcal{N})$,

$$P_{\text{sym}}(\mathcal{N}) = \max_{(S_{\text{sym}},\ldots,S_{\text{sym}})\in \text{P}} S_{\text{sym}}$$
$$= \frac{1}{K}\left[\max_{\forall i}\left[\sum_l \log_2\left(1 + \frac{\sigma_{\omega_i}^2 |F(T_i(\mathcal{N}_i))|^2}{\sigma_{\mathcal{N}}^2}\right) - \sum_K I_{\text{MQA}}(y:E_k)\right]\right] \quad (201)$$
$$= \frac{1}{K}\max_{\forall i}\left[\sum_l \left(\log_2\left(1 + \frac{\sigma_{\omega_i}^2 |F(T_i(\mathcal{N}_i))|^2}{\sigma_{\mathcal{N}_i^*}^2}\right)\right)\right].$$

Particularly, the AMQD-MQA secret key rates $S_1$ and $S_2$ of users $U_1$ and $U_2$ are the corresponding secret key rates obtained in Sections 3.1–3.2.

Specifically, the corner points are the $P_k$ private classical capacities, for $U_1$ and $U_2$ evaluated as

$$P_1 = \max_{\forall \rho_i} S_1 = \max_{\forall \rho_i}(R_1 - I_{\text{MQA}}(y:E_1)), \quad (202)$$

and

$$P_2 = \max_{\forall \rho_i} S_2 = \max_{\forall \rho_i}(R_2 - I_{\text{MQA}}(y:E_2)), \quad (203)$$

where $R_k$ is the rate of classical communication of user $U_k$—see (190)—measured by either the $I(A:B_k)$ or $\chi(A:B_k)$ between $A$ (Alice) and the $k$-th Bob, $B_k$, while $I_{\text{MQA}}(y:E_k)$ is Eve's classical communication rate, quantified by $I(B:E_k)$, $\chi(B:E_k)$ or $I(A:E_k)$, $\chi(A:E_k)$, respectively.

∎

## 3.4 SVD-Assisted AMQD-MQA Secret Key Rates

The SVD-assisted AMQD-MQA secret key rates $S_1'$ and $S_2'$ of users $U_1$ and $U_2$ are the related secret key rates obtained in Section 3.3 and in the Appendix.

The corresponding transmission rates are evaluated as follows:

$$C_{\text{sum}}'(\mathcal{N}) = \max_{\forall i}\sum_l \log_2\left(1 + \frac{\sigma_{\omega_i''}^2 |F(T_i(\mathcal{N}_i))|^2}{\sigma_{\mathcal{N}}^2}\right), \quad (204)$$

$$C_{\text{sym}}'(\mathcal{N}) = \frac{1}{K}\max_{\forall i}\left[\sum_l \log_2\left(1 + \frac{\sigma_{\omega_i''}^2 |F(T_i(\mathcal{N}_i))|^2}{\sigma_{\mathcal{N}}^2}\right)\right], \quad (205)$$

thus

$$S'_{\text{sum}}(\mathcal{N}) \leq \max_{\forall i}\left[\sum_l \log_2\left(1 + \frac{\sigma^2_{\omega''_i}|F(T_i(\mathcal{N}_i))|^2}{\sigma^2_{\mathcal{N}}}\right) - \sum_K I'_{\text{MQA}}(y:E_k)\right], \tag{206}$$

$$S'^{\text{MQA}}_{\text{sym}} \leq \tfrac{1}{K}\max_{\forall i}\left[\sum_l \log_2\left(1 + \frac{\sigma^2_{\omega''_i}|F(T_i(\mathcal{N}_i))|^2}{\sigma^2_{\mathcal{N}}}\right) - \sum_K I'_{\text{MQA}}(y:E_k)\right], \tag{207}$$

where

$$\sum_K I'_{\text{MQA}}(y:E_k) \leq \sum_K I_{\text{MQA}}(y:E_k), \tag{208}$$

which precisely follows from (165). Particularly, for $K$ users $U_1,\ldots U_K$, the sum private classical $P'_{\text{sum}}(\mathcal{N})$ and symmetric private classical $P'_{\text{sym}}(\mathcal{N})$ capacities in the SVD-assisted AMQD-MQA are expressed precisely as

$$\begin{aligned}
P'_{\text{sum}}(\mathcal{N}) &= \max_{\forall i}\left[\sum_l \log_2\left(1 + \frac{\sigma^2_{\omega''_i}|F(T_i(\mathcal{N}_i))|^2}{\sigma^2_{\mathcal{N}}}\right) - \sum_K I'_{\text{MQA}}(y:E_k)\right] \\
&= \max_{\forall i}\left[\sum_l \left(\log_2\left(1 + \frac{\sigma^2_{\omega''_i}|F(T_i(\mathcal{N}_i))|^2}{\sigma^2_{\mathcal{N}}}\right) - \sum_K I'_{\text{MQA}}(y:E_k)\right)\right] \\
&= \max_{\forall i}\left[\sum_l \left(\log_2\left(1 + \frac{\sigma^2_{\omega_i}|F(T'_i(\mathcal{N}_i))|^2}{\sigma^2_{\mathcal{N}}}\right) - \sum_K I'_{\text{MQA}}(y:E_k)\right)\right] \\
&= \max_{\forall i}\left[\sum_l \left(\log_2\left(1 + \frac{\sigma^2_{\omega_i}|F(T'_i(\mathcal{N}_i))|^2}{\sigma^2_{\mathcal{N}^*_i}}\right)\right)\right],
\end{aligned} \tag{209}$$

where $\sigma^2_{\mathcal{N}^*_i}$ was shown in (69). Thus for any $v_i > 1$, without loss of generality,

$$P'_{\text{sum}}(\mathcal{N}) > P_{\text{sum}}(\mathcal{N}), \tag{210}$$

which follows from (168) for $\mathcal{N}_i$, since

$$|F(T'_i(\mathcal{N}_i))|^2 = |F(T_i(\mathcal{N}_i))|^2 \frac{\sigma^2_{\omega''_i}}{\sigma^2_{\omega_i}} > |F(T_i(\mathcal{N}_i))|^2. \tag{211}$$

In particular, the $P'_{\text{sym}}(\mathcal{N})$ SVD-assisted symmetric private classical capacity is precisely as

$$\begin{aligned}
P'_{\text{sym}}(\mathcal{N}) &= \tfrac{1}{K}\max_{\forall i}\left[\sum_l \log_2\left(1 + \frac{\sigma^2_{\omega''_i}|F(T_i(\mathcal{N}_i))|^2}{\sigma^2_{\mathcal{N}}}\right) - \sum_K I_{\text{MQA}}(y:E_k)\right] \\
&= \tfrac{1}{K}\max_{\forall i}\left[\sum_l \left(\log_2\left(1 + \frac{\sigma^2_{\omega''_i}|F(T_i(\mathcal{N}_i))|^2}{\sigma^2_{\mathcal{N}}}\right) - \sum_K I_{\text{MQA}}(y:E_k)\right)\right] \\
&= \tfrac{1}{K}\max_{\forall i}\left[\sum_l \left(\log_2\left(1 + \frac{\sigma^2_{\omega_i}|F(T'_i(\mathcal{N}_i))|^2}{\sigma^2_{\mathcal{N}}}\right) - \sum_K I_{\text{MQA}}(y:E_k)\right)\right] \\
&= \tfrac{1}{K}\max_{\forall i}\left[\sum_l \left(\log_2\left(1 + \frac{\sigma^2_{\omega_i}|F(T'_i(\mathcal{N}_i))|^2}{\sigma^2_{\mathcal{N}^*}}\right)\right)\right],
\end{aligned} \tag{212}$$

where for any $v_i > 1$, without loss of generality,

$$P'_{\text{sym}}(\mathcal{N}) > P_{\text{sym}}(\mathcal{N}). \tag{213}$$

The $P_k'$ corner point coincidences to the private classical capacity of user $U_k$:

$$P_1' = \max_{\forall \rho_i} S_1' = \max_{\forall \rho_i}\left(R_1 - I_{\text{MQA}}(y:E_1')\right) \tag{214}$$

and

$$P_2' = \max_{\forall \rho_i} S_2' = \max_{\forall \rho_i}\left(R_2 - I_{\text{MQA}}(y:E_2')\right), \tag{215}$$

where $R_k$ is the rate of classical communication between Alice and Bob $k$, measured by either the $I(A:B_k)$ or $\chi(A:B_k)$, while $I_{\text{MQA}}(y:E_k') < I_{\text{MQA}}(y:E_k)$ is the SVD-assisted rate of Eve, quantified by $I(B:E_k')$, $\chi(B:E_k')$ or $I(A:E_k')$, $\chi(A:E_k')$ (see Section 3.3).

The AMQD-MQA secret key rate ratios for users $U_1$ and $U_2$ are depicted in Figure 2. The classical and the private classical capacity regions are denoted by C and P. The $S_1$ and $S_2$ represent the ratio of the corresponding secret key formulas, precisely $R_{one-way}^{RR,M_{\text{hom}}}$, $R_{one-way}^{RR,M_{\text{het}}}$, $R_{two-way}^{RR,M_{\text{hom}}}$, or $R_{two-way}^{RR,M_{\text{het}}}$ between users $U_1$ and $U_2$, obtained in Section 3.1–3.3, depending on the actual setting of AMQD-MQA.

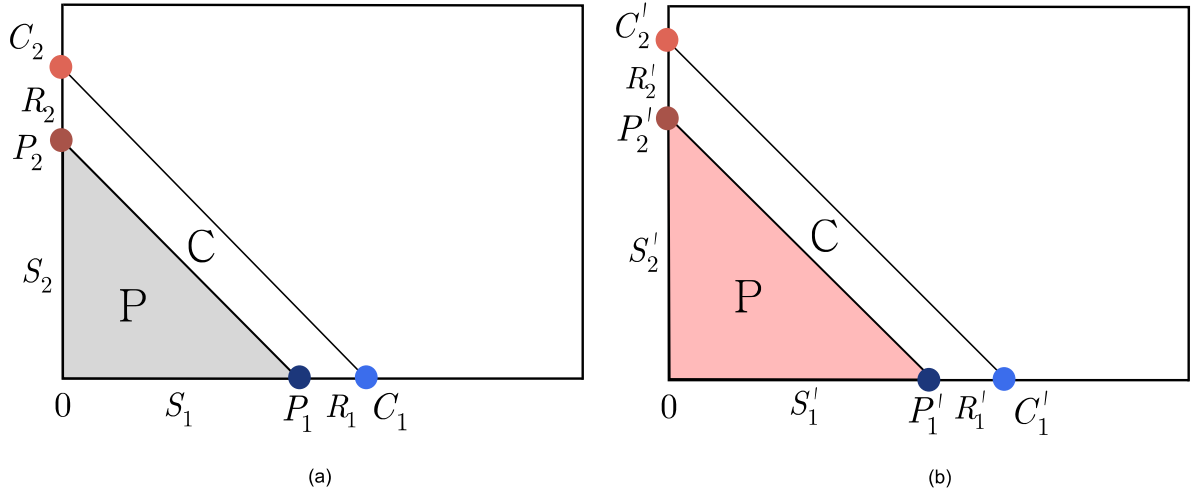

**Figure 2.** The P private classical capacity region of AMQD-MQA (a) and of SVD-assisted AMQD-MQA (b).

The SVD-assistance leads to an improved $I'(A:B_k)$ or $\chi'(A:B_k)$ between Alice and Bob. See (165). In particular, this enhancement, overall, results in an increased secret key rates $S_1'$ and $S_2'$ for users $U_1$ and $U_2$.

∎

These results conclude the proof of Theorem 2.

∎

# 4 Physical Boundaries of an Optimal Gaussian Attack

**Theorem 3** (Tolerable excess noise of AMQD). *The AMQD leads to an improved amount of tolerable excess noise in comparison to single-carrier CVQKD.*

*Proof.*

As we have seen in Section 3, from the variance $\sigma_X^2$ of the Gaussian link, the excess noise is evaluated as

$$N = \sigma_X^2 - \sigma_o^2 = \frac{\left(\frac{1}{l}\sum_l W_i - 1\right)\left(\frac{1}{l}\sum_l |F(T_{Eve,i})|^2\right)}{1 - \frac{1}{l}\sum_l |F(T_{Eve,i})|^2}, \tag{216}$$

where $\sigma_o^2$ is the vacuum noise. From [1–3] for the RR of AMQD, the following inequalities bring up for the $N_{tol,AMQD}^{RR,coh.}$ maximum tolerable excess noise, in a coherent state transmission:

$$N_{tol,AMQD}^{RR,coh.}$$
$$= \left(\frac{1}{l}\sum_l |F(T_i(\mathcal{N}_i))|^2 \sigma_X^2 + \frac{1}{l}\sum_l |F(T_i(\mathcal{N}_i))|^2\right)\left(\frac{1}{l}\sum_l |F(T_i(\mathcal{N}_i))|^2 \sigma_X^2 + \frac{\frac{1}{l}\sum_l |F(T_i(\mathcal{N}_i))|^2}{\sigma_{\omega_0}^2}\right) < 1. \tag{217}$$

Thus from (216),

$$N_{tol,AMQD}^{RR,coh.}$$
$$= \left[\left(\sigma_o^2 + \left(\frac{\left(\frac{1}{l}\sum_l W_i - 1\right)\left(\frac{1}{l}\sum_l |F(T_{Eve,i})|^2\right)}{1 - \frac{1}{l}\sum_l |F(T_{Eve,i})|^2}\right)\right)\left(1 - \frac{1}{l}\sum_l |F(T_{Eve,i}(\mathcal{N}_i))|^2\right) + \left(1 - \frac{1}{l}\sum_l |F(T_{Eve,i}(\mathcal{N}_i))|^2\right)\right]$$
$$\cdot \left[\left(1 - \frac{1}{l}\sum_l |F(T_{Eve,i}(\mathcal{N}_i))|^2\right)\left(\sigma_o^2 + \left(\frac{\left(\frac{1}{l}\sum_l W_i - 1\right)\left(\frac{1}{l}\sum_l |F(T_{Eve,i})|^2\right)}{1 - \frac{1}{l}\sum_l |F(T_{Eve,i})|^2}\right)\right) + \frac{\left(1 - \frac{1}{l}\sum_l |F(T_{Eve,i}(\mathcal{N}_i))|^2\right)}{\sigma_{\omega_0}^2}\right] < 1. \tag{218}$$

Thus, for squeezed states [1–3], the tolerable excess noise is precisely as follows:

$$N_{tol,AMQD}^{RR,squ.}$$
$$= \left(1 - \frac{1}{\sigma_{\omega_0}^2} - \frac{1}{\left(1 - \frac{1}{l}\sum_l |F(T_{Eve,i}(\mathcal{N}_i))|^2\right)}\right) - \frac{1}{2}\left(s - \frac{1}{\sigma_{\omega_0}^2}\right)$$
$$+ \sqrt{\frac{1}{\left(1 - \frac{1}{l}\sum_l |F(T_{Eve,i}(\mathcal{N}_i))|^2\right)^2} + \frac{1}{4}\left(s - \frac{1}{\sigma_{\omega_0}^2}\right)^2} > N_{tol,AMQD}^{RR,coh.} < 1. \tag{219}$$

Specifically, the inequality of (219) can be sharpened for the RR AMQD settings as follows. Using coherent states for the transmission, the upper bound on the $N_{tol,AMQD}$ tolerable excess noise in an AMQD modulation in one-way CVQKD, for the case of RR, is as follows.

In particular, the upper bound on the tolerable excess noise, $N_{tol,single}^{RR,one-way}$, is precisely evaluated as [3]

$$N_{tol,AMQD}^{RR,one-way} = \alpha N_{tol,single}^{RR,one-way}, \tag{220}$$

where $\alpha > 1$, and

$$N_{tol,single}^{RR,one-way} = \tfrac{1}{2}\left(\sqrt{1+\tfrac{16}{e^2}}-1\right) \approx 0.39, \tag{221}$$

thus

$$\begin{aligned}N_{tol,AMQD}^{RR,one-way} &= \alpha\left(\tfrac{1}{2}\left(\sqrt{1+\tfrac{16}{e^2}}-1\right)\right) \approx \alpha 0.39 \\ &= \tfrac{1}{2}\left(\sqrt{1+\tfrac{16}{e^2}}-1\right) + c^{RR},\end{aligned} \tag{222}$$

where $c^{RR} > 0$ is a constant, expressed precisely as

$$\begin{aligned}c^{RR} &= N_{tol,AMQD}^{RR,coh.} - N_{tol}^{RR,coh.} \\ &= \left(\left(\tfrac{1}{l}\sum_l |F(T_i(\mathcal{N}_i))|^2 \sigma_X^2 + \tfrac{1}{l}\sum_l |F(T_i(\mathcal{N}_i))|^2\right)\left(\tfrac{1}{l}\sum_l |F(T_i(\mathcal{N}_i))|^2 \sigma_X^2 + \tfrac{\tfrac{1}{l}\sum_l |F(T_i(\mathcal{N}_i))|^2}{\sigma_{\omega_0}^2}\right)\right) \\ &\quad -\left(|T(\mathcal{N}_i)|^2 \sigma_X^2 + |T(\mathcal{N}_i)|^2\right)\left(|T(\mathcal{N}_i)|^2 \sigma_X^2 + \tfrac{|T(\mathcal{N}_i)|^2}{\sigma_{\omega_0}^2}\right).\end{aligned} \tag{223}$$

From a security argument linked to the entangling cloning attack [1], the following upper bound can be derived for the maximal tolerable excess noise for the DR of AMQD:

$$N_{tol,AMQD}^{DR} < 2 - \frac{1}{\tfrac{1}{l}\sum_l |F(T_i(\mathcal{N}_i))|^2}. \tag{224}$$

Without loss of generality, the inequality of can be sharpened for the DR AMQD settings as follows [3]:

$$N_{tol,AMQD}^{DR,one-way} = \beta N_{tol,single}^{DR,one-way} = 0.8 + c^{DR}, \tag{225}$$

where $\beta > 1$, and

$$N_{tol,single}^{DR,one-way} \approx 0.8, \tag{226}$$

where $c^{DR} > 0$ is a constant, expressed precisely as

$$\begin{aligned}c^{DR} &= N_{tol,AMQD}^{DR,one-way} - N_{tol}^{DR,one-way} \\ &= \frac{1}{|T(\mathcal{N})|^2} - \frac{1}{\tfrac{1}{l}\sum_l |F(T_i(\mathcal{N}_i))|^2}.\end{aligned} \tag{227}$$

The result in follows from the following equation [3]:

$$\frac{1}{1+N_{tol,single}^{DR,one-way}}\left(\frac{\sqrt{1+N_{tol,single}^{DR,one-way}}+1}{\sqrt{1+N_{tol,single}^{DR,one-way}}-1}\right)^{\sqrt{1+N_{tol,single}^{DR,one-way}}} = e^2, \tag{228}$$

thus,

$$\frac{1}{1+N_{tol,AMQD}^{DR,one-way}}\left(\frac{\sqrt{1+N_{tol,AMQD}^{DR,one-way}}+1}{\sqrt{1+N_{tol,AMQD}^{DR,one-way}}-1}\right)^{\sqrt{1+N_{tol,AMQD}^{DR,one-way}}} = e^2 - \wp, \tag{229}$$

for some $\wp > 0$.

In particular, for two-way CVQKD, these upper bounds change as follows:

$$N_{tol,AMQD}^{RR,two-way} = \alpha' N_{tol,single}^{RR,two-way} \approx \alpha' N_{tol,single}^{DR,one-way} \approx \alpha' 0.8, \tag{230}$$

and

$$N_{tol,AMQD}^{DR,two-way} = \beta' N_{tol,single}^{DR,two-way} \approx \beta' 0.75. \tag{231}$$

Without loss of generality, the improvement in the tolerable excess noise can be approached by the ratio $\kappa \geq 1$ between the excess noise $N_{single} = (W-1)\left(\left|T_{Eve}\right|^2\right)\Big/1-\left|T_{Eve}\right|^2$ and $N_{AMQD}$ as

$$\begin{aligned}\kappa &= \frac{N_{single}}{N_{AMQD}}\\ &= \frac{(W-1)\left(\left|T_{Eve}\right|^2\right)}{1-\left|T_{Eve}\right|^2}\frac{1-\frac{1}{l}\sum_l\left|F(T_{Eve,i})\right|^2}{\left(\frac{1}{l}\sum_l W_i -1\right)\left(\frac{1}{l}\sum_l\left|F(T_{Eve,i})\right|^2\right)}\\ &= \frac{(W-1)\left|T_{Eve}\right|^2 - \left|T_{Eve}\right|^2\frac{1}{l}\sum_l\left|F(T_{Eve,i})\right|^2}{\left(\frac{1}{l}\sum_l W_i -1\right)\frac{1}{l}\sum_l\left|F(T_{Eve,i})\right|^2 - \left|T_{Eve}\right|^2\frac{1}{l}\sum_l\left|F(T_{Eve,i})\right|^2}\\ &= \frac{(W-1)\left|T_{Eve}\right|^2 - \left|T_{Eve}\right|^2\frac{1}{l}\sum_l\left|F(T_{Eve,i})\right|^2}{\left(\frac{1}{l}\sum_l W_i -1\right)\left|T\right|^2\frac{1}{l}\sum_l\left|F(T_{Eve,i})\right|^2}.\end{aligned} \quad (232)$$

In a precise form, the maximal amount of tolerable excess noise $N_{tol,AMQD}^{RR,one-way}$, $N_{tol,AMQD}^{DR,one-way}$ and $N_{tol,AMQD}^{RR,two-way}$, $N_{tol,AMQD}^{DR,two-way}$ in an AMQD modulation is determined at zero secret key rates [1-3]. At a homodyne measurement $M_{\text{hom}}$, the tolerable excess noise is derived at

$$S_{one-way}^{RR,M_{\text{hom}}} = S_{one-way}^{DR,M_{\text{hom}}} = S_{two-way}^{RR,M_{\text{hom}}} = S_{two-way}^{RR,M_{\text{hom}}} = 0, \quad (233)$$

while for the heterodyne measurement $M_{\text{het}}$, it is determined from

$$S_{one-way}^{RR,M_{\text{het}}} = S_{one-way}^{DR,M_{\text{het}}} = S_{two-way}^{RR,M_{\text{het}}} = S_{two-way}^{RR,M_{\text{het}}} = 0. \quad (234)$$

For the corresponding rate formulas, see Section 3.

The results are summarized in Figure 3. The tolerable excess noise is $N_{tol,AMQD} = \chi N_{tol,single}$, where $\chi = x\kappa = x\,N_{single}\big/N_{AMQD} \geq 1$ and $N_{AMQD}\big/N_{single} \leq x \leq 1$.

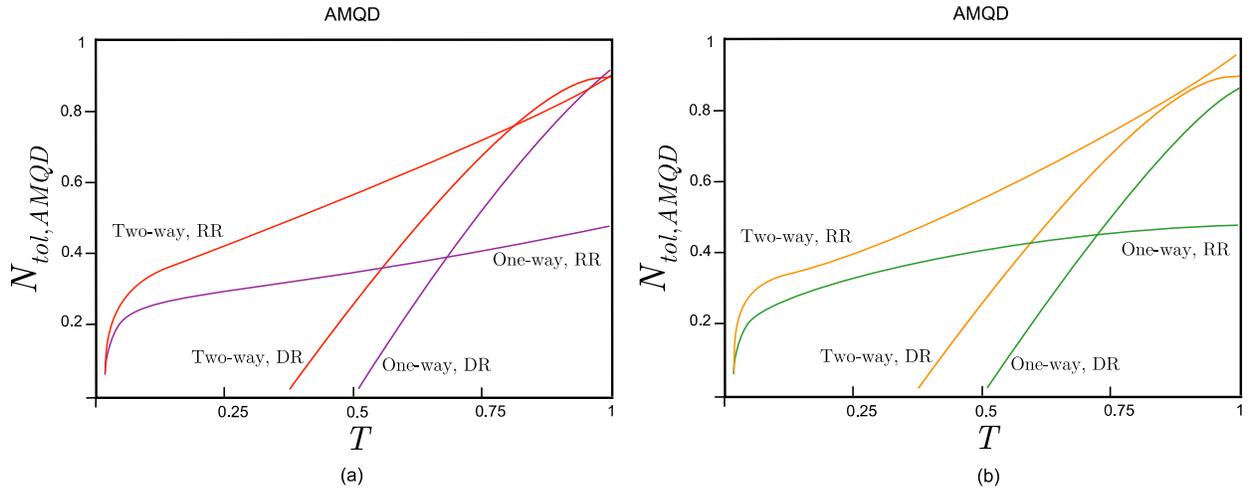

**Figure 3.** Tolerable excess noise as a function of channel transmittance in an AMQD modulation with homodyne (a) and heterodyne measurements (b). Abbreviations: DR – Direct Reconciliation, RR – Reverse Reconciliation.

Specifically, the additional degree of freedom injected by SVD into the transmission leads to an increased tolerable excess noise, as

$$N_{AMQD}^{SVD} = \frac{1-\frac{1}{l}\sum_{l}\left|F\left(T'_{Eve,i}\right)\right|^{2}}{\left(\frac{1}{l}\sum_{l}W_{i}-1\right)\left(\frac{1}{l}\sum_{l}\left|F\left(T'_{Eve,i}\right)\right|^{2}\right)}, \tag{235}$$

where

$$\tfrac{1}{l}\sum_{l}\left|F\left(T'_{Eve,i}\right)\right|^{2} < \tfrac{1}{l}\sum_{l}\left|F\left(T_{Eve,i}\right)\right|^{2}, \tag{236}$$

thus

$$N_{tol,AMQD}^{SVD} > \mathrm{M}\chi N_{tol,single}, \tag{237}$$

where $\mathrm{M} > 1$. In particular, the improvement in the amount of tolerable excess noise is present more significantly in two-way CVQKD, since the multicarrier-based transmission enhances the rates of private classical communication for each channel uses, $\mathcal{M}_1$ and $\mathcal{M}_2$, between Alice to Bob.

The physically allowed $\frac{1}{l}\sum_{l}W_{i}$ variances of an optimal Gaussian collective attack as a function of $\frac{1}{l}\sum_{l}\left|F\left(T_{i}\left(\mathcal{N}_{i}\right)\right)\right|^{2}$ of the $l$ Gaussian sub-channels $\mathcal{N}_{i}$ are depicted in Figure 4. Specifically, the upper bounds on the $\frac{1}{l}\sum_{l}W_{i}$ variances reveal those physical boundaries at which a non-zero secret rate is possible through a noisy Gaussian quantum link $\mathcal{N}$. Precisely, by exceeding these physical boundaries on the variance of Eve's EPR state, no secret transmission is possible between the legal parties; thus Eve's variance $W$ is restricted to the range of $0 < W_i \leq \frac{1}{l}\sum_{l}W_{i}$.

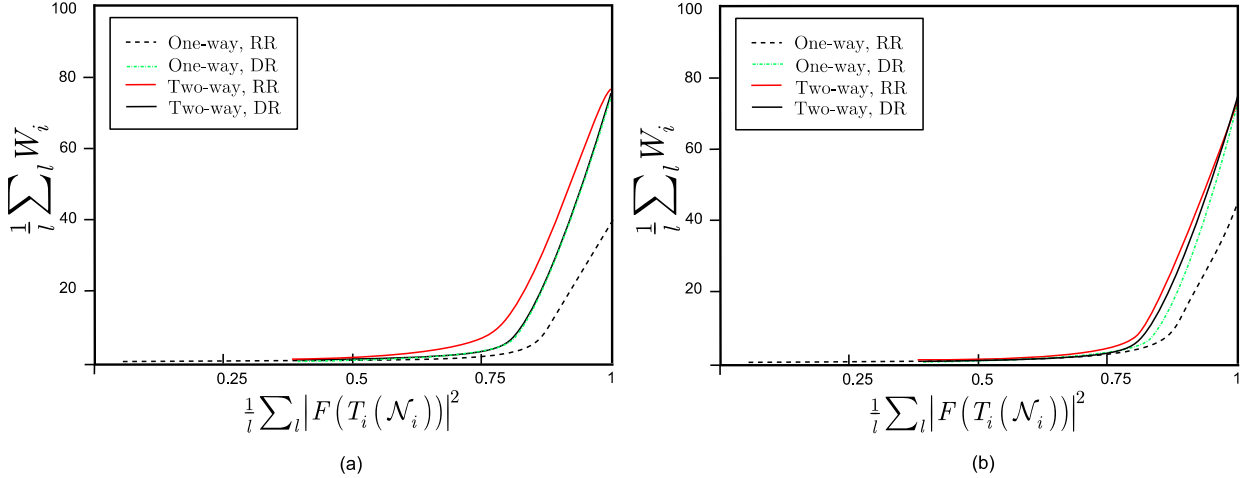

**Figure 4.** The physical boundaries of Eve's variance in an optimal Gaussian attack in AMQD modulation for homodyne (a) and heterodyne measurements (b). Abbreviations: DR – Direct Reconciliation, RR – Reverse Reconciliation.

Particularly, the boundaries on $\frac{1}{l}\sum_{l}W_{i}$ confirm that AMQD tolerates higher amounts of excess noise in case of $M_{\mathrm{hom}}$, in comparison to $M_{\mathrm{het}}$.

∎

# 5 Conclusions

The multicarrier CVQKD extends the possibilities of current, standard single-carrier CVQKD. We studied the security thresholds of multicarrier CVQKD, through AMQD modulation, and its extended SVD-assisted version. The results have also been extended to a multiuser environment, through the analysis of AMQD-MQA. We proved the secret key rate formulas of multicarrier CVQKD and determined the security parameters of the transmission. We investigated the private classical capacity of a Gaussian sub-channel and characterized the private classical capacity domain. The physical boundaries of optimal Gaussian attacks against multicarrier transmission have also been revealed, along with the maximal tolerable variance parameters. The results revealed that multicarrier CVQKD provides a considerably improved framework for the experimental implementation of an unconditionally secure communication over standard telecommunication networks.

# Acknowledgements

The author would like to thank Professor Sandor Imre for useful discussions. This work was partially supported by the GOP-1.1.1-11-2012-0092 (*Secure quantum key distribution between two units on optical fiber network*) project sponsored by the EU and European Structural Fund, and by the COST Action MP1006.

# References

[1] F. Grosshans, N.J. Cerf, J. Wenger, R. Tualle-Brouri, P. Grangier, Virtual entanglement and reconciliation protocols for quantum cryptography with continuous variables. *Quant. Info. and Computation 3, 535-552* (2003).

[2] S. Pirandola, S. Mancini, S. Lloyd, and S. L. Braunstein, Continuous-variable Quantum Cryptography using Two-Way Quantum Communication, *arXiv:quant-ph/0611167v3* (2008).

[3] M. Navascues, A Acin, Security bounds for continuous variables quantum key distribution. *Phys. Rev. Lett.* 94, 020505 (2005).

[4] L. Gyongyosi, Adaptive Multicarrier Quadrature Division Modulation for Continuous-variable Quantum Key Distribution, *arXiv:1310.1608* (2013).

[5] L. Gyongyosi, Multiuser Quadrature Allocation for Continuous-Variable Quantum Key Distribution, *arXiv:1312.3614* (2013).

[6] L. Gyongyosi, Singular Layer Transmission for Continuous-Variable Quantum Key Distribution, *arXiv:1402.5110* (2014).

[7] F. Grosshans, P. Grangier, Reverse reconciliation protocols for quantum cryptography with continuous variables, *arXiv:quant-ph/0204127v1* (2002).

[8] F. Grosshans, et al. Quantum key distribution using Gaussian-modulated coherent states. *Nature* 421, 238-241 (2003).

[9] S. Pirandola, R. Garcia-Patron, S. L. Braunstein and S. Lloyd. *Phys. Rev. Lett.* 102 050503. (2009).

[10] S. Pirandola, A. Serafini and S. Lloyd. *Phys. Rev. A* 79 052327. (2009).

[11] S. Pirandola, S. L. Braunstein and S. Lloyd. *Phys. Rev. Lett.* 101 200504 (2008).

[12] C. Weedbrook, S. Pirandola, S. Lloyd and T. Ralph. *Phys. Rev. Lett.* 105 110501 (2010).

[13] C. Weedbrook, S. Pirandola, R. Garcia-Patron, N. J. Cerf, T. Ralph, J. Shapiro, and S. Lloyd. *Rev. Mod. Phys.* 84, 621 (2012).

[14] William Shieh and Ivan Djordjevic. *OFDM for Optical Communications.* Elsevier (2010).

[15] L. Gyongyosi, Scalar Reconciliation for Gaussian Modulation of Two-Way Continuous-variable Quantum Key Distribution, *arXiv:1308.1391* (2013).

[16] P. Jouguet, S. Kunz-Jacques, A. Leverrier, P. Grangier, E. Diamanti, Experimental demonstration of long-distance continuous-variable quantum key distribution, *arXiv:1210.6216v1* (2012).

[17] M. Navascues, F. Grosshans, and A. Acin. Optimality of Gaussian Attacks in Continuous-variable Quantum Cryptography, *Phys. Rev. Lett. 97, 190502* (2006).

[18] R. Garcia-Patron and N. J. Cerf. Unconditional Optimality of Gaussian Attacks against Continuous-Variable Quantum Key Distribution. *Phys. Rev. Lett. 97, 190503* (2006).

[19] F. Grosshans, Collective attacks and unconditional security in continuous variable quantum key distribution. *Phys. Rev. Lett. 94,* 020504 (2005).

[20] M R A Adcock, P Høyer, and B C Sanders, Limitations on continuous-variable quantum algorithms with Fourier transforms, *New Journal of Physics 11 103035* (2009)

[21] L. Hanzo, H. Haas, S. Imre, D. O'Brien, M. Rupp, L. Gyongyosi. Wireless Myths, Realities, and Futures: From 3G/4G to Optical and Quantum Wireless, *Proceedings of the IEEE*, Volume: 100, *Issue: Special Centennial Issue*, pp. 1853-1888. (2012).

[22] S. Imre and L. Gyongyosi. *Advanced Quantum Communications - An Engineering Approach.* Wiley-IEEE Press (New Jersey, USA), (2012).

[23] D. Tse and P. Viswanath. *Fundamentals of Wireless Communication*, Cambridge University Press, (2005).

[24] David Middlet, *An Introduction to Statistical Communication Theory: An IEEE Press Classic Reissue*, Hardcover, IEEE, ISBN-10: 0780311787, ISBN-13: 978-0780311787 (1960)

[25] Steven Kay, *Fundamentals of Statistical Signal Processing, Volumes I-III*, Prentice Hall, (2013)

[26] S. Imre, F. Balazs: *Quantum Computing and Communications – An Engineering Approach*, John Wiley and Sons Ltd, ISBN 0-470-86902-X, 283 pages (2005).

[27] D. Petz, *Quantum Information Theory and Quantum Statistics*, Springer-Verlag, Heidelberg, Hiv: 6. (2008).

[28] L. Gyongyosi, S. Imre: Properties of the Quantum Channel, *arXiv:1208.1270* (2012).

# Supplemental Information

## S.1 Homodyne and Heterodyne Measurements

In a *homodyne* measurement $M_{\text{hom}}$, only one quadrature, the position or the momentum quadrature is measured. The variances of Alice's quadratures are $\langle x_j^2 \rangle = \sigma_{\omega_0}^2$ and $\langle p_j^2 \rangle = \sigma_{\omega_0}^2$. Bob's quadratures have variances $\langle x_j'^2 \rangle = \langle p_j'^2 \rangle = \frac{1}{l}\sum_l |F(T_i(\mathcal{N}_i))|^2 \sigma_{\omega_0}^2 + \sigma_{\mathcal{N}}^2$, where $\frac{1}{l}\sum_l |F(T_i(\mathcal{N}_i))|^2$ is the Fourier transformed transmittance of $\mathcal{N}$, while $\sigma_{\mathcal{N}}^2$ is the noise variance of $\mathcal{N}$. The systems of Alice and Eve are modeled as independent coherent states, and using the estimator operators $e_{x_j'}$ and $e_{p_j'}$ on Alice's side, the following relations hold for the estimation of Bob's noisy quadrature $x_j'$ from $x_j$:

$$e_{x_j'} = \varepsilon_{A,x_j} x_j, \tag{S.1}$$

where

$$\varepsilon_{A,x_j} = \frac{\langle x_j' x_j \rangle}{\langle x_j^2 \rangle}, \tag{S.2}$$

$$\hat{\sigma}_{x_j'|e_{x_j'}}^2 = \left\langle \left( x_j' - e_{x_j'} \right)^2 \right\rangle = \langle x_j'^2 \rangle - \frac{|\langle x_j x_j' \rangle|^2}{\langle x_j^2 \rangle}, \tag{S.3}$$

and

$$e_{p_j'} = \varepsilon_{A,p_j} p_j \tag{S.4}$$

$$\varepsilon_{A,p_j} = \frac{\langle p_j' p_j \rangle}{\langle p_j^2 \rangle}, \tag{S.5}$$

$$\hat{\sigma}_{p_j'|e_{p_j'}}^2 = \left\langle \left( p_j' - e_{p_j'} \right)^2 \right\rangle = \langle p_j'^2 \rangle - \frac{|\langle p_j p_j' \rangle|^2}{\langle p_j^2 \rangle}. \tag{S.6}$$

From the estimators $e_{x_j'}$, $e_{p_j'}$ and the differences of $\left( x_j' - e_{x_j'} \right)$, $\left( p_j' - e_{p_j'} \right)$, Bob's quadratures $x_j'$ and $p_j'$ can be expressed as follows:

$$\begin{aligned} x_j' &= \left( x_j' - e_{x_j'} \right) + e_{x_j'}, \\ p_j' &= \left( p_j' - e_{p_j'} \right) + e_{p_j'}. \end{aligned} \tag{S.7}$$

For these quantities, the following inequalities hold

$$|\langle x_j x'_j \rangle|^2 \leq \langle x_j^2 \rangle \langle x_j'^2 \rangle - N_0^2 \frac{\langle x_j^2 \rangle}{\langle p_j'^2 \rangle}$$
$$= \sigma_{\omega_0}^2 \left( \tfrac{1}{l} \sum_l |F(T_i(\mathcal{N}_i))|^2 \sigma_{\omega_0}^2 + \sigma_{\mathcal{N}}^2 \right) - \frac{N_0^2 \sigma_{\omega_0}^2}{\tfrac{1}{l}\sum_l |F(T_i(\mathcal{N}_i))|^2 \sigma_{\omega_0}^2 + \sigma_{\mathcal{N}}^2}, \quad \text{(S.8)}$$

and

$$|\langle p_j p'_j \rangle|^2 \leq \langle p_j^2 \rangle \langle p_j'^2 \rangle - N_0^2 \frac{\langle p_j^2 \rangle}{\langle x_j'^2 \rangle}$$
$$= \sigma_{\omega_0}^2 \left( \tfrac{1}{l} \sum_l |F(T_i(\mathcal{N}_i))|^2 \sigma_{\omega_0}^2 + \sigma_{\mathcal{N}}^2 \right) - \frac{N_0^2 \sigma_{\omega_0}^2}{\tfrac{1}{l}\sum_l |F(T_i(\mathcal{N}_i))|^2 \sigma_{\omega_0}^2 + \sigma_{\mathcal{N}}^2}. \quad \text{(S.9)}$$

In case of *heterodyne* measurement $M_{\text{het}}$, both of the position and momentum quadratures are measured. Alice has both of the estimators $e_{x'_j}$ and $e_{p'_j}$ from the measurement, thus both quadratures are known for Alice. The commutation relations for Alice's quadrature pair $(x_j, p_j)$ are

$$[x_j, p_j] = [x'_j, p'_j] = 0, \quad \text{(S.10)}$$
$$[x_j - e_{x'_j}, p_j - e_{p'_j}] = -[x_j, p_j]. \quad \text{(S.11)}$$

Assuming that the estimators have variances $\langle e_{x'_j}^2 \rangle = \langle e_{p'_j}^2 \rangle = \tfrac{1}{l}\sum_l |F(T_i(\mathcal{N}_i))|^2 \sigma_{\omega_0}^2 + \sigma_{\mathcal{N}}^2$, the conditional variances $\hat{\sigma}^2_{x'_j|e_{x'_j}}$ and $\hat{\sigma}^2_{p'_j|e_{p'_j}}$ are reevaluated as follows:

$$\hat{\sigma}^2_{x'_j|e_{x'_j}} = \left\langle \left(x'_j - e_{x'_j}\right)^2 \right\rangle = \langle x_j'^2 \rangle - \frac{|\langle x_j x'_j \rangle|^2}{\langle x_j^2 \rangle}$$
$$= \left( \tfrac{1}{l}\sum_l |F(T_i(\mathcal{N}_i))|^2 \sigma_{\omega_0}^2 + \sigma_{\mathcal{N}}^2 - \frac{\left(\tfrac{1}{l}\sum_l |F(T_i(\mathcal{N}_i))|^2 \sigma_{\omega_0}^2 + \sigma_{\mathcal{N}}^2\right)^2 - 1}{\tfrac{1}{l}\sum_l |F(T_i(\mathcal{N}_i))|^2 \sigma_{\omega_0}^2 + \sigma_{\mathcal{N}}^2 + \text{H}} \right) N_0 \quad \text{(S.12)}$$
$$= \frac{\text{H}\left(\tfrac{1}{l}\sum_l |F(T_i(\mathcal{N}_i))|^2 \sigma_{\omega_0}^2 + \sigma_{\mathcal{N}}^2\right) + 1}{\tfrac{1}{l}\sum_l |F(T_i(\mathcal{N}_i))|^2 \sigma_{\omega_0}^2 + \sigma_{\mathcal{N}}^2 + \text{H}} N_0,$$

and

$$\hat{\sigma}^2_{p'_j|e_{p'_j}} = \left\langle \left(p'_j - e_{p'_j}\right)^2 \right\rangle = \langle p_j'^2 \rangle - \frac{|\langle p_j p'_j \rangle|^2}{\langle p_j^2 \rangle}$$
$$= \left( \tfrac{1}{l}\sum_l |F(T_i(\mathcal{N}_i))|^2 \sigma_{\omega_0}^2 + \sigma_{\mathcal{N}}^2 - \frac{\left(\tfrac{1}{l}\sum_l |F(T_i(\mathcal{N}_i))|^2 \sigma_{\omega_0}^2 + \sigma_{\mathcal{N}}^2\right)^2 - 1}{\tfrac{1}{l}\sum_l |F(T_i(\mathcal{N}_i))|^2 \sigma_{\omega_0}^2 + \sigma_{\mathcal{N}}^2 + \tfrac{1}{\text{H}}} \right) N_0 \quad \text{(S.13)}$$
$$= \frac{\tfrac{1}{l}\sum_l |F(T_i(\mathcal{N}_i))|^2 \sigma_{\omega_0}^2 + \sigma_{\mathcal{N}}^2 + \text{H}}{\text{H}\left(\tfrac{1}{l}\sum_l |F(T_i(\mathcal{N}_i))|^2 \sigma_{\omega_0}^2 + \sigma_{\mathcal{N}}^2\right) + 1} N_0 = \frac{N_0^2}{\hat{\sigma}^2_{x'_j|e_{x'_j}}},$$

where

$$\text{H} = \frac{1 - |T_A|^2}{|T_A|^2}, \quad \text{(S.14)}$$

and where $|T_A|^2$ is the transmittance of Alice's internal beam splitter used for the separation of the quadratures.

## S.2 AMQD Secret Key Rates in Two-Way CVQKD

In case of the two-way CVQKD, Bob starts the communication by sending via AMQD modulation a quantum system to Alice, who adds her internal secret to the received system, which is then sent back via AMQD modulation to Bob. We use the formalisms of [2] throughout to derive the security parameters.

The corresponding covariance matrix $\mathbf{K}_{\rho_{B_i}}$ of the density matrix $\rho_{B_i}$ of the noisy Gaussian subcarrier CV $\left|\phi_i''\right\rangle = \tilde{x}_i'' + \mathrm{i}\tilde{p}_i''$, where $\left|\phi_i''\right\rangle$ refers to a Gaussian subcarrier CV $\left|\phi_i\right\rangle$ which is affected by the two channel uses, specifically by $\mathcal{M}_1$ and $\mathcal{M}_2$, is precisely

$$
\mathbf{K}_{\rho_{B_i}} = \begin{pmatrix} \sigma_\omega^2 I & \left|F\left(T_i\left(\mathcal{N}_i\right)\right)\right|^2 \left(\sqrt{\left(\sigma_\omega^2\right)^2 - 1}\right) Z \\ \left|F\left(T_i\left(\mathcal{N}_i\right)\right)\right|^2 \left(\sqrt{\left(\sigma_\omega^2\right)^2 - 1}\right) Z & \lambda_{B_i} \end{pmatrix}, \quad (S.15)
$$

where

$$
\lambda_{B_i} = \left[\left|F\left(T_i\left(\mathcal{N}_i\right)\right)\right|^2 \sigma_\omega^2 + \left(1 - \left(\left|F\left(T_i\left(\mathcal{N}_i\right)\right)\right|^2\right)^2\right) W_i\right] I \\
+ \left|F\left(T_i\left(\mathcal{N}_i\right)\right)\right|^2 \sigma_\omega^2 I. \quad (S.16)
$$

Particularly, the covariance matrix $\mathbf{K}_{\rho_{E_i}}$ of Eve's density matrix $\rho_{E_i}$ is precisely evaluated as [2]

$$
\mathbf{K}_{\rho_{E_i}} = \begin{pmatrix} \sigma_{E_i|A_i}^2 I & \kappa_i Z & \mu'' I & 0 \\ \kappa_i Z & W_i I & \theta'' Z & 0 \\ \mu'' I & \theta'' Z & \lambda_{E_i} & \kappa_i \partial \\ 0 & 0 & \kappa_i Z & W_i I \end{pmatrix}, \quad (S.17)
$$

where

$$
\kappa_i = \sqrt{\left(1 - \left|F\left(T_{Eve,i}\right)\right|^2\right)\left(W_i^2 - 1\right)} \\
= \sqrt{\left|F\left(T_i\left(\mathcal{N}_i\right)\right)\right|^2 \left(W_i^2 - 1\right)}. \quad (S.18)
$$

while

$$
\lambda_{E_i} = \xi_i I + \left(1 - \left|F\left(T_i\left(\mathcal{N}_i\right)\right)\right|^2\right) \sigma_\omega^2 I, \quad (S.19)
$$

and

$$
\xi_i = \left|F\left(T_i\left(\mathcal{N}_i\right)\right)\right|^2 \left(1 - \left|F\left(T_i\left(\mathcal{N}_i\right)\right)\right|^2\right) \sigma_\omega^2 \\
+ \left(1 - \left|F\left(T_i\left(\mathcal{N}_i\right)\right)\right|^2\right)^2 W_i + \left|F\left(T_i\left(\mathcal{N}_i\right)\right)\right|^2 W_i, \quad (S.20)
$$

$$\mu_i'' = -\sqrt{1 - \left|F\left(T_i\left(\mathcal{N}_i\right)\right)\right|^2} \mu_i, \tag{S.21}$$

where

$$\mu_i = \left(W_i - \sigma_\omega^2\right)\sqrt{\left(\left|F\left(T_{Eve,i}\right)\right|^2 \left|F\left(T_i\left(\mathcal{N}_i\right)\right)\right|^2\right)}, \tag{S.22}$$

$$\theta_i'' = -\sqrt{1 - \left|F\left(T_i\left(\mathcal{N}_i\right)\right)\right|^2} \theta_i, \tag{S.23}$$

where

$$\theta_i = \sqrt{\left(1 - \left|F\left(T_i\left(\mathcal{N}_i\right)\right)\right|^2\right)\left(W_i^2 - 1\right)}. \tag{S.24}$$

Without loss of generality, the conditional covariance matrices of Bob's $B_i$ and Eve's $E_i$ systems with respect to Alice's Gaussian subcarrier variables $x_i, p_i$ are precisely as follows:

$$\mathbf{K}_{x_i'|x_i} = \mathbf{K}_{p_i'|p_i} = \mathbf{K}\left(0, \sigma_\omega^2\right), \tag{S.25}$$

$$\mathbf{K}_{E_i|x_i} = \mathbf{K}_{E_i|p_i} = \mathbf{K}\left(0, \sigma_\omega^2\right), \tag{S.26}$$

and

$$\mathbf{K}_{x_i',p_i'|x_i,p_i} = \mathbf{K}_{E_i|x_i,p_i} = \mathbf{K}\left(0,0\right). \tag{S.27}$$

The covariance matrices (S.25)–(S.27), the symplectic spectra for the single-carrier Gaussians will be constructed next.

### S.2.1 Homodyne Measurement, Reverse Reconciliation

Specifically, the related covariance matrices of the recovered single-carrier Gaussians are

$$\mathbf{K}_{x_j'|x_j} = \mathbf{K}_{p_j'|p_j} = \mathbf{K}\left(0, \sigma_{\omega_0}^2\right), \tag{S.28}$$

$$\mathbf{K}_{E_j|x_j} = \mathbf{K}_{E_j|p_j} = \mathbf{K}\left(0, \sigma_{\omega_0}^2\right). \tag{S.29}$$

From the covariance matrices (S.28)–(S.29), the symplectic spectra are evaluated as follows. For each channel uses $\mathcal{M}_1$ and $\mathcal{M}_2$, the condition $0 < \frac{1}{l}\sum_l \left|F\left(T_i\left(\mathcal{N}_i\right)\right)\right|^2 < 1$ holds for the $l$ Gaussian sub-channels $\mathcal{N}_i$ of $\mathcal{M}_1$, $\mathcal{M}_2$, and $\sigma_{\omega_0}^2 \gg 1$ holds, the symplectic spectra $\mathcal{S}_{x_j'}, \mathcal{S}_{p_j'}$ for Bob's single-carrier Gaussian CV, $B_j$, are precisely as follows:

$$\mathcal{S}_{x_j'} = \mathcal{S}_{p_j'} = \left(\frac{1}{\wp_2}\frac{1}{l}\sum_l\left|F\left(T_i\left(\mathcal{N}_i\right)\right)\right|^2 \sigma_{\omega_0}^2, \frac{1}{\wp_1}\frac{1}{l}\sum_l\left|F\left(T_i\left(\mathcal{N}_i\right)\right)\right|^2 \sigma_{\omega_0}^2\right), \tag{S.30}$$

where

$$\wp_1\wp_2 = \tfrac{1}{l}\sum_l \left|F\left(T_i\left(\mathcal{N}_i\right)\right)\right|^2, \tag{S.31}$$

$$\begin{aligned}\mathcal{S}_{x'_j|x_j} &= \mathcal{S}_{p'_j|p_j} \\ &= \left(\gamma\sigma^2_{\omega_0},\gamma^{-1}\sqrt{\tfrac{1}{l}\sum_l\left|F\left(T_i\left(\mathcal{N}_i\right)\right)\right|^2\left(1-\left(\tfrac{1}{l}\sum_l\left|F\left(T_i\left(\mathcal{N}_i\right)\right)\right|^2\right)^2\right)\tfrac{1}{l}\sum_l W_i \sigma^2_{\omega_0}}\right),\end{aligned} \tag{S.32}$$

where

$$\gamma = \sqrt{1+\left(\tfrac{1}{l}\sum_l\left|F\left(T_i\left(\mathcal{N}_i\right)\right)\right|^2\right)^2\left[\left(\tfrac{1}{l}\sum_l\left|F\left(T_i\left(\mathcal{N}_i\right)\right)\right|^2\right)^2 + \tfrac{1}{l}\sum_l\left|F\left(T_i\left(\mathcal{N}_i\right)\right)\right|^2 - 2\right]}. \tag{S.33}$$

Without loss of generality, for Eve's system, $E_j$, the symplectic spectra $\mathcal{S}_{E_j}$ is precisely as follows:

$$\mathcal{S}_{E_j} = \left(\tfrac{1}{\Omega_2}\left(1-\tfrac{1}{l}\sum_l\left|F\left(T_i\left(\mathcal{N}_i\right)\right)\right|^2\right)^2\sigma^2_{\omega_0}, \tfrac{1}{\Omega_1}\left(1-\tfrac{1}{l}\sum_l\left|F\left(T_i\left(\mathcal{N}_i\right)\right)\right|^2\right)^2\sigma^2_{\omega_0}, \tfrac{1}{l}\sum_l W_i, \tfrac{1}{l}\sum_l W_i\right), \tag{S.34}$$

where

$$\Omega_1\Omega_2 = \left(1-\tfrac{1}{l}\sum_l\left|F\left(T_i\left(\mathcal{N}_i\right)\right)\right|^2\right)^2. \tag{S.35}$$

In particular, in case of reverse reconciliation, the symplectic spectra $\mathcal{S}_{E_j|x'_j}$ and $\mathcal{S}_{E_j|p'_j}$ of $E_j$ are evaluated precisely as follows:

$$\begin{aligned}\mathcal{S}_{E_j|x'_j} &= \mathcal{S}_{E_j|p'_j} \\ &= \begin{pmatrix} \tfrac{1}{\Pi_2}\sqrt{\tfrac{1}{\tfrac{1}{l}\sum_l|F(T_i(\mathcal{N}_i))|^2}\left(1-\tfrac{1}{l}\sum_l\left|F\left(T_i\left(\mathcal{N}_i\right)\right)\right|^2\right)^3\left(1+\left(\tfrac{1}{l}\sum_l\left|F\left(T_i\left(\mathcal{N}_i\right)\right)\right|^2\right)^3\right)\tfrac{1}{l}\sum_l W_i\sigma^2_{\omega_0}}, \\ \tfrac{1}{\Pi_1}\sqrt{\tfrac{1}{\tfrac{1}{l}\sum_l|F(T_i(\mathcal{N}_i))|^2}\left(1-\tfrac{1}{l}\sum_l\left|F\left(T_i\left(\mathcal{N}_i\right)\right)\right|^2\right)^3\left(1+\left(\tfrac{1}{l}\sum_l\left|F\left(T_i\left(\mathcal{N}_i\right)\right)\right|^2\right)^3\right)\tfrac{1}{l}\sum_l W_i\sigma^2_{\omega_0}}, \\ \tfrac{1}{l}\sum_l W_i, 1 \end{pmatrix},\end{aligned} \tag{S.36}$$

where

$$\Pi_1\Pi_2 = \sqrt{\tfrac{1}{\tfrac{1}{l}\sum_l|F(T_i(\mathcal{N}_i))|^2}\left(1-\tfrac{1}{l}\sum_l\left|F\left(T_i\left(\mathcal{N}_i\right)\right)\right|^2\right)^3\left(1+\left(\tfrac{1}{l}\sum_l\left|F\left(T_i\left(\mathcal{N}_i\right)\right)\right|^2\right)^3\right)\tfrac{1}{l}\sum_l W_i}. \tag{S.37}$$

Particularly, from the symplectic spectra of (S.30)–(S.34) and (S.36), the $S^{RR,M_{\text{hom}}}_{two-way}$ asymptotic rate for the RR, $M_{\text{hom}}$, two-way AMQD modulation is expressed as follows:

$$S_{two-way}^{RR,M_{\text{hom}}} = I(A:B) - \chi(B:E)$$
$$= \frac{1}{2}\log_2 \frac{\left[1-\left(\frac{1}{l}\sum_l|F(T_i(\mathcal{N}_i))|^2\right)+\left(\frac{1}{l}\sum_l|F(T_i(\mathcal{N}_i))|^2\right)^2\right]}{\left(1-\frac{1}{l}\sum_l|F(T_i(\mathcal{N}_i))|^2\right)^2} \quad \text{(S.38)}$$
$$-\left(\frac{\frac{1}{l}\sum_l W_i+1}{2}\log_2\frac{\frac{1}{l}\sum_l W_i+1}{2} - \frac{\frac{1}{l}\sum_l W_i-1}{2}\log_2\frac{\frac{1}{l}\sum_l W_i-1}{2}\right).$$

∎

### S.2.2 Homodyne Measurement, Direct Reconciliation

Without loss of generality, the covariance matrices of the recovered single-carrier Gaussians are

$$\mathbf{K}_{x'_j|x_j} = \mathbf{K}_{p'_j|p_j} = \mathbf{K}\left(0, \sigma^2_{\omega_0}\right), \quad \text{(S.39)}$$

$$\mathbf{K}_{E_j|x_j} = \mathbf{K}_{E_j|p_j} = \mathbf{K}\left(0, \sigma^2_{\omega_0}\right). \quad \text{(S.40)}$$

Specifically, using (S.39)–(S.40), the corresponding symplectic spectra of $B_j$ and $E_j$ are as shown in (S.30)–(S.34), while Eve's symplectic spectra $\mathcal{S}_{E_j|x_j} = \mathcal{S}_{E_j|p_j}$ in the direct reconciliation are evaluated precisely as follows:

$$\mathcal{S}_{E_j|x_j} = \mathcal{S}_{E_j|p_j}$$
$$= \left(\ell\left(1-\frac{1}{l}\sum_l|F(T_i(\mathcal{N}_i))|^2\right)\sigma^2_{\omega_0}, \frac{1}{\ell}\sqrt{\left[1-\left(\frac{1}{l}\sum_l|F(T_i(\mathcal{N}_i))|^2\right)^2\right]\sigma^2_{\omega_0}\frac{1}{l}\sum_l W_i, \frac{1}{l}\sum_l W_i, 1}\right), \quad \text{(S.41)}$$

where

$$\ell = \sqrt{1 + 3\frac{1}{l}\sum_l|F(T_i(\mathcal{N}_i))|^2 + \left(\frac{1}{l}\sum_l|F(T_i(\mathcal{N}_i))|^2\right)^2}. \quad \text{(S.42)}$$

In particular, from the symplectic spectra of (S.30)–(S.34) and (S.41), the $S_{two-way}^{DR,M_{\text{hom}}}$ asymptotic rate of the DR, $M_{\text{hom}}$, two-way AMQD modulation is

$$S_{two-way}^{DR,M_{\text{hom}}} = \chi(A:B) - \chi(A:E)$$
$$= \frac{1}{2}\log_2 \frac{\frac{1}{l}\sum_l|F(T_i(\mathcal{N}_i))|^2}{\left(1-\frac{1}{l}\sum_l|F(T_i(\mathcal{N}_i))|^2\right)^2} - \left(\frac{\frac{1}{l}\sum_l W_i+1}{2}\log_2\frac{\frac{1}{l}\sum_l W_i+1}{2} - \frac{\frac{1}{l}\sum_l W_i-1}{2}\log_2\frac{\frac{1}{l}\sum_l W_i-1}{2}\right). \quad \text{(S.43)}$$

∎

### S.2.3 Heterodyne Measurement, Reverse Reconciliation

Particularly, the covariance matrices of the recovered single-carrier Gaussians are

$$\mathbf{K}_{x'_j,p'_j|x_i,p_i} = \mathbf{K}_{E_j|x_j,p_j} = \mathbf{K}(0,0). \quad \text{(S.44)}$$

The symplectic spectra $\mathcal{S}_{x'_j}, \mathcal{S}_{p'_j}$ have been evaluated in (S.30). On the other hand, in a heterodyne measurement $M_{\text{het}}$, the symplectic spectra $\mathcal{S}_{x'_j, p'_j}$ of Bob's system is precisely as follows:

$$\mathcal{S}_{x'_j, p'_j | x_j, p_j} = \left( \left[ 1 - \left( \tfrac{1}{l} \sum_l |F(T_i(\mathcal{N}_i))|^2 \right)^2 \right] \sigma^2_{\omega_0}, \tfrac{1}{l} \sum_l W_i \right). \tag{S.45}$$

Eve's symplectic spectra $\mathcal{S}_{E_j | x'_j, p'_j}$ is evaluated exactly as

$$\mathcal{S}_{E_j | x'_j, p'_j} = \left( \Gamma_1, \Gamma_2, \Gamma_3, \left[ 1 - \left( \tfrac{1}{l} \sum_l |F(T_i(\mathcal{N}_i))|^2 \right)^2 \right] \sigma^2_{\omega_0} \right), \tag{S.46}$$

where

$$\Gamma_1 \Gamma_2 \Gamma_3 = \frac{1 + \tfrac{1}{l} \sum_l W_i \left[ 1 + \left( \tfrac{1}{l} \sum_l |F(T_i(\mathcal{N}_i))|^2 \right)^3 \right] + \left( 1 - \tfrac{1}{l} \sum_l |F(T_i(\mathcal{N}_i))|^2 \right) \left[ 1 + \left( \tfrac{1}{l} \sum_l |F(T_i(\mathcal{N}_i))|^2 \right)^2 \right] \tfrac{1}{l} \sum_l W_i}{\tfrac{1}{l} \sum_l |F(T_i(\mathcal{N}_i))|^2 \left( 1 + \tfrac{1}{l} \sum_l |F(T_i(\mathcal{N}_i))|^2 \right)}. \tag{S.47}$$

Without loss of generality, using the symplectic spectra of (S.34) and (S.45)-(S.46), the $S^{RR, M_{\text{het}}}_{two-way}$ asymptotic rate for the RR, $M_{\text{het}}$, two-way AMQD modulation is expressed as

$$\begin{aligned}
S^{RR, M_{\text{het}}}_{two-way} &= I(A:B) - \chi(B:E) \\
&= \log_2 \frac{2 \tfrac{1}{l} \sum_l |F(T_i(\mathcal{N}_i))|^2 \left( 1 + \tfrac{1}{l} \sum_l |F(T_i(\mathcal{N}_i))|^2 \right)}{\hat{\sigma}^2_{E_j | A_j} \left( 1 - \tfrac{1}{l} \sum_l |F(T_i(\mathcal{N}_i))|^2 \right) \left[ 1 + \left( \tfrac{1}{l} \sum_l |F(T_i(\mathcal{N}_i))|^2 \right)^2 + \tfrac{1}{l} \sum_l W_i - \left( \tfrac{1}{l} \sum_l |F(T_i(\mathcal{N}_i))|^2 \right)^2 \tfrac{1}{l} \sum_l W_i \right]} \\
&\quad + \sum_i \left( \frac{\Gamma_i + 1}{2} \log_2 \frac{\Gamma_i + 1}{2} - \frac{\Gamma_i - 1}{2} \log_2 \frac{\Gamma_i - 1}{2} \right) \\
&\quad - 2 \left( \frac{\tfrac{1}{l} \sum_l W_i + 1}{2} \log_2 \frac{\tfrac{1}{l} \sum_l W_i + 1}{2} - \frac{\tfrac{1}{l} \sum_l W_i - 1}{2} \log_2 \frac{\tfrac{1}{l} \sum_l W_i - 1}{2} \right).
\end{aligned} \tag{S.48}$$

∎

### S.2.4 Heterodyne Measurement, Direct Reconciliation

Specifically, the corresponding covariance matrices of the recovered single-carrier Gaussians are

$$\mathbf{K}_{x'_j, p'_j | x_j, p_j} = \mathbf{K}_{E_j | x_j, p_j} = \mathbf{K}(0, 0). \tag{S.49}$$

Particularly, Eve's corresponding symplectic spectra is $\mathcal{S}_{E_j | x_j, p_j}$, which is expressed as follows:

$$\mathcal{S}_{E_j | x_j, p_j} = \left( \left[ 1 - \left( \tfrac{1}{l} \sum_l |F(T_i(\mathcal{N}_i))|^2 \right)^2 \right] \sigma^2_{\omega_0}, \tfrac{1}{l} \sum_l W_i, 1, 1 \right). \tag{S.50}$$

Without loss of generality, from the symplectic spectra of (S.45)–(S.46) and (S.50), the $S^{DR, M_{\text{het}}}_{two-way}$ asymptotic rate for the RR, $M_{\text{het}}$, two-way AMQD modulation is expressed precisely as

$$\begin{aligned}
S_{two-way}^{DR,M_{\text{het}}} &= \chi(A:B) - \chi(A:E) \\
&= 2R_{two-way}^{DR,\text{hom}} \\
&= 2(\chi(A:B) - \chi(A:E)) \\
&= \log_2 \frac{\frac{1}{l}\sum_l |F(T_i(\mathcal{N}_i))|^2}{\left(1-\frac{1}{l}\sum_l |F(T_i(\mathcal{N}_i))|^2\right)^2} - 2\left(\frac{\frac{1}{l}\sum_l W_i+1}{2}\log_2 \frac{\frac{1}{l}\sum_l W_i+1}{2} - \frac{\frac{1}{l}\sum_l W_i-1}{2}\log_2 \frac{\frac{1}{l}\sum_l W_i-1}{2}\right).
\end{aligned} \quad (S.51)$$

∎

## S.3 Notations

The notations of the manuscript are summarized in Table S.1.

**Table S.1.** The summary of the notations of the manuscript.

| Notation | Description |
|---|---|
| $\mathbf{K}$ | Correlation matrix. |
| $diag(\cdot)$ | Diagonal matrix. |
| $i$ | Index for the $i$-th subcarrier Gaussian CV, $\left|\phi_i\right\rangle = x_i + \mathrm{i}p_i$. |
| $j$ | Index for the $j$-th Gaussian single-carrier CV, $\left|\varphi_j\right\rangle = x_j + \mathrm{i}p_j$. |
| $l$ | Number of Gaussian sub-channels $\mathcal{N}_i$ for the transmission of the Gaussian subcarriers. The overall number of the sub-channels is $n$. The remaining $n-l$ sub-channels do not transmit valuable information. |
| $(x_i, p_i)$ | Position and momentum quadratures of the $i$-th Gaussian subcarrier, $\left|\phi_i\right\rangle = x_i + \mathrm{i}p_i$. |
| $(x'_i, p'_i)$ | Noisy position and momentum quadratures of Bob's $i$-th noisy subcarrier Gaussian CV, $\left|\phi'_i\right\rangle = x'_i + \mathrm{i}p'_i$. |
| $(x_j, p_j)$ | Position and momentum quadratures of the $j$-th Gaussian single-carrier $\left|\varphi_j\right\rangle = x_j + \mathrm{i}p_j$. |
| $(x'_j, p'_j)$ | Noisy position and momentum quadratures of Bob's $j$-th recovered single-carrier Gaussian CV $\left|\varphi'_j\right\rangle = x'_j + \mathrm{i}p'_j$. |
| $x_{A,i}, p_{A,i}$ | Alice's quadratures in the transmission of the $i$-th subcarrier. |
| $x_{B,i}, p_{B,i}$ | Bob's quadratures in the transmission of the $i$-th subcarrier. |
| $x_{E,i}, p_{E,i}$ | Eve's quadratures in the transmission of the $i$-th subcarrier. |
| $\mathcal{Cl}_{Eve}$ | Eve's entangling cloner. |
| $\text{SNR}_i^*$ | Signal to noise ratio of the $i$-th Gaussian sub-channel $\mathcal{N}_i$. Used for the evaluation of the private classical capacity $P(\mathcal{N}_i)$, expressed as $\text{SNR}_i^* = \frac{\sigma_{\omega_i}^2}{\sigma_{\mathcal{N}_i^*}^2}$, where $\sigma_{\omega_i}^2$ is the modulation variance of the subcarrier quadratures, and $\sigma_{\mathcal{N}_i^*}^2$ is the noise variance. |
| $\mathcal{S}_\rho$ | Symplectic spectra of $\rho$, $\mathcal{S}_{\rho_\mathcal{G}} = (s_1...,s_n)$, where $s_i$-s are real elements. |
| S | Superset. |
| $I$ | Identity matrix, $I = diag(1,1)$. |
| $Z$ | Pauli $Z$ matrix, $Z = diag(1,-1)$. |
| $S(\cdot)$ | The von Neumann entropy function. For a Gaussian density $\rho_\mathcal{G}$ it is expressed as $S(\rho_\mathcal{G}) = \sum_{i=1}^n g(s_i)$, where $g(s_i) = \frac{s_i+1}{2}\log_2\frac{s_i+1}{2} - \frac{s_i-1}{2}\log_2\frac{s_i-1}{2}$, and the $s_i$-s are real elements of the symplectic spectra $\mathcal{S}_{\rho_\mathcal{G}} = (s_1...,s_n)$. |
| $\langle x_i^2 \rangle, \langle p_i^2 \rangle$ | Variances of Alice's subcarrier quadratures, $\langle x_i^2 \rangle = \langle p_i^2 \rangle = \sigma_{\omega_i}^2$. |
| $\langle x_i'^2 \rangle, \langle p_i'^2 \rangle$ | Variances of Bob's noisy subcarrier quadratures, $\langle x_i'^2 \rangle = \langle p_i'^2 \rangle = |F(T_i(\mathcal{N}_i))|^2 \sigma_{\omega_i}^2 + \sigma_{\mathcal{N}_i}^2$. |
| $e_{x_i'}$ | Alice's estimator on Bob's noisy $x_i'$ subcarrier quadrature from her $x_i$ quadrature, evolved as $e_{x_i'} = \varepsilon_{A,x_i} x_i$, where $\varepsilon_{A,x_i'} = \frac{\langle x_i' x_i \rangle}{\langle x_i^2 \rangle}$. |
| $e_{p_i'}$ | Alice's estimator on Bob's noisy $p_i'$ subcarrier quadrature from her $p_i$, expressed as $e_{p_i'} = \varepsilon_{A,p_i'} p_i$, where $\varepsilon_{A,p_i'} = \frac{\langle p_i' p_i \rangle}{\langle p_i^2 \rangle}$. |
| $e_{x_i}, e_{p_i}$ | Bob's estimators on the quadratures of Alice's $i$-th subcarrier. |
| $e_{x_i'}^E, e_{p_i'}^E$ | Eve's estimators on the quadratures of Bob's $i$-th subcarrier in the RR direction. |
| $e_{x_i}^E, e_{p_i}^E$ | Eve's estimators on the quadratures of Alice's $i$-th subcarrier in the DR direction. |
| $e_{x_{E,j}}^A, e_{p_{E,j}}^A$ | Alice's estimators on the quadratures of Eve's $j$-th single- |
| | carrier in the DR direction. |
| $e^B_{x_{E,j}}, e^B_{p_{E,j}}$ | Bob's estimators on the quadratures of Eve's $j$-th single-carrier in the RR direction. |
| $\hat{\sigma}^2_{x'_i \vert e_{x'_i}}$ | Conditional variance evaluated for the position subcarrier quadratures $x'_i, e_{x'_i}$, where $e_{x'_i}$ is Alice's estimator on Bob's noisy $x'_i$ subcarrier quadrature from her $x_i$, expressed as $\hat{\sigma}^2_{x'_i \vert e_{x'_i}} = \left\langle \left( x'_i - e_{x'_i} \right)^2 \right\rangle = \left\langle x'^2_i \right\rangle - \frac{\vert \langle x_i x'_i \rangle \vert^2}{\langle x_i^2 \rangle}$. |
| $\hat{\sigma}^2_{p'_i \vert e_{p'_i}}$ | Conditional variance evaluated for the momentum subcarrier quadratures $p'_i, e_{p'_i}$, where $e_{p'_i}$ is Alice's estimator on Bob's noisy $p'_i$ subcarrier quadrature from her $p_i$, expressed as $\hat{\sigma}^2_{p'_i \vert e_{p'_i}} = \left\langle \left( p'_i - e_{p'_i} \right)^2 \right\rangle = \left\langle p'^2_i \right\rangle - \frac{\vert \langle p_i p'_i \rangle \vert^2}{\langle p_i^2 \rangle}$. |
| $\chi_x(A:B), \chi_p(A:B)$ | Holevo quantities between Alice and Bob, evaluated with respect to the $x$ and $p$ quadratures. |
| $\chi_x(B:E), \chi_p(B:E)$ | Holevo quantities between Bob and Eve, evaluated with respect to the $x$ and $p$ quadratures. |
| $\chi_x(A:E), \chi_p(A:E)$ | Holevo quantities between Alice and Eve, evaluated with respect to the $x$ and $p$ quadratures. |
| $S_x^{RR}, S_p^{RR}$ | Secret key rates for reverse reconciliation, with respect to the $x$ and $p$ quadratures. |
| $S_x^{DR}, S_p^{DR}$ | Secret key rates for direct reconciliation, with respect to the $x$ and $p$ quadratures. |
| $\sigma^2_{\mathcal{N}_i}$ | Noise variance of the Gaussian sub-channel $\mathcal{N}_i$ for the characterization of classical rates and classical capacity $C(\mathcal{N}_i)$. |
| $\sigma^2_{\mathcal{N}_i^*}$ | Noise variance of the Gaussian sub-channel $\mathcal{N}_i$ for the characterization of private classical capacity $P(\mathcal{N}_i)$, $\sigma^2_{\mathcal{N}_i^*} = \sigma^2_{\omega_i} \left( \frac{\sigma^2_{\omega_i} \vert F(T_i(\mathcal{N}_i)) \vert^2 + \sigma^2_{X_i}}{1 + \sigma^2_{X_i} \sigma^2_{\omega_i} \vert F(T_i(\mathcal{N}_i)) \vert^2} - 1 \right)^{-1}$, where $\sigma^2_{X_i} = \sigma^2_0 + N_i$, and $\sigma^2_0$ is the vacuum noise. |
| $N_i$ | Excess noise of the Gaussian sub-channel $\mathcal{N}_i$, $N_i = \frac{(W_i - 1)\left( \vert F(T_{Eve,i}) \vert^2 \right)}{1 - \vert F(T_{Eve,i}) \vert^2}$, where $\vert F(T_{Eve,i}) \vert^2 = 1 - \vert F(T_i) \vert^2$ is Eve's Fourier-transformed transmittance, while $W_i$ is the variance of Eve's $i$-th EPR state. |
| $P(\mathcal{N}_i)$ | Private classical capacity of the Gaussian sub-channel $\mathcal{N}_i$. |
| $P^{RR}(\mathcal{N}_i)$ | Private classical capacity of the Gaussian sub-channel $\mathcal{N}_i$ in a RR CVQKD protocol run. |
| $P^{DR}(\mathcal{N}_i)$ | Private classical capacity of the Gaussian sub-channel $\mathcal{N}_i$ in a DR CVQKD protocol run. |
| $\tilde{x}_i, \tilde{p}_i$ | Modulated quadratures, expressed as $\tilde{x}_i = x_i + \tilde{x}_i \vert x_i$, and $\tilde{p}_i = p_i + \tilde{p}_i \vert p_i$, where the variables $x_i$ and $p_i$ refer to the modulated position and momentum quadratures. |
| $\sigma_{\tilde{x}_i}^2, \sigma_{\tilde{p}_i}^2$ | The modulation variances of subcarrier quadratures $\tilde{x}_i$ and $\tilde{p}_i$, expressed as $\sigma_{\tilde{x}_i}^2 = \sigma_{\tilde{p}_i}^2 = \sigma_\omega^2 - 1 + \left(\sigma_{\tilde{x}_i \vert x_i}^2 + \sigma_{\tilde{p}_i \vert p_i}^2\right)$, where $\sigma_{\tilde{x}_i \vert x_i}^2 = \sigma_{\tilde{p}_i \vert p_i}^2 = 1$, and $\sigma_{\tilde{x}_i}^2 = \sigma_{\tilde{p}_i}^2 = \sigma_\omega^2$. |
| $s$ | Squeezing factor, $s = 1$ for coherent states, constrained into the range of $\frac{1}{\sigma_{\omega_i}^2} < s < \sigma_{\omega_i}^2$. |
| $N_0$ | Shot-noise variance. |
| $W_i$ | Variance of Eve's EPR state, used in the attack of the $i$-th Gaussian sub-channel $\mathcal{N}_i$. |
| $W$ | Averaged (i.e., taken for a single-carrier Gaussian CV) value of Eve's EPR state variances, $W = \frac{1}{l}\sum_l W_i$. |
| $\hat{\sigma}_{B_j \vert A_j}^2, \hat{\sigma}_{E_j \vert A_j}^2$ | Averaged (i.e., taken for a single-carrier Gaussian CV) conditional variances for Alice and Bob, and Alice and Eve, expressed as $\hat{\sigma}_{B_j \vert A_j}^2 = \frac{1}{l}\sum_{i=1}^{l}\sigma_{B_i \vert A_i}^2$, $\hat{\sigma}_{E_j \vert A_j}^2 = \frac{1}{l}\sum_{i=1}^{l}\sigma_{E_i \vert A_i}^2$. |
| $\frac{1}{l}\sum_{i=1}^{l}\left(\left\vert F(T_{Eve,i})\right\vert^2\right)$ | Eve's averaged Fourier-transformed transmittance parameter, $\frac{1}{l}\sum_{i=1}^{l}\left(\left\vert F(T_{Eve,i})\right\vert^2\right) = 1 - \frac{1}{l}\sum_{i=1}^{l}\left\vert F(T_i(\mathcal{N}_i))\right\vert^2$. |
| $\frac{1}{l}\sum_{i=1}^{l}\left\vert F(T_i(\mathcal{N}_i))\right\vert^2$ | Averaged Fourier-transformed transmittance parameter of $\mathcal{N}$, $\frac{1}{l}\sum_{i=1}^{l}\left\vert F(T_i(\mathcal{N}_i))\right\vert^2 = 1 - \frac{1}{l}\sum_{i=1}^{l}\left(\left\vert F(T_{Eve,i})\right\vert^2\right)$. |
| $\left\vert F(T'_{Eve,i})\right\vert^2$ | Eve's Fourier-transformed transmittance parameter in an SVD-assisted AMQD multicarrier setting, expressed as $\left\vert F(T'_{Eve,i})\right\vert^2 = 1 - \upsilon_i\left(1 - \left\vert F(T_{Eve,i})\right\vert^2\right)$, where $\upsilon_i = \frac{\nu_{Eve} - \left(\sigma_\mathcal{N}^2 \big/ \max_{n_{\min}} \lambda_i^2\right)}{\nu_{Eve} - \left(\sigma_\mathcal{N}^2 \big/ \max_{\forall i}\left\vert F(T_i(\mathcal{N}_i))\right\vert^2\right)}$, $\left\vert F(T'_{Eve,i})\right\vert^2 < \left\vert F(T_{Eve,i})\right\vert^2$. |
| $\sigma_{\omega''}^2$ | Virtual modulation variance of the subcarrier quadratures in an SVD-assisted AMQD multicarrier transmission, ex- |
| | pressed as $\sigma_{\omega''}^2 = \sigma_\omega^2(1+c) > \sigma_\omega^2$, where $c > 0$ is a constant. |
| $\frac{1}{l}\sum_l \left|F\left(T_i'(\mathcal{N}_i)\right)\right|^2$ | Averaged Fourier-transformed transmittance parameter for the SVD-assisted multicarrier CVQKD transmission, evaluated as $\frac{1}{l}\sum_l \left|F(T_i(\mathcal{N}_i))\right|^2 \frac{\nu_{Eve} - \left(\sigma_\mathcal{N}^2 / \max\lambda_i^2 \atop n_{\min}\right)}{\nu_{Eve} - \left(\sigma_\mathcal{N}^2 / \max_{\forall i}\left|F(T_i(\mathcal{N}_i))\right|^2\right)}$. |
| $M_{\hom}$ | Homodyne measurement. |
| $M_{\het}$ | Heterodyne measurement. |
| $R_{one-way}^{RR, M_{\hom}}$ | Secret key rate in one-way CVQKD, at homodyne measurement and reverse reconciliation. |
| $R_{one-way}^{RR, M_{\het}}$ | Secret key rate in one-way CVQKD, at heterodyne measurement and reverse reconciliation. |
| $R_{two-way}^{RR, M_{\hom}}$ | Secret key rate in two-way CVQKD, at homodyne measurement and reverse reconciliation. |
| $R_{two-way}^{RR, M_{\het}}$ | Secret key rate in two-way CVQKD, at heterodyne measurement and reverse reconciliation. |
| $R_{one-way}^{DR, M_{\hom}}$ | Secret key rate in one-way CVQKD, at homodyne measurement and direct reconciliation. |
| $R_{one-way}^{DR, M_{\het}}$ | Secret key rate in one-way CVQKD, at heterodyne measurement and direct reconciliation. |
| $R_{two-way}^{DR, M_{\hom}}$ | Secret key rate in two-way CVQKD, at homodyne measurement and direct reconciliation. |
| $R_{two-way}^{DR, M_{\het}}$ | Secret key rate in two-way CVQKD, at heterodyne measurement and direct reconciliation. |
| P | The private classical capacity domain of $\mathcal{N}$, $\mathrm{P} = \mathcal{H}\left(\bigcup_{z_1, z_2} \mathrm{P}(z_1, z_2)\right)$, where $\mathcal{H}$ refers to the convex hull of independent input distributions. |
| $P_1, ..., P_K$ | Corner points of the private classical capacity region P of $K$ users, $U_{1,...,K}$. |
| $U_1, ... U_K$ | $K$ users in a multiuser CVQKD scenario. |
| $I_{\mathrm{MQA}}(\cdot)$ | The corresponding correlation measure function between Alice and Bob, Bob and Eve, or Alice and Eve in AMQD-MQA. It identifies a mutual information function or a Holevo quantity, depending on the direction of the reconciliation scheme and the measurement attributes of the actual CVQKD protocol run. |
| $E_k$ | Eve's variable in the attack of user's $U_k$ transmission. |
| $\sum_K I_{\text{MQA}}(y:E_k)$ | Characterizes the eavesdropped information from the $K$ users in a multiuser CVQKD scenario, quantified by the mutual information function or the Holevo information functions, depending on the CVQKD protocol. |
| $S_k^{\text{MQA}}$ | The secret key rate of user $U_k$. |
| $S_{\text{sum}}^{\text{MQA}}$ | The sum secret key rate of $K$ users, characterizes the overall secret key rate of the users, $S_{\text{sum}}^{\text{MQA}} = \sum_K S_k^{\text{MQA}}$. |
| $S_{\text{sym}}^{\text{MQA}}$ | The symmetric secret key rate of $K$ users, characterizes the common rate at which all $K$ users can have a simultaneous reliable secret communication through $\mathcal{N}$. |
| $P_{\text{sum}}(\mathcal{N})$ | The sum private classical capacity of $\mathcal{N}$, characterizes the overall maximized secret key rate of the users. |
| $P_{\text{sym}}(\mathcal{N})$ | The symmetric private classical capacity of $\mathcal{N}$, characterizes the maximized common rate at which all $K$ users can have a simultaneous reliable secret communication through $\mathcal{N}$. |
| $S'_{\text{sum}}(\mathcal{N})$ | The sum secret key rate of $K$ users in an SVD-assisted multicarrier CVQKD setting. |
| $S'^{\text{MQA}}_{\text{sym}}$ | The symmetric secret key rate of $K$ users in an SVD-assisted multicarrier CVQKD setting. |
| $P'_{\text{sum}}(\mathcal{N})$ | The sum private classical capacity of $\mathcal{N}$, characterizes the overall maximized secret key rate of the users in SVD-assisted AMQD-MQA. |
| $P'_{\text{sym}}(\mathcal{N})$ | The symmetric private classical capacity of $\mathcal{N}$, characterizes the maximized common rate at which all $K$ users can have a simultaneous reliable secret communication in SVD-assisted AMQD-MQA. |
| $N_{tol,AMQD}^{RR,coh.}$ | The maximal tolerable excess noise in AMQD, for RR and coherent state transmission. |
| $N_{tol,AMQD}^{RR,squ.}$ | The maximal tolerable excess noise in AMQD, for RR and squeezed state transmission. |
| $N_{tol,single}^{RR,one-way}$ | The maximal tolerable excess noise in single-carrier transmission, for RR and one-way CVQKD. |
| $N_{tol,single}^{DR,one-way}$ | The maximal tolerable excess noise in single-carrier transmission, for DR and one-way CVQKD. |
| $N_{tol,AMQD}^{DR,one-way}$ | The maximal tolerable excess noise in AMQD, for DR and one-way CVQKD. |
| $N_{tol,single}^{RR,one-way}$ | The maximal tolerable excess noise in AMQD, for RR and one-way CVQKD. |
| $N_{tol,AMQD}^{SVD}$ | The maximal tolerable excess noise in SVD-assisted AMQD. |
| $\mathcal{M}_1, \mathcal{M}_2$ | Identifies the two channel uses of a two-way CVQKD protocol. First channel use, $\mathcal{M}_1$, from Bob to Alice, the second one, $\mathcal{M}_2$, from Alice to Bob. |
| H | Expressed as $H = \frac{1-|T_A|^2}{|T_A|^2}$, where $|T_A|^2$ is the transmittance of Alice's internal beam splitter used for the separation of the position and momentum quadratures. |
| $z \in \mathcal{CN}(0, \sigma_z^2)$ | The variable of a single-carrier Gaussian CV state, $|\varphi_i\rangle \in \mathcal{S}$. Zero-mean, circular symmetric complex Gaussian random variable, $\sigma_z^2 = \mathbb{E}[|z|^2] = 2\sigma_{\omega_0}^2$, with i.i.d. zero mean, Gaussian random quadrature components $x, p \in \mathbb{N}(0, \sigma_{\omega_0}^2)$, where $\sigma_{\omega_0}^2$ is the variance. |
| $\Delta \in \mathcal{CN}(0, \sigma_\Delta^2)$ | The noise variable of the Gaussian channel $\mathcal{N}$, with i.i.d. zero-mean, Gaussian random noise components on the position and momentum quadratures $\Delta_x, \Delta_p \in \mathbb{N}(0, \sigma_\mathcal{N}^2)$, $\sigma_\Delta^2 = \mathbb{E}[|\Delta|^2] = 2\sigma_\mathcal{N}^2$. |
| $d \in \mathcal{CN}(0, \sigma_d^2)$ | The variable of a Gaussian subcarrier CV state, $|\phi_i\rangle \in \mathcal{S}$. Zero-mean, circular symmetric Gaussian random variable, $\sigma_d^2 = \mathbb{E}[|d|^2] = 2\sigma_\omega^2$, with i.i.d. zero mean, Gaussian random quadrature components $x_d, p_d \in \mathbb{N}(0, \sigma_\omega^2)$, where $\sigma_\omega^2$ is the modulation variance of the Gaussian subcarrier CV state. |
| $F^{-1}(\cdot) = \text{CVQFT}^\dagger(\cdot)$ | The inverse CVQFT transformation, applied by the encoder. Continuous-variable unitary operation. |
| $F(\cdot) = \text{CVQFT}(\cdot)$ | The CVQFT transformation, applied by the decoder. Continuous-variable unitary operation. |
| $F^{-1}(\cdot) = \text{IFFT}(\cdot)$ | Inverse FFT transform, applied by the encoder. |
| $\sigma_{\omega_0}^2$ | Single-carrier modulation variance. |
| $\sigma_\omega^2 = \frac{1}{l}\sum_l \sigma_{\omega_i}^2$ | Multicarrier modulation variance for the $l$ Gaussian subchannels $\mathcal{N}_i$. Determined as $\sigma_\omega^2 = \nu_{Eve} - \nu_{\min}$, where $\nu_{Eve} = \frac{1}{\lambda}$, $\lambda = |F(T_\mathcal{N}^*)|^2 = \frac{1}{n}\sum_{i=1}^n \left|\sum_{k=1}^n T_k^* e^{\frac{-i2\pi ik}{n}}\right|^2$ and $T_\mathcal{N}^*$ is the expected transmittance of the Gaussian subchannels under an optimal Gaussian collective attack. |

| | |
|---|---|
| $\lvert\phi_i\rangle = \lvert\mathrm{IFFT}(z_{k,i})\rangle$ $= \lvert F^{-1}(z_{k,i})\rangle = \lvert d_i\rangle.$ | The $i$-th Gaussian subcarrier CV of user $U_k$, where IFFT stands for the Inverse Fast Fourier Transform, $\lvert\phi_i\rangle \in \mathcal{S}$, $d_i \in \mathcal{CN}(0, \sigma_{d_i}^2)$, $\sigma_{d_i}^2 = \mathbb{E}[\lvert d_i\rvert^2]$, $d_i = x_{d_i} + \mathrm{i}p_{d_i}$, $x_{d_i} \in \mathbb{N}(0, \sigma_{\omega_F}^2)$, $p_{d_i} \in \mathbb{N}(0, \sigma_{\omega_F}^2)$ are i.i.d. zero-mean Gaussian random quadrature components, and $\sigma_{\omega_F}^2$ is the variance of the Fourier transformed Gaussian state. |
| $\lvert\varphi_{k,i}\rangle = \mathrm{CVQFT}(\lvert\phi_i\rangle)$ | The decoded single-carrier CV of user $U_k$ from the subcarrier CV, expressed as $F(\lvert d_i\rangle) = \lvert F(F^{-1}(z_{k,i}))\rangle = \lvert z_{k,i}\rangle$. |
| $\mathcal{N}$ | Gaussian quantum channel. |
| $\mathcal{N}_i, i = 1,\ldots,n$ | Gaussian sub-channels. |
| $T(\mathcal{N})$ | Channel transmittance, normalized complex random variable, $T(\mathcal{N}) = \mathrm{Re}\,T(\mathcal{N}) + \mathrm{i}\,\mathrm{Im}\,T(\mathcal{N}) \in \mathcal{C}$. The real part identifies the position quadrature transmission, the imaginary part identifies the transmittance of the position quadrature. |
| $T_i(\mathcal{N}_i)$ | Transmittance coefficient of Gaussian sub-channel $\mathcal{N}_i$, $T_i(\mathcal{N}_i) = \mathrm{Re}(T_i(\mathcal{N}_i)) + \mathrm{i}\,\mathrm{Im}(T_i(\mathcal{N}_i)) \in \mathcal{C}$, quantifies the position and momentum quadrature transmission, with (normalized) real and imaginary parts $0 \leq \mathrm{Re}\,T_i(\mathcal{N}_i) \leq 1/\sqrt{2}$, $0 \leq \mathrm{Im}\,T_i(\mathcal{N}_i) \leq 1/\sqrt{2}$, where $\mathrm{Re}\,T_i(\mathcal{N}_i) = \mathrm{Im}\,T_i(\mathcal{N}_i)$. |
| $T_{Eve}$ | Eve's transmittance, $T_{Eve} = 1 - T(\mathcal{N})$. |
| $T_{Eve,i}$ | Eve's transmittance for the $i$-th subcarrier CV. |
| $\mathcal{A} \subseteq K$ | The subset of allocated users, $\mathcal{A} \subseteq K$. Only the allocated users can transmit information in a given (particularly the $j$-th) AMQD block. The cardinality of subset $\mathcal{A}$ is $\lvert\mathcal{A}\rvert$. |
| $U_k,\ k = 1,\ldots,\lvert\mathcal{A}\rvert$ | An allocated user from subset $\mathcal{A} \subseteq K$. |
| $\mathbf{z} = \mathbf{x} + \mathrm{i}\mathbf{p} = (z_1,\ldots,z_d)^T$ | An $d$-dimensional, zero-mean, circular symmetric complex random Gaussian vector that models $d$ Gaussian CV input states, $\mathcal{CN}(0, \mathbf{K_z})$, $\mathbf{K_z} = \mathbb{E}[\mathbf{z}\mathbf{z}^\dagger]$, where $z_i = x_i + \mathrm{i}p_i$, $\mathbf{x} = (x_1,\ldots,x_d)^T$, $\mathbf{p} = (p_1,\ldots,p_d)^T$, with $x_i \in \mathbb{N}(0, \sigma_{\omega_0}^2)$, $p_i \in \mathbb{N}(0, \sigma_{\omega_0}^2)$ i.i.d. zero-mean Gaussian random variables. |
| $\mathbf{d} = F^{-1}(\mathbf{z})$ | An $l$-dimensional, zero-mean, circular symmetric complex random Gaussian vector of the $l$ Gaussian subcarrier CVs, $\mathcal{CN}(0, \mathbf{K_d})$, $\mathbf{K_d} = \mathbb{E}[\mathbf{dd}^\dagger]$, $\mathbf{d} = (d_1, ..., d_l)^T$, $d_i = x_i + \mathrm{i} p_i$, $x_i, p_i \in \mathbb{N}(0, \sigma^2_{\omega_F})$ are i.i.d. zero-mean Gaussian random variables, $\sigma^2_{\omega_F} = 1/\sigma^2_{\omega_0}$. The $i$-th component is $d_i \in \mathcal{CN}(0, \sigma^2_{d_i})$, $\sigma^2_{d_i} = \mathbb{E}[|d_i|^2]$. |
| $\mathbf{y}_k \in \mathcal{CN}\left(0, \mathbb{E}[\mathbf{y}_k \mathbf{y}_k^\dagger]\right)$ | A $d$-dimensional zero-mean, circular symmetric complex Gaussian random vector. |
| $y_{k,m}$ | The $m$-th element of the $k$-th user's vector $\mathbf{y}_k$, expressed as $y_{k,m} = \sum_l F(T_i(\mathcal{N}_i)) F(d_i) + F(\Delta_i)$. |
| $F(\mathbf{T}(\mathcal{N}))$ | Fourier transform of $\mathbf{T}(\mathcal{N}) = [T_1(\mathcal{N}_1)..., T_l(\mathcal{N}_l)]^T \in \mathcal{C}^l$, the complex transmittance vector. |
| $F(\Delta)$ | Complex vector, expressed as $F(\Delta) = e^{\frac{-F(\Delta)^T \mathbf{K}_{F(\Delta)} F(\Delta)}{2}}$, with covariance matrix $\mathbf{K}_{F(\Delta)} = \mathbb{E}\left[F(\Delta) F(\Delta)^\dagger\right]$. |
| $\mathbf{y}[j]$ | AMQD block, $\mathbf{y}[j] = F(\mathbf{T}(\mathcal{N})) F(\mathbf{d})[j] + F(\Delta)[j]$. |
| $\tau = \|F(\mathbf{d})[j]\|^2$ | An exponentially distributed variable, with density $f(\tau) = (1/2\sigma^{2n}_\omega) e^{-\tau/2\sigma^2_\omega}$, $\mathbb{E}[\tau] \leq n 2\sigma^2_\omega$. |
| $T_{Eve,i}$ | Eve's transmittance on the Gaussian sub-channel $\mathcal{N}_i$, $T_{Eve,i} = \mathrm{Re}\, T_{Eve,i} + \mathrm{i}\, \mathrm{Im}\, T_{Eve,i} \in \mathcal{C}$, $0 \leq \mathrm{Re}\, T_{Eve,i} \leq 1/\sqrt{2}$, $0 \leq \mathrm{Im}\, T_{Eve,i} \leq 1/\sqrt{2}$, $0 \leq |T_{Eve,i}|^2 < 1$. |
| $H_{diff}(x)$ | Differential entropy of the continuous-variable $x$. |
| $H_{diff}(x|y)$ | Conditional differential entropy for input continuous-variable $x$, and output continuous-variable $y$. |
| $C_{sum}(\mathcal{N})$ | Sum capacity, the total throughput over the $l$ sub-channels of $\mathcal{N}$ at a constant modulation variance $\sigma^2_\omega$. |
| $C_{sym}(\mathcal{N})$ | Symmetric capacity, the maximum common rate at which all users can reliably transmit information over the $l$ sub-channels of $\mathcal{N}$. |
| $R_k$ | Transmission rate of user $U_k$. |
| $R_{sum}(\mathcal{N})$ | Sum rate, the total rate over the $l$ sub-channels of $\mathcal{N}$ at a constant modulation variance $\sigma^2_\omega$. |
| $R_{sym}(\mathcal{N})$ | Symmetric rate, the common rate at which all users can |

| | | |
|---|---|---|
| | | reliably transmit information over the $l$ sub-channels of $\mathcal{N}$. |
| | $\mathsf{C}$ | The $\mathcal{H}$ convex hull of independent input distributions. |
| | $C_1,...,C_K$ | Corner points of the capacity region $\mathsf{C}$ of $K$ users, $U_{1,...,K}$. |
| | $d_i$ | A $d_i$ subcarrier in an AMQD block. For subset $\mathcal{A} \subseteq K$ with $|\mathcal{A}|$ users and $l$ Gaussian sub-channels, can be rewritten as $d_i = \frac{1}{\sqrt{n}} \sum_{k=1}^{|\mathcal{A}|} z_k e^{\frac{-i2\pi ik}{n}}, i = 1,...,l$. |
| | $\nu_{\min}$ | The $\min\{\nu_1,...,\nu_l\}$ minimum of the $\nu_i$ sub-channel coefficients, where $\nu_i = \sigma_{\mathcal{N}}^2 / |F(T_i(\mathcal{N}_i))|^2$ and $\nu_i < \nu_{Eve}$. |

## S.4 Abbreviations

| | |
|---|---|
| AMQD | Adaptive Multicarrier Quadrature Division |
| BS | Beam Splitter |
| CV | Continuous-Variable |
| CVQFT | Continuous-Variable Quantum Fourier Transform |
| CVQKD | Continuous-Variable Quantum Key Distribution |
| DR | Direct Reconciliation |
| DV | Discrete Variable |
| FFT | Fast Fourier Transform |
| HOM | Homodyne (measurement) |
| HET | Heterodyne (measurement) |
| IFFT | Inverse Fast Fourier Transform |
| MQA | Multiuser Quadrature Allocation |
| QKD | Quantum Key Distribution |
| RR | Reverse Reconciliation |
| SNR | Signal to Noise Ratio |
| SVD | Singular Value Decomposition |